\documentclass[useAMS,usenatbib]{mn2e}

\usepackage{amssymb,amsmath,amsfonts}
\usepackage{graphicx}
\usepackage{epsfig,color} 
\usepackage{natbib}
\usepackage{lscape}
\usepackage[T1]{fontenc}
\usepackage{aecompl}
\usepackage{times}
\usepackage{mathtools}

\def\lsim{ \lower .75ex \hbox{$\sim$} \llap{\raise .27ex \hbox{$<$}} }

\title[]{A weak gravitational lensing recalibration of the scaling
  relations linking the gas properties of dark halos to their mass}

\author[Wang et al.]{Wenting Wang$^{1}$\thanks{E-mail:bilinxing.wenting@gmail.com}, Simon D.M. White$^{2}$, Rachel Mandelbaum$^{3}$, Bruno Henriques$^{2,4}$, \and
Michael E. Anderson$^{2}$, Jiaxin Han$^{1}$\\
  {}$^{1}$Institute for Computational Cosmology, University of Durham, South Road, Durham, DH1 3LE, UK\\
  {}$^{2}$Max Planck Institut fur Astrophysik, Karl-Schwarzschild-Str. 1, 85741 Garching b. M\"unchen, Germany\\
  {}$^{3}$McWilliams Center for Cosmology, Department of Physics, Carnegie Mellon University, Pittsburgh, PA 15213, USA\\
  {}$^{4}$Institute for Astronomy, Department of Physics, ETH Zurich, 8093 Zurich, Switzerland\\
}

\begin{document}

\maketitle
\begin{abstract}
We use weak gravitational lensing to measure mean mass profiles around
Locally Brightest Galaxies (LBGs). These are selected from the
SDSS/DR7 spectroscopic and photometric catalogues to be brighter than
any neighbour projected within 1.0 Mpc and differing in redshift by
$<1000$~km/s. Most ($> 83\%$) are expected to be the central galaxies
of their dark matter halos. Previous stacking analyses have used this
LBG sample to measure mean Sunyaev-Zeldovich flux and mean X-ray
luminosity as a function of LBG stellar mass.  In both cases, a
simulation of the formation of the galaxy population was used to
estimate effective halo mass for LBGs of given stellar mass, allowing
the derivation of scaling relations between the gas properties of
halos and their mass. By comparing results from a variety of
simulations to our lensing data, we show that this procedure has
significant model dependence reflecting: (i) the failure of any given
simulation to reproduce observed galaxy abundances exactly; (ii) a
dependence on the cosmology underlying the simulation; and (iii) a
dependence on the details of how galaxies populate halos. We use our
lensing results to recalibrate the scaling relations, eliminating most
of this model dependence and explicitly accounting both for residual
modelling uncertainties and for observational uncertainties in the
lensing results. The resulting scaling relations link the mean gas
properties of dark halos to their mass over an unprecedentedly wide
range, $10^{12.5}<M_{500}/ \mathrm{M_\odot}<10^{14.5}$, and should
fairly and robustly represent the full halo population.

\end{abstract}

\nokeywords

\section{Introduction}

Recent measurements of fluctuations in the microwave background have
determined the fraction of all matter that is baryonic to high
accuracy, $\Omega_b/\Omega_m = 0.158 \pm 0.002$, according to
\cite{2015arXiv150201589P}. At low redshift, the baryon fraction of
the richest galaxy clusters is close to this value, with $\sim$85\% 
of the material in the form of X-ray emitting gas and only 
$\sim$15\% in the form of stars \citep[e.g.][]{2008MNRAS.383..879A}. 
In less massive systems, the detected baryon fraction is lower and 
is dominated, for individual galaxy halos, by the stars in the central 
galaxy. This contribution maximises at about 4\% within the virial radius 
for galaxies similar in mass to the Milky Way \citep[e.g.][]
{Mandelbaum2006a,2010MNRAS.402.1796W,2010MNRAS.404.1111G}. In the low 
redshift universe as a whole, stars account for only a few percent of 
the expected baryons \citep{2009MNRAS.398.2177L}. Most remain undetected 
and are thought to be in the intergalactic medium, either photoionised 
by the metagalactic UV background, or heated, enriched and ejected from
galaxy halos by feedback from star formation and/or AGN activity
\citep[e.g.][]{1999ApJ...514....1C,2010MNRAS.405.1025M,
2012ApJ...754..116M,2014MNRAS.444.1518V,2015MNRAS.446..521S}.

Recent stacking results based on the Planck sky maps
\citep{2013A&A...557A..52P} and on the Rosat All Sky Survey
\citep{2015MNRAS.449.3806A} have substantially extended the halo 
mass range over which signals from associated hot gas have been
detected. Both studies stacked signals as a function of stellar 
mass around a sample of Locally Brightest Galaxies (LBGs) taken 
from the Seventh Data Release of the Sloan Digital Sky Survey
\citep[SDSS/DR7][]{2009ApJS..182..543A}. An independent 
stacking analysis of Planck data by \cite{2015ApJ...808..151G} 
recently confirmed the  original estimates of mean SZ signal as 
a function of LBG stellar mass.
These galaxies are selected to be brighter than any of their 
neighbours and are predominantly the central galaxies of their 
dark halos. 

The measured Sunyaev-Zeldovich (SZ) and X-ray signals as a function 
of LBG stellar mass can be converted into scaling relations between 
halo mass and halo gas properties provided that effective halo mass 
is known as a function of LBG stellar mass. Both \cite{2013A&A...557A..52P} 
and \cite{2015MNRAS.449.3806A} estimated this relation by forward 
modelling of sample selection and signal measurement using a simulation 
of galaxy population evolution from \cite{2013MNRAS.428.1351G}. 
This simulation was tuned to reproduce the observed stellar mass 
function in the SDSS, so this procedure amounts to an 
abundance-matching calibration of halo mass. Both scaling relations 
were found to be unbroken power-laws over the range 
$10^{12.5}<M_{\mathrm halo}/\mathrm{M_\odot}<10^{14.5}$, 
but whereas the relation for SZ flux was consistent with detection 
of the cosmic baryon fraction in hot gas at all halo masses
\citep[e.g.][]{2011A&A...536A..11P, 2012ApJ...754..119M}, that
for the X-ray flux was substantially steeper than self-similar
behaviour would require \citep[e.g.][]{2007ApJ...658..917D,
2008MNRAS.387L..28R,2009ApJ...692.1033V,2014MNRAS.439..611W}, 
suggesting that the concentration of baryons within halos must 
decrease with decreasing halo mass, i.e., the hot gas 
distribution within lower mass halos is less centrally 
concentrated than those in massive clusters. 
\cite{2015MNRAS.451.3868L} showed that when resolution effects in 
the observations are taken into account, this behaviour can be 
reproduced by simulations with plausible amounts of AGN feedback.

The results of \cite{2013A&A...557A..52P} and \cite{2015MNRAS.449.3806A} 
are intriguing because they seem to indicate the current state of a 
significant fraction of the baryons which had previously been ``missing''. 
It is, however, a matter of concern that the mass calibration of their 
scaling relations rests on a specific simulation of the formation and 
evolution of the galaxy population.  The goal of this paper is 
four-fold: (i) to check the halo mass calibration provided by the model 
of \cite{2013MNRAS.428.1351G} by comparing its predictions with weak
gravitational lensing measurements for the LBG sample used in the
stacking analyses; (ii) to transfer the mass calibration of these
scaling relations to one coming directly from the lensing
measurements, thus substantially reducing its model dependence; (iii)
to evaluate the residual model dependence of this calibration and its
origin by comparing results for a large number of different galaxy
population simulations; and (iv) to combine the resulting estimate of
residual systematic uncertainties with uncertainties coming from noise
in the lensing measurements to obtain realistic error bars on the
parameters of the recalibrated scaling relations.

The plan of our paper is the following. In Sec.~\ref{sec:obs}, we
describe the observational samples and the methods we use to measure
differential lensing surface density profiles and two-point
autocorrelation functions. Our N-body simulations and the galaxy
formation models based on them are introduced in
Sec.~\ref{sec:sim}. We present results for differential lensing
surface density profiles for SDSS LBGs in Sec.~\ref{sec:results} and
compare them with predictions from models with varying cosmologies,
N-body realisations and physical models for galaxy formation.  Galaxy
clustering measurements for the SDSS LBGs are compared with various
models in Sec.~\ref{sec:clustering}. Finally, we recalibrate the
scaling relations of \cite{2013A&A...557A..52P} and
\cite{2015MNRAS.449.3806A} in Sec.~\ref{sec:scaling}, providing new
relations which account consistently for the uncertainties both in
effective halo mass and in stacked SZ and X-ray flux for each bin of
LBG stellar mass. In a companion paper, \cite{2015M} compare lensing
signals for active and passive (e.g. blue and red) subsamples of our
LBG sample, finding red LBGs to have significantly more massive halos
than blue ones of the same stellar mass, while the same conclusion
is reached by \cite{ZandM2015b} through halo modelling to both 
lensing and clustering measurements. When quoting observational
results, we adopt as our fiducial cosmological model the
first-year Planck cosmology \citep[with present values of the 
Hubble constant $H_0=67.3\mathrm{\ km\ s^{-1}/Mpc}$, of the matter 
density $\Omega_m=0.315$ and of the cosmological constant 
$\Omega_\Lambda=0.685$]{2014A&A...571A..16P}. Our simulations 
assume a variety of other cosmologies as described 
in Sec.~\ref{sec:sim}. In the following we will define the 
reduced Hubble parameter h as 
$\mathrm{h}=H_0/100\mathrm{\ km\ s^{-1}/Mpc}$.

\section{Observational Samples and Techniques}
\label{sec:obs}

\begin{figure}
\epsfig{figure=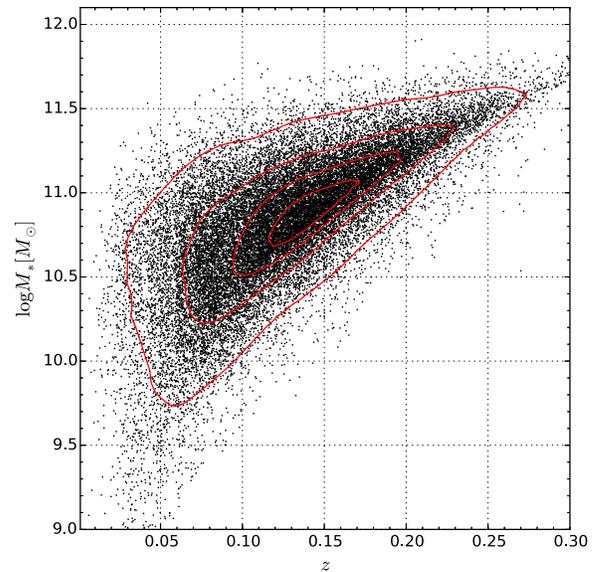,width=0.49\textwidth}
\caption{A scatter plot of our sample of locally brightest galaxies in
  the redshift -- stellar mass plane. Red equidensity contours enclose
  10\%, 30\%, 60\% and 90\% of the galaxies. To avoid overcrowding only
  10\% of the sample galaxies are plotted.}
\label{fig:massz}
\end{figure}

\subsection{The locally brightest galaxy sample}
In order to define a large sample of galaxies that are almost 
all centrally located and dominant within their dark matter halos, we
select locally brightest galaxies (hereafter LBGs) from the New York
University Value Added Galaxy Catalogue
\citep[][VAGC]{2005AJ....129.2562B}, which is based on the seventh
data release of the Sloan Digital Sky Survey
\citep[][SDSS/DR7]{2009ApJS..182..543A}.  This parent catalogue
contains 602,251 galaxies with high quality spectra and is flux
limited at $r=17.7$ ($r$-band, extinction-corrected Petrosian magnitude).

LBGs are selected to be galaxies that are brighter in $r$ than all
other catalogue members projected within 1.0 Mpc and with redshift
differing by less than 1000 km/s. The SDSS spectroscopic sample is
incomplete due to exclusion effects when placing fibres in crowded
fields.  To ensure that no candidates have brighter companions without
spectroscopy, we have used a photometric redshift catalogue
\citep[][photoz2]{2009MNRAS.396.2379C} based on SDSS/DR7 photometry to 
look for additional companions missed in the spectroscopic sample. We 
eliminate any candidate with a companion in this catalogue of equal or 
brighter $r$-band magnitude and projected within 1.0 Mpc, unless 
the photometric redshift distribution of the potential companion is 
inconsistent with the spectroscopic redshift of the candidate.\footnote{
Instead of providing a single value of photometric redshift, the 
photoz2 catalogue provides the probability distribution over
$0<z<1.47$. We consider the photometry of the neighbour to be 
inconsistent with its being a true companion if the probability 
of being at a redshift less than or equal to that of the candidate 
is less than 0.1, which means the candidate lies in the low 
redshift tail of the photometric redshift probability distribution 
for the companion, and it is thus unlikely to be a true companion. 
We did not consider the case when the candidate lies in the upper 
tail of the distribution because the spectroscopic galaxies mostly 
have $z<0.3$.}

After this selection 279,343 LBGs remain. Of these, 161,791 have 
stellar mass greater than $10^{10.8}\mathrm{M_\odot}$ (see 
Sec.~\ref{sec:mass} for details about how the stellar masses are 
measured). This is the limit to which SZ and X-ray signals have been 
clearly detected in the stacking analyses of \cite{2013A&A...557A..52P} 
and \cite{2015MNRAS.449.3806A}, hereafter P13 and A15, respectively. 
The LBG sample used here differs slightly from that used in these 
earlier studies, despite being selected according to the same criteria, 
because in this paper we interpret all data using the cosmology of 
\cite{2014A&A...571A..16P} whereas the earlier papers adopted that 
of WMAP7 \citep{2011ApJS..192...18K}. This slightly changes the 
isolation criterion in projected separation, and also increases the 
stellar mass assigned to each SDSS galaxy by about 10\% (because 
of the switch from ${\mathrm h}=0.704$ to ${\mathrm h}=0.673$).

Redshift and colour distributions of the selected LBG sample are shown
in Fig.~1 of P13. These are very similar to the corresponding
distributions for the parent sample, except that the blue fraction of
LBGs is significantly enhanced for objects with stellar mass below
that of the Milky Way, shifting the population to slightly higher
redshift. Redshift distributions for the specific stellar mass bins
used for SZ and X-ray stacking are shown in Fig.~2 of A15, and the
numbers, mean distances, estimated mean halo masses and other
properties of these subsamples are given in Table 1 of that
paper. The mean redshift rises from $z\sim 0.07$ at $M_*\sim
10^{10}\mathrm{M_\odot}$ to $z\sim 0.3$ at $M_*\sim
10^{12}\mathrm{M_\odot}$. In Fig.~\ref{fig:massz}, we show a scatter
plot of our sample in the redshift -- stellar mass plane. Red
equidensity contours enclose 10\%, 30\%, 60\% and 90\% of the
points. Clearly more massive LBGs are biased to higher redshifts as a
result of the SDSS flux limit, and at any given stellar mass fewer 
than half of the objects lie at redshifts where the sample can be
considered complete. We will discuss in Sec.~\ref{sec:sdssclustering}
and Sec.~\ref{sec:modelstack}. how we account for such incompleteness.

P13 and A15 used a galaxy formation simulation (denoted G13--W7$^\prime$ 
below) from \cite{2013MNRAS.428.1351G} to explore the expected relation 
of the LBGs to the dark halos in which they reside. Fig.~2 
of P13 shows that at every stellar mass, more than 83\% of LBGs should 
be the central galaxies of their halos -- of the remaining ``satellites'' 
two thirds are included because they are more luminous than the central 
galaxy of their halo. A scatter plot of main halo mass against 
stellar mass in Fig.~3 of P13 demonstrates that while a well-defined 
relation is predicted, it has substantial scatter, a significant 
high-mass tail due to the remaining satellites, and a noticeable offset 
between star-forming and passive systems. Histograms of the expected 
halo mass at each LBG stellar mass are given in Fig.~B.1 of P13. These 
are roughly lognormal with a FWHM of about an order of magnitude. 
Histograms of the projected offsets of satellite systems from halo 
centre are shown as a function of stellar mass in Fig.~C.2 of P13. 
These are typically a few hundred kpc.

The distribution of offsets from the true halo centre for 
satellite LBGs are presented in Fig.~C2 of P13, where it is also  
shown that a more strictly selected sample of LBGs that are locally 
brightest within $r_p=$2Mpc (2000km/s along the line-of-sight) gives 
almost the same distribution of offsets. With this more strictly selected 
sample of LBGs, the total number has decreased by 30\% for LBGs that are 
more massive than $\log_{10}M_\ast=11$. However, the sample purity of 
true central galaxies is only slightly increased to about 87\% at 
$\log_{10}M_\ast \sim11$. The selection criteria in P13, A15 and this 
paper are thus a compromise between sample size and purity. The 
robustness of the scaling relations against changes in the isolation 
criteria has been tested in P13 using this more strictly selected 
sample, showing that the scaling relation is insensitive 
to the isolation criteria.

\subsection{Stellar mass estimation and a reference galaxy sample}
\label{sec:mass}

The stellar masses used in our analysis were all estimated with the
K-correct software of \cite{2007AJ....133..734B} by fitting stellar
population synthesis models to K-corrected, five-band photometry from
SDSS assuming a \cite{2003PASP..115..763C} stellar initial mass
function. K-correct\footnote{http://howdy.physics.nyu.edu/index.php/Kcorrect} 
uses more than 400 spectral templates, of which most correspond to 
instantaneous burst models from \cite{2003MNRAS.344.1000B} based on 
the Padova1994 isochrones \citep{1994A&AS..104..365F,1994A&AS..105...29F,
1994A&AS..105...39F}.

\cite{2013ApJ...767...50M} compared these stellar masses to
independently derived values based on 12-band UV to mid-infrared
photometry.  Comparison has also been made with MPA/JHU stellar masses
\footnote{The data were produced by a collaboration of researchers 
(currently or formerly) from the Max-Planck Institute for Astrophysics 
(MPA) and the John Hopkins University (JHU).} based on the 
$H\delta_A$ and $D_n(4000)$ spectral indices \citep{2003MNRAS.341...33K} 
in addition to SDSS photometry, as well as with masses from an extension 
of this method by \cite{2007ApJS..173..267S}. Differences between 
these methods are almost always less than 0.1 dex, with K-correct 
tending to give systematically slightly higher (lower) values than 
the other schemes at low (high) stellar mass. A fitting formula 
quantifying the difference between K-correct and MPA/JHU stellar 
masses is available in \cite{2009MNRAS.398.2177L}. Note P13 and 
A15 used the same K-correct stellar masses to derive the effective 
halo mass for the SZ/X-ray scaling relations. Any systematic 
uncertainties beyond the stellar mass, is supposed to be self-consistently 
removed through the lensing recalibration (see Sec.~\ref{sec:scaling}).

In Sec.~5 of this paper we analyse the clustering of LBGs by computing
cross-correlations between stellar mass-limited subsamples and a
reference galaxy sample. We choose the latter to be all galaxies in
the parent spectroscopic sample that are more massive than
$10^{10}\mathrm{M_\odot}$.

\subsection{The source catalogue for lensing}
The source galaxy catalogue used for our gravitational lensing study
has been described in detail in \cite{2012MNRAS.425.2610R}, and thus
we introduce it only very briefly here. Catalogue construction uses
the re-Gaussianization method \citep{2003MNRAS.343..459H} to correct
for the point spread function (PSF) which affects the observed 
shape of galaxies. Thanks to this PSF correction scheme, 
\cite{2012MNRAS.425.2610R} and \cite{2012MNRAS.420.1518M,2013MNRAS.432.1544M} 
were able to reduce systematic errors significantly compared with 
previous SDSS source catalogues. More recently, \cite{2015MNRAS.446.1356H}
developed an improved maximum-likelihood approach, and with the same 
source catalogue they measured relations between halo mass and group 
luminosity with high accuracy. The source galaxy catalogue of 
\cite{2012MNRAS.425.2610R} covers 9243 deg$^2$ and fully overlaps 
the SDSS/DR7 footprint within which our LBGs are selected. It includes 
all objects with extinction corrected $r$-band model magnitudes brighter 
than 21.8. Photometric redshifts of the sources have been estimated 
using the SDSS five-band photometry and the Zurich Extragalactic 
Bayesian Redshift Analyzer \citep[ZEBRA;][]{2006MNRAS.372..565F}. 
The effect of using photometric redshifts in lensing analyses like 
ours has been quantified by \cite{2012MNRAS.420.3240N}.

\subsection{Measuring the differential density profile}
\label{sec:lensing}

In this study we follow the method described in
\cite{2005MNRAS.361.1287M} and \cite{2006MNRAS.370.1008M} to calculate
mean differential projected density profiles for the total mass
distribution stacked around our selected LBGs. We point the reader to
\cite{2005MNRAS.361.1287M} for full details of this method, and only
summarise its main features in this section. Some further details are
given in the companion paper, \cite{2015M}.

The mean differential surface mass density profile, $\Delta
\Sigma(r_p)$, is defined as the difference between the mean surface
density enclosed by projected radius $r_p$ (denoted
$\bar{\Sigma}(<r_p)$ and the mean surface density at that radius
(denoted $\Sigma(r_p)$). The quantity $\Delta \Sigma(r_p)$ can be
related to the mean tangential shear, $\gamma_t$, and the lensing
critical density, $\Sigma_c$,
\begin{equation}
\Delta \Sigma=\gamma_t \Sigma_c, 
\end{equation}
where $\Sigma_c$ is defined as
\begin{equation}
 \Sigma_c=\frac{c^2}{4\pi G}\frac{D_s}{D_{ls} D_l}.
\end{equation}
$D_l$ and $D_s$ refer to the angular diameter distances of lens and
source, respectively, and $D_{ls}$ is the angular diameter distance
between lens and source.  We use physical separations in our analysis
rather than comoving separations.

The mean tangential shear can be related to the directly measurable
mean tangential ellipticity, $e_t$, of source galaxies, the two
differing by a factor of twice the shear responsivity, defined as the
response of the ensemble averaged ellipticity to a small shear 
\citep[see][]{2002AJ....123..583B}. Thus, the mean differential
surface mass density profile can be estimated from the mean tangential
ellipticity as follows:
\begin{equation}
 \Delta\Sigma=\frac{\sum_i w_i (e_t\Sigma_c)_i}{2R \sum_i w_i}, 
\end{equation}
where $R$ is the shear responsivity. This is essentially a weighted
average of $e_t\Sigma_c$ over the set of stacking centres. The weight,
$w_i$, involves the inverse square of the total noise, composed of 
measurement noise, $\sigma_{e_t,i}$ and shape noise, $\sigma_{SN}$, 
\begin{equation}
 w_i=\frac{\Sigma_{c,i}^{-2}}{\sigma_{e_t,i}^2+\sigma_{SN}^2}.
\label{eqn:weight}
\end{equation}

Signals are measured around both real galaxies and random points.
The signal around random points should be subtracted from that 
around real galaxies to remove contributions from systematic shear, 
though on the scales probed in our analysis such systematics are 
not significant. Moreover, the signal must be multiplied 
by a factor of $B(r)=n(r)/n_\mathrm{rand}(r)$, which is the 
the ratio of the number density of sources around real galaxies 
relative to that around random points. This accounts for dilution 
of the signal by physically-associated pairs. In the end, we 
estimate the errors of $\Delta \Sigma(r_p)$ through jackknife 
resampling.

For all the lensing measurements in this paper, we follow the
convention of this subfield and quote values for $r_p$ in physical
units; this contrasts with the next subsection and with Section 5
where we follow the convention of the galaxy clustering subfield and
quote values for $r_p$ in comoving units.

\subsection{Clustering measurements}
\label{sec:sdssclustering}

We measure the projected cross-correlation function between LBGs and
our reference galaxy sample using the Landy-Szalay estimator 
\citep{1993ApJ...412...64L},
\begin{equation}
 \xi(r_p,r_\pi)=\frac{CD(r_p,r_\pi)-CR_D(r_p,r_\pi)-DR_C(r_p,r_\pi)}{R_C R_D(r_p,r_\pi)}+1,
\end{equation}
where $CD$ is the pair count between central LBGs and the reference
sample, $CR_D$ is the pair count between LBGs and the random sample
corresponding to the reference sample, $DR_C$ is the pair count
between the reference sample and the random sample corresponding to
LBGs, and $R_C R_D$ is the pair count between the two random samples.
$r_p$ and $r_\pi$ are the projected and the line-of-sight 
components of the (comoving) separations, respectively.

This estimator thus uses separate random samples corresponding to the
LBGs and the reference galaxies. These are constructed to have exactly
the same sky coverage, number density, redshift and magnitude
distribution as the corresponding real galaxies, but the sky
coordinates are randomised within the SDSS footprint.

The SDSS galaxy samples are flux-limited in the $r$-band. As a result,
fainter galaxies are only included at lower redshifts
(Fig.~\ref{fig:massz}). In order to get volume-limited statistics to
allow comparison between samples and with the simulations, we weight
individual pairs in the above estimator by
$1/\mathrm{min}(V_\mathrm{max,1,i},V_\mathrm{max,1,j})$, where
$V_\mathrm{max,1,i}$ and $V_\mathrm{max,1,j}$ are the total volumes
within which galaxies $i$ and $j$ could be observed, given the SDSS
footprint and apparent magnitude limit. This weighting is adopted for
all variations of pair counts, i.e. for $CD$, $CR_D$, $DR_C$ and $R_C
R_D$. It results in volume-weighted statistics provided the mean
number per unit volume of correlated pairs with given physical
properties (e.g. individual galaxy luminosities and stellar masses and
pair separation) varies at most weakly with redshift over the range
$0<z<0.15$ which contributes most of the weight to the
cross-correlation measurements. We have checked that this is indeed
the case for the models we discuss in the next section.

After obtaining $\xi(r_p,r_\pi)$, we integrate it to get the projected
correlation function,
\begin{equation}
 w_p(r_p)=\int \xi(r_p,r_\pi)\mathrm{d}r_\pi,
\end{equation}
using a depth of $\pm$ 40 Mpc/h for the line-of-sight integral. This
is adequate for projected distances of 10~Mpc or less, as used in our
analysis \citep[see][]{2009MNRAS.397.1862P}. We estimate error bars
for $w_p(r_p)$ through bootstrap resampling 100 subsamples from the
full LBG sample. Note, for each subsample the contribution to the
counts from individual LBG-reference galaxy pairs can change due to
the redistribution of weights assigned to the LBGs. We did not
bootstrap companions in the reference sample. As noted above, when
compiling pair counts and quoting results for correlation functions we
will take $r_p$ to refer to a particular {\it comoving} projected
separation.


\section{Galaxy Formation Simulations}
\label{sec:sim}

Realistic simulations of cosmic structure formation are very valuable
when interpreting stacking measurements like those of this paper. To
the extent that they represent faithfully both the population of
objects used as stacking centres and the environments in which they
form, such simulations can characterise the selection effects
introduced by observational sample construction and the averaging
process inherent in the subsequent stacking. P13 and A15 interpreted
their SZ and X-ray stacking analyses of LBGs with a simulation from
\cite{2013MNRAS.428.1351G} that used semi-analytic methods to follow
the formation and evolution of galaxies in a high-resolution N-body
simulation of a $\Lambda$CDM cosmology. Simulations of this type can
be carried out in large volumes and tuned to match the properties of
the observed galaxy population quite closely. Here we use this and a
number of other simulations to assess how the quantitative
interpretation of our results is affected by uncertainties in
cosmological parameters and in galaxy modelling assumptions.

\subsection{N-body simulations}
We use two very large dark-matter-only simulations to represent the
distribution of dark matter.  These are the Millennium Simulation
\citep[][MS]{2005Natur.435..629S} which in its original version
adopted cosmological parameters from the first-year data of WMAP
\citep[][WMAP1]{2003ApJS..148..175S}, and a second simulation
\citep[][MS-W7]{2013MNRAS.428.1351G} which is identical to the MS
except that the cosmological parameters were taken from the seventh
year analysis of WMAP \citep[][WMAP7]{Komatsu2011}. The fluctuation
phases for the MS-W7 initial conditions are taken from the public,
multi-scale, Gaussian, white-noise field (Panphasia) described in
\cite{2013MNRAS.434.2094J}, and are different from those used for the
MS. Table~\ref{tbl:sim} provides basic numerical and cosmological
parameters for the two simulations. Both follow the same number of
particles in a cube of the same side length from redshift 127 to the
present day. The dark matter particle mass is slightly larger in the
MS-W7 than in the MS.

\begin{table}
\caption{Numerical and cosmological parameters for the simulations 
used in our analysis}
\begin{center}
\begin{tabular}{lccc}\hline\hline
           & \multicolumn{1}{c}{MS} & \multicolumn{1}{c}{MS-W7} & \multicolumn{1}{c}{MS-P$^\prime$} \\ \hline
box size [Mpc/h]  &  500    &   500    &  480\\
softening [kpc/h] &  5  &   5   & 4.8  \\
particle mass [$\mathrm{M_\odot/h}$] &  $8.61\times10^8$ & $9.36\times10^8$ & $9.62\times10^8$ \\
particle number &  $2160^3$  &  $2160^3$  &  $2160^3$  \\
$\Omega_m$ &  0.25  & 0.272  & 0.315   \\
$\Omega_\Lambda$ & 0.75   &   0.728 & 0.685  \\
$\Omega_b$ &  0.045  &  0.0455 & 0.0487   \\
$H_0 [\mathrm{km s^{-1}/Mpc}]$ &  73.0  & 70.4 &  67.3   \\
$n_s$ &  1  & 0.961  & 0.96  \\
$\sigma_8$ &  0.9  &  0.807 & 0.829   \\
$\sigma_8 \Omega_m^{0.6}$ &  0.392  &  0.370 & 0.415   \\
\hline
\label{tbl:sim}
\end{tabular}
\end{center}
\end{table}

\subsection{Scaling to other cosmologies}
In addition to the cosmologies adopted for the original MS and MS-W7,
we would like the first-year Planck cosmology
\citep{2014A&A...571A..16P} to be represented among our galaxy
formation simulations. The scaling algorithm developed by
\cite{2010MNRAS.405..143A} and improved by \cite{2015MNRAS.448..364A}
allows the stored output of an N-body simulation to be scaled from the
cosmology in which it was originally carried out to a different
cosmology. This procedure involves three steps: rescaling
the box size, particle mass and velocities, relabelling the output
times, and rescaling the amplitudes of individual large-scale
fluctuation modes. For this paper we scale the original MS simulation
to both the WMAP7 and the Planck cosmologies, in the first case using
the parameters of Table~1 in \cite{2013MNRAS.428.1351G}, and in the
second the parameters of Fig.~1 of \cite{2015MNRAS.451.2663H}. We
scale the MS to the WMAP7 cosmology to enable direct comparison with
MS-W7 for the properties of interest in this paper.  Further
comparison of the direct and scaled simulations can be found in
\cite{2013MNRAS.428.1351G}.

Table~\ref{tbl:output} lists available simulation redshifts over the
range needed for our analysis. The outputs we use for the original MS
and MS-W7 are the same. After scaling the MS to the WMAP7 and Planck
cosmologies, the time coordinate is relabelled and the available
redshifts are different; for WMAP7, the new $z=0$ corresponds to
$z=0.28$ in the original simulation, while for Planck it corresponds
to $z=0.12$. The scaled simulation size and particle mass, and the new
cosmological parameters are given for the Planck case in
Table~\ref{tbl:sim}. For the WMAP7 case, the cosmological parameters
are the same as for MS-W7 but the scaled box size and particle mass
are $521.55\ \mathrm{Mpc/h}$ and $1.0062\times10^{9}\ \mathrm{M_\odot/h}$, 
respectively. In the next section, we introduce the galaxy formation 
models used in our analysis, and all models based on a scaled N-body 
simulation will be indicated by a prime ($'$) after their names.

\begin{table}
\caption{Output redshifts used for the MS (columns 1, 3 and 4) and for the
MS-W7 (column 2) for the various cosmologies adopted in our analysis}
\begin{center}
\begin{tabular}{cccc}\hline\hline
\multicolumn{1}{c}{WMAP1} & \multicolumn{1}{c}{WMAP7} & \multicolumn{1}{c}{WMAP7-scaled} & \multicolumn{1}{c}{Planck-scaled} \\ \hline
   0      &  0      &  0      &  0     \\
   0.04   &  0.04   &  0.04   &  0.05  \\
   0.09   &  0.09   &  0.08   &  0.08  \\
   0.12   &  0.12   &  0.12   &  0.11  \\
   0.17   &  0.17   &  0.17   &  0.18  \\
   0.24   &  0.24   &  0.22   &  0.22  \\
   0.28   &  0.28   &  0.27   &  0.26  \\
   0.32   &  0.32   &  0.33   &  0.31  \\
\hline
\label{tbl:output}
\end{tabular}
\end{center}
\end{table}

\subsection{Galaxy formation models}

\begin{table*}
\caption{The galaxy formation simulations used in our analysis, showing
their different combinations of astophysical modelling (rows) and 
background cosmology (columns).} 

\begin{center}
\begin{tabular}{lccc}\hline\hline
           & \multicolumn{1}{c}{WMAP1} & \multicolumn{1}{c}{WMAP7} & \multicolumn{1}{c}{Planck} \\ \hline
Guo et al. (2011) & $\mathrm{G11-W1}$ & $\mathrm{G11-W7'}$ & $\mathrm{G11-P'}$ \\
Guo et al. (2013) &     & $\mathrm{G13-W7}$,~$\mathrm{G13-W7'}$ & \\
Henriques et al. (2015)-p & $\mathrm{H15p-W1}$ &   & $\mathrm{H15p-P'}$ \\
Henriques et al. (2015)-f &      &      & $\mathrm{H15f-P'}$ \\
\hline
\label{tbl:model}
\end{tabular}
\end{center}
\end{table*}

We have used four recent versions of the Munich semi-analytic models
to follow the formation and evolution of galaxies in the evolving
halo/subhalo population of our four N-body simulations (the original
MS and MS-W7, together with the two scaled versions of the
MS). Altogether, we consider eight different galaxy formation
simulations, which are summarised in Table~\ref{tbl:model}. Among
these, $\mathrm{G11-W1}$,~$\mathrm{G13-W7}$ and $\mathrm{G13-W7'}$, 
and $\mathrm{H15f-P'}$ are the models published in
\cite{2011MNRAS.413..101G}, \cite{2013MNRAS.428.1351G} and
\cite{2015MNRAS.451.2663H}, respectively. For each of these, the
uncertain star formation and feedback efficiencies were tuned to
produce close fits to SDSS data on the stellar mass functions of
low-redshift galaxies, and the resulting model also reproduces
observed luminosity and autocorrelation functions quite well.  In the
case of $\mathrm{H15f-P'}$, \cite{2015MNRAS.451.2663H} used the
abundances and passive fractions of galaxies at $0.4 \leq z \leq 3$ as
additional constraints on model parameters.  Details of how well these
models reproduce the observed galaxy population can be found in the
original papers.

Our model suite thus includes three different cosmologies, WMAP1,
WMAP7 and Planck. The Planck cosmology is available only by using the
scaling algorithm introduced above, whereas for the WMAP7 cosmology we
have both a direct run and one obtained through scaling the MS.  These
three cosmologies are indicated by the columns of
Table~\ref{tbl:model}. 

Our model suite includes four different sets of galaxy formation
parameters following the recipes of \cite{2011MNRAS.413..101G},
\cite{2013MNRAS.428.1351G} and \cite{2015MNRAS.451.2663H}. These
correspond to the rows of Table~\ref{tbl:model}.  Note that we have
two versions of the \cite{2015MNRAS.451.2663H} model. H15p is a
preliminary version corresponding to that originally submitted and
posted on the preprint archive, while H15f is the final version
revised and retuned in response to comments from a referee.  Compared
with the preliminary version, the threshold between major and minor
mergers was shifted to produce a better match to the observed
morphology distribution of galaxies, and a minor problem with the
merger trees was fixed which had led to unphysically rapid growth of
the central galaxy of the most massive halo in the simulation.

Thus, $\mathrm{G11-W1}$,~$\mathrm{G11-W7'}$ and $\mathrm{G11-P'}$ have
exactly the same galaxy formation physics and parameters, but
different cosmologies. $\mathrm{G13-W7}$ and $\mathrm{G13-W7'}$ are
based on the same cosmology and galaxy formation physics (including
all associated parameters) but the former is implemented on MS-W7 and
the latter on a scaled version of the MS.  $\mathrm{H15p-W1}$ and
$\mathrm{H15p-P'}$ share the same galaxy formation physics, including
all parameters, but are based on WMAP1 and Planck cosmologies
respectively.  $\mathrm{H15f-P'}$ is an updated version of
$\mathrm{H15p-P'}$, as just noted.

\begin{figure}
\epsfig{figure=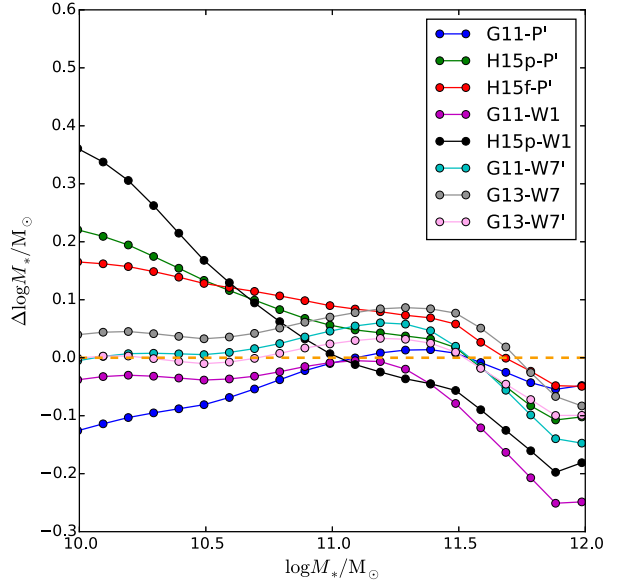,width=0.49\textwidth}
\caption{The correction to stellar masses, $\Delta \log
  M_\ast/\mathrm{M_\odot}$, needed to bring model stellar mass
  functions into exact agreement with the observed SDSS stellar mass
  function. The $x$-axis is the stellar mass after correction, and the
  correction is defined to be negative when it reduces the stellar
  masses assigned to galaxies. A horizontal orange dashed line marks 
  zero correction to guide the eye.}
\label{fig:cmpshiftMF}
\end{figure}

\begin{figure}
\epsfig{figure=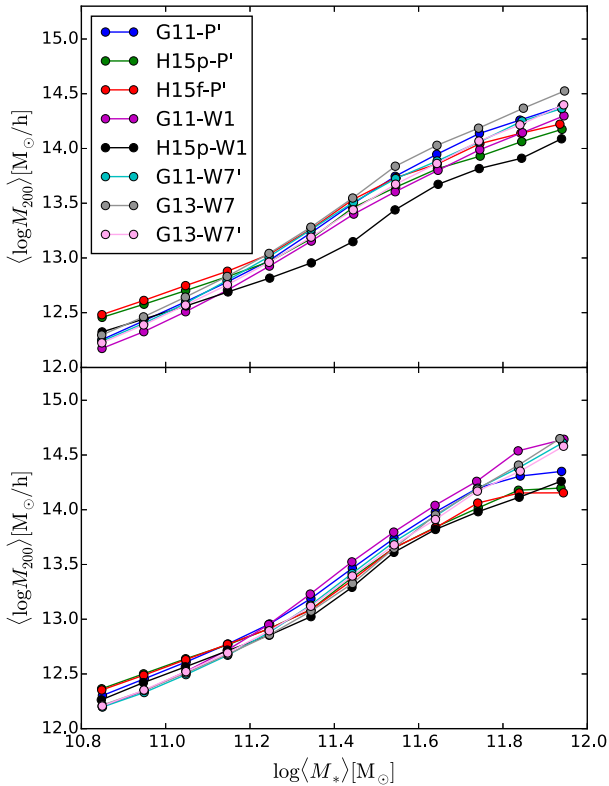,width=0.49\textwidth}
\caption{Mean halo mass, calculated as $\langle\log M_{200}\rangle$,
  as a function of stellar mass for LBGs in our eight galaxy formation
  simulations. The upper panel shows results from the original
  simulations, while the lower panel shows results after the
  corrections of Fig.~\ref{fig:cmpshiftMF} have been used to bring the
  stellar mass function of each model into exact coincidence with that
  of the SDSS.}
\label{fig:shiftMFplot}
\end{figure}

\subsection{Corrections to stellar mass through abundance matching}
\label{sec:corr}

Although the published semi-analytic models of the last section were
all tuned to give reasonable matches to the SDSS stellar mass
function, a number of other observational constraints such as their
luminosity functions, their size, morphology and kinematics
distributions, and their clustering were also considered when setting
up the physical modelling and determining parameters. In particular,
the parameters of the model of \cite{2015MNRAS.451.2663H} were
adjusted by MCMC sampling to fit not only the low redshift stellar
mass function, but also the abundance and passive fractions of
galaxies over the redshift range $0<z\leq 3$. Because of these
multiple constraints, the quality of the fit to the observed
low-redshift SDSS stellar mass function varies between models. This
can significantly affect the mass of the halos assigned to galaxies of
a specific stellar mass, particularly at high mass.

In order to separate differences in model predictions caused purely by
deviations from the observed stellar mass function, from differences
caused by cosmology or by other aspects of how (sub)halos are
populated with galaxies, we use an ``abundance matching'' method to
correct the stellar masses in each model so that it reproduces the
SDSS mass function exactly. We preserve the {\it ranking} of the model
galaxies in stellar mass, but shift them to match the mass function of
\cite{2009MNRAS.398.2177L}.  Explicitly, we take the cumulative number
density of galaxies at each value of $M_\ast$ from SDSS, and find the
stellar mass for which the model predicts the same cumulative
abundance.  The difference between SDSS and model stellar masses at
this abundance then gives the necessary correction as a function of
stellar mass. Fig.~\ref{fig:cmpshiftMF} gives the result of this
exercise. Notice that over most of the mass range shown the
corrections are quite small, demonstrating that the galaxy formation
simulations have indeed reproduced the observations quite
accurately. In particular, this is true for $\mathrm{G13-W7'}$, the
model used for comparison in the stacking analyses of P13 and A15.

Halo mass -- stellar mass relations before and after these corrections
to the stellar mass are shown in the top and bottom panels of
Fig.~\ref{fig:shiftMFplot}, respectively. Twelve stellar mass bins have
been chosen with a width of 0.1 dex in $\log M_\ast$. For each bin, we
calculate the mean log halo mass, $\langle \log M_{200}
\rangle$ and plot it against the log mean stellar mass, 
$\log\langle M_\ast \rangle$, for all eight galaxy formation 
simulations\footnote{For central BCGs, $M_{200}$ denotes the mass of 
a surrounding spherical region with mean enclosed density 200 times
the critical density of the Universe. For satellite BCGs, we use the
$M_{200}$ value of the associated central galaxy.}. We choose to use 
$\langle \log M_{200} \rangle$ instead of $\log \langle M_{200}
\rangle$ throughout the paper because, as we will see, halo mass
distributions at fixed stellar mass are broad and closer to lognormal
than to normal. In addition, this definition of mean mass is closer to
the effective halo mass implied by SZ, X-ray and lensing observations
of stacked BCGs than is $\langle M_{200}\rangle$ (see P13, A15 and
Sect.~\ref{sec:scaling} below). On the other hand, we choose 
to use $\log \langle M_\ast \rangle$ because for the narrow 
stellar mass bins used in this paper the halo mass distribution 
changes slowly over the bin. We have checked that using 
$\langle\log M_\ast \rangle$, $\log \langle M_\ast \rangle$ or the 
middle bin values gives very similar results.

For the bottom panel of Fig.~\ref{fig:shiftMFplot}, the stellar mass
of each simulated galaxy has been corrected by an amount taken from
Fig.~\ref{fig:cmpshiftMF}. The simulations compared here thus have
identical stellar mass functions. It is encouraging to see that the
scatter between simulations, although not entirely eliminated, is
significantly less than in the top panel. The standard deviation 
among the eight models in the upper panel and for the twelve stellar 
mass  bins are 0.104, 0.090, 0.073, 0.060, 0.068, 0.098, 0.120, 0.111, 
0.098, 0.109, 0.130 and 0.131 (from the least to the most massive bin), 
while for the bottom panel the values become 0.065, 0.065, 0.054, 
0.039, 0.036, 0.061, 0.070, 0.056, 0.073, 0.093, 0.136 and 0.198.
This demonstrates that much of the scatter between the original models 
can indeed be ascribed to imperfect matching of the observed stellar 
mass function. At the most massive end, however, the scatter remains 
substantial and even becomes bigger, while the models split 
into two groups after the correction: the Guo et al. group and
Henriques et al. group. This indicates that while models with similar
physical recipes differ mainly because of the varying degree to which
they fit the observed SDSS stellar mass functions, models with
differing physical recipes can produce different predictions, even
when they are applied to the same N-body simulation and produce the
same stellar mass function. This is because they are differently
sensitive to the detailed assembly histories of massive halos.
Note Fig.~\ref{fig:shiftMFplot} only shows the mean halo mass at 
fixed stellar mass. In Sec.~\ref{sec:results}, we will show the 
distribution of $\log M_{200}$, finding that for models with 
similar mean halo mass at given LBG stellar mass, the shape of 
the halo mass distribution is very broad and can differ substantially 
between models.

\begin{figure*}
\epsfig{figure=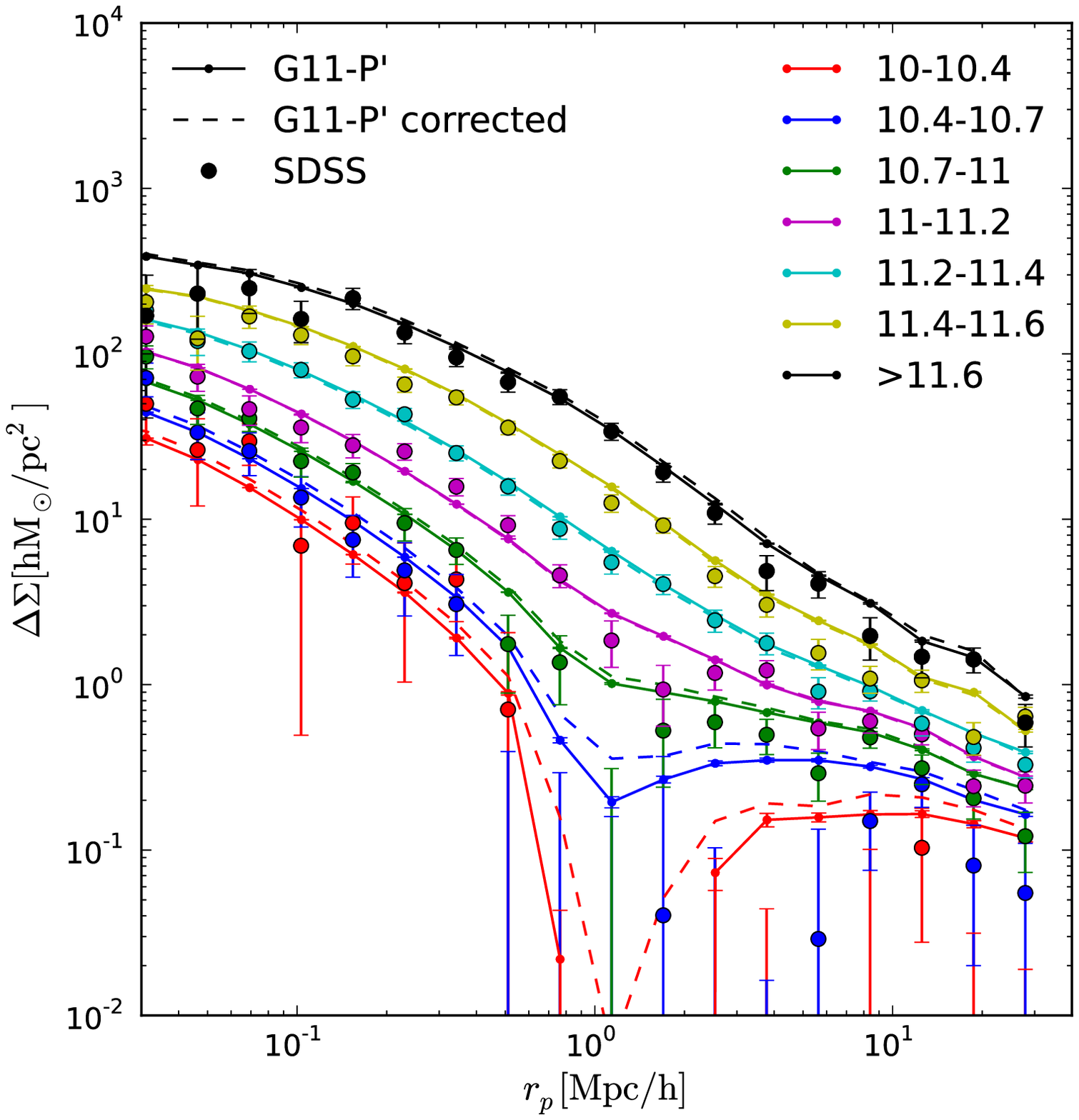,width=0.49\textwidth}%
\epsfig{figure=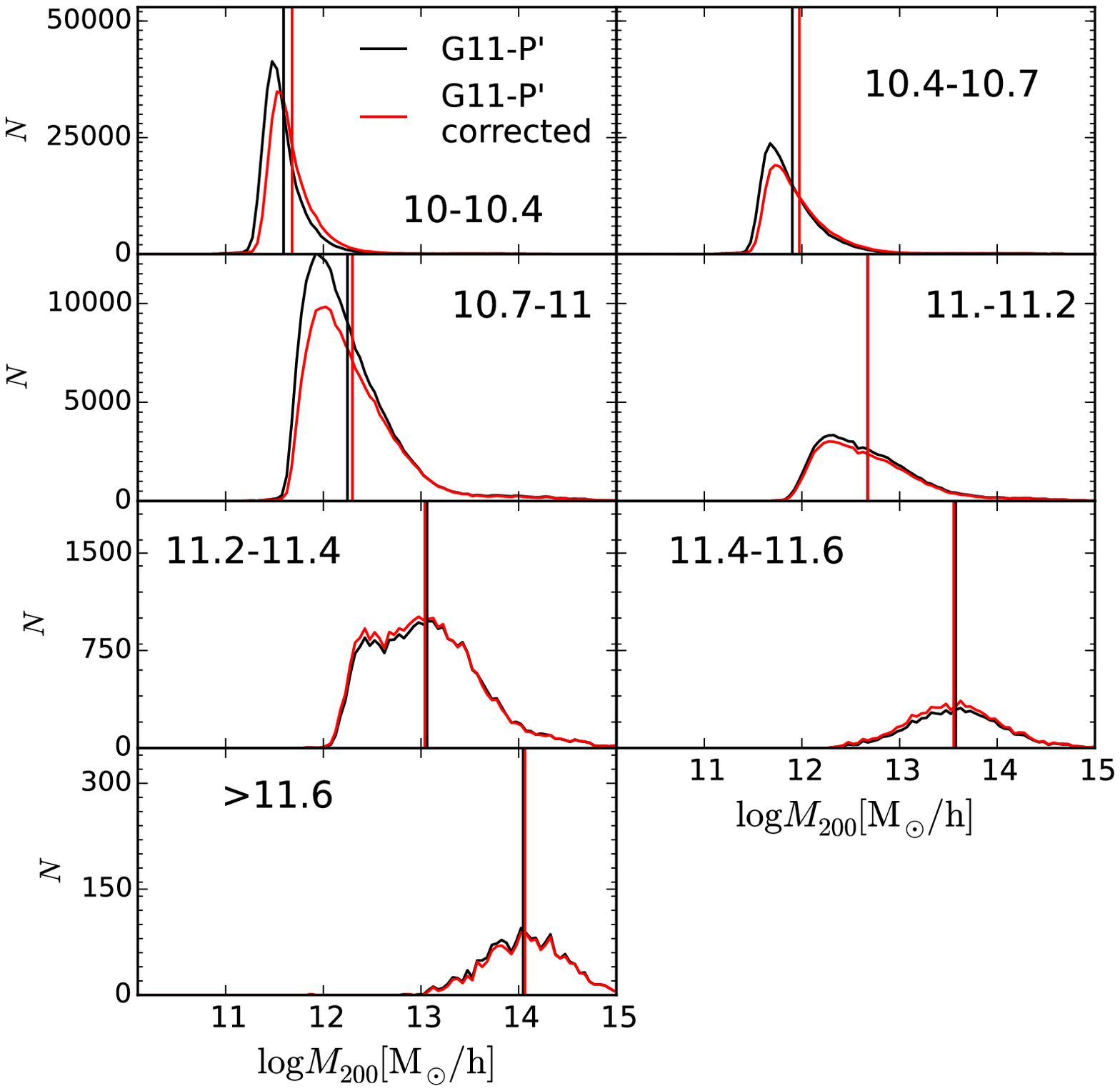,width=0.49\textwidth}
\caption{{\bf Left:} Differential surface density profiles centred on
  LBGs in the $\mathrm{G11-P'}$ model are compared to our
  observational results for SDSS (filled circles with error
  bars). Solid lines are for the original model, while dashed lines
  include corrections to bring the stellar mass function into exact
  agreement with SDSS. Lines and symbols are coloured based on the
  logarithmic range in stellar mass (in units of $\mathrm{M_\odot}$)
  as indicated by the legends. $r_p$ is given in physical (not
  co-moving) units in this and all subsequent plots of this type. 
  The large simulation box size enables us to achieve very 
  small errors for the model (comparable to the symbol size). Note 
  we only show the errors for solid lines. Errors for the  
  dashed lines are of comparable size. {\bf
    Right:} The host halo mass distributions for LBGs at $z=0$ within
  these same stellar mass bins and for the same model,
  $\mathrm{G11-P'}$.  Vertical lines mark the characteristic halo
  mass, $\langle \log M_{200} \rangle$ for each histogram. Black lines
  correspond to the original model, and red lines to the model after
  correction to reproduce the SDSS stellar mass function exactly.}
\label{fig:g11planck}
\end{figure*}

\subsection{Selecting locally brightest galaxies in the models}

We select LBGs in our simulations as in P13 and A15 using criteria
patterned closely on those used to select SDSS LBGs (see also
\citealt{2012MNRAS.424.2574W} and \citealt{2014MNRAS.442.1363W}).
Specifically, we project the simulation box along its $z$-axis
(representing the line-of-sight) and assign each model galaxy a
redshift based on its $z$-coordinate and velocity. Isolation criteria
in projected separation and redshift difference can then be applied to
the simulation in the same way as to the observational data.  Such
direct projection of the simulation box maximises statistical signal,
but fails to represent effects due to the flux limit of the real
survey, to the K-corrections needed to obtain rest-frame magnitudes
from the observations, to the incompleteness for close pairs caused by
fibre exclusion effects, to the complex geometry of the real survey,
and to the evolution of the real population across the redshift range
surveyed. \cite{2012MNRAS.424.2574W} and \cite{2014MNRAS.442.1363W}
have tested that, for many of these effects, the results obtained by
simply projecting the simulation box are unbiased compared to results
obtained from a full light-cone mock catalogue. As described below, we
combine data from a range of simulation outputs (see
Table~\ref{tbl:output}) to account consistently for the effects of
evolution and of the flux limit of the observational survey.

In the simulations we know whether each selected LBG is truly a
central galaxy.  As noted in Sect.~\ref{sec:obs}, the fraction of LBGs
that are central galaxies (i.e. the purity of the sample) is
predicted by $\mathrm{G13-W7'}$ to exceed 83\% at all stellar
masses. The maximum contamination ($\sim 17\%$) by satellites is
predicted to occur at stellar masses slightly above
$10^{11}\mathrm{M_\odot}$.  P13 checked those LBGs that are
satellites, finding that for $M_\ast \geq 10^{11}\mathrm{M_\odot}$
about two-thirds are brighter than the true central galaxies of their
halos. The remainder are fainter than their centrals, and are
considered locally brightest because they are projected more than 1
Mpc from their centrals (60\%) or have redshifts differing by more
than 1,000 km/s. This high level of purity allows us to consider
our LBG sample as a sample of central galaxies with only minor
contamination by satellites.

The projected offsets of satellite LBGs from true halo centre are
analysed in Appendix C of P13 and are typically
a few hundred kpc. It is shown there that a more strictly selected
sample of LBGs (selected to be locally brightest within a projected
separation of 2 Mpc and a redshift difference of 2000 km/s) has a very
similar offset distribution and, despite a 30\% reduction in sample
size, only a slightly improved level of purity, about 87\% for $M_\ast
\geq 10^{11}\mathrm{M_\odot}$.

\subsection{Stacked lensing density profiles in the simulations}
\label{sec:modelstack}

Our SDSS sample of LBGs is flux-limited in the $r$-band and hence
strongly biased towards higher stellar mass (actually higher
luminosity) at higher redshift (see Fig.~\ref{fig:massz}). The
redshift range spanned is approximately $0 < z < 0.35$, so that high
mass LBGs are seen at a systematically slightly earlier stage of
cosmic evolution, thus on a slightly different $M_\ast$ -- $M_{200}$
relation, than low mass systems. Moreover, the lensing signal measured
around each galaxy in a stack is weighted by the inverse square of the
(redshift-dependent) lensing critical density, $\Sigma_c$ (see
\cite{2006MNRAS.370.1008M} and Sec.~\ref{sec:lensing}). To compare our
simulations appropriately to our SDSS measurements, we thus need to
reproduce the stellar mass -- redshift distribution of the
observations and to assign the same weights to objects when stacking.
Our procedure to achieve this is as follows.

\noindent{(1) Each SDSS LBG is matched to ten simulated LBGs based on
  redshift, stellar mass and luminosity. Explicitly, an SDSS LBG is
  first matched to the simulation output which is closest to its
  redshift. Ten simulated LBGs are then picked from this output, with
  stellar mass\footnote{In this paper, stellar masses are always quoted
    in units of the solar mass, $\mathrm{M_\odot}$. The original
    stellar mass in the VAGC catalogue was given in units of
    $\mathrm{M_\odot} \mathrm{h}^{-2}$. To eliminate the Hubble
    constant, we adopt the Planck cosmology. The cosmologies of some
    of our simulations differ, however, so we scale the stellar masses
    (in units of $\mathrm{M_\odot}$) of simulated galaxies by
    $({\mathrm{h_{sim. cosm.}}} /{\mathrm{h_{planck}}})^2$ when
    matching to the SDSS galaxies.} and $r$-band absolute
  magnitude\footnote{The term $-5\log \mathrm{h}$ in the SDSS
    magnitudes is eliminated assuming the Planck cosmology. For
    stellar-mass-corrected models, the $r$-band luminosity is corrected
    by the same factor as the stellar mass in order to keep the
    stellar mass-to-light ratio unchanged.}  differing from the SDSS values
  by less than 0.02 dex and 0.05 mag, respectively.  If fewer than ten
  simulated galaxies satisfy these requirements, the tolerances are
  increased iteratively by factors of 1.5 until ten simulated galaxies
  are matched. In this way, the SDSS and simulation LBG samples end up
  with identical joint distributions of stellar mass, $r$-band
  luminosity and redshift.}

\noindent{(2) Each matched LBG in the simulation is assigned the
  redshift of the corresponding SDSS LBG, and is given a ``weight''
  equal to the mean weight (Eqn.~\ref{eqn:weight}) averaged over 
  the background source population of SDSS galaxies} at
  that redshift.

\noindent{(3) Matched LBGs are grouped by stellar mass in the simulation,
  after scaling by a factor of
  $({\mathrm{h_{sim. cosm.}}}/{\mathrm{h_{planck}}})^2$.  This is to
  treat the simulation as though we do not know its true cosmology,
  and so compare it to observation assuming our fiducial Planck
  cosmology. The differential surface density profiles of dark matter
  particles around the matched LBGs are calculated as a function of
  physical (not comoving) projected radius, $r_p$, and are averaged
  using the weights assigned in the previous step.}

\section{Stacked lensing surface density profiles}
\label{sec:results}

\subsection{Fit quality for a ``good'' model}
\label{sec:g11planck}

We start by presenting our measurements of differential surface
density profiles for stacks of SDSS locally brightest galaxies. We
compare these with predictions from $\mathrm{G11-P'}$, which turns out
to be the simulation that gives the best overall fit to the
observational data. $\mathrm{G11-P'}$ is also special in that it
requires the smallest correction of all our models (see 
Fig.~\ref{fig:cmpshiftMF}) to bring its high-mass stellar mass 
function ($\log M_\ast/\mathrm{M_\odot} > 10.7$) into exact agreement with 
that derived for the SDSS by \cite{2009MNRAS.398.2177L}. Its
behaviour at high mass in the lower panel of Fig.~\ref{fig:shiftMFplot} 
is notable for dropping below that of the other Guo et al.~models.
  
Points with error bars in the left-hand plot of
Fig.~\ref{fig:g11planck} show observational results from SDSS for LBGs
stacked into seven bins of $\log M_\ast/\mathrm{M_\odot}$ with boundaries as
indicated by the legend. Here and in all later similar plots, $\Delta
\Sigma $ is given as a function of projected radius, $r_p$, in
physical (rather than comoving) units. Solid lines are the predictions
for $\Delta \Sigma(r_p)$ for these same LBG stellar mass bins from
$\mathrm{G11-P'}$.  Dashed lines are based on the same model, but
after correcting to bring its stellar mass function into exact
agreement with SDSS (see Sec.~\ref{sec:corr}). Because of the large
number of dark matter particles in the simulation and the large number
of LBGs stacked, the statistical uncertainties on the model
predictions are very small, and are indicated only for the solid
lines. For SDSS also the uncertainties are small at most radii in
all but the lowest stellar mass bins.

The agreement of $\mathrm{G11-P'}$ with the SDSS measurements is very
good, both in amplitude and in shape over almost three orders of
magnitude in projected radius and 1.5 orders of magnitude in stellar
mass. Both theory and observation show a clear transition between a
small-scale ``one-halo'' term, where the signal is dominated by the
LBGs' own halos, and a large-scale ``two-halo'' term, where the
signal is dominated by the correlated environment surrounding the LBG
halos. The transition region is distorted by a feature at $\sim
1~\mathrm{Mpc}$ induced by our requirement that LBGs should have no
brighter companion projected within this radius. The stellar mass
corrected and uncorrected cases are very similar at high mass, where
the corrections are small; differences are more noticeable in the
lowest mass bins. For the 18 data points shown, the $\chi^2$
differences between the observed profiles and the uncorrected 
model are  20.45, 52.40, 14.95, 36.48, 44.79, 48.93 and 54.79 for the 
most massive to the least massive bin. For the stellar mass corrected
model, the corresponding values are 32.74, 44.14, 12.29, 36.10, 54.03,
67.35 and 66.12. While many of these values are formally sufficient to
exclude the model, they actually represent remarkably good fits given
the small size of many of the formal (bootstrap) error bars and the
fact that no model parameters were adjusted to fit these data.

As we will see later through comparison with other simulations, the
excellent agreement of $\mathrm{G11-P'}$ with SDSS is especially
notable for the three most massive bins. The model also agrees
reasonably well with the observations for the four lower mass bins,
but it seems to have some difficulty reproducing the one-halo/two-halo
transition and the 1.0 Mpc/h feature induced by our isolation
criterion.  The one-halo/two halo transition has often proved
difficult to fit accurately with physical or phenomenological models
of clustering \citep[see, e.g.][for Halo Occupation Distribution fits
to galaxy autocorrelations in SDSS]{2011ApJ...736...59Z,2013MNRAS.430..725V,
2015MNRAS.454.1161Z}. The 1.0 Mpc/h feature is due to the requirement 
of no brighter companions within 1 Mpc, which translates to a plummeting 
of the differential surface density profiles at this scale because 
galaxies trace the underlying mass distribution. This is especially 
true for low mass LBGs, as our method makes us preferentially select 
rather isolated systems.

\begin{figure*}
\epsfig{figure=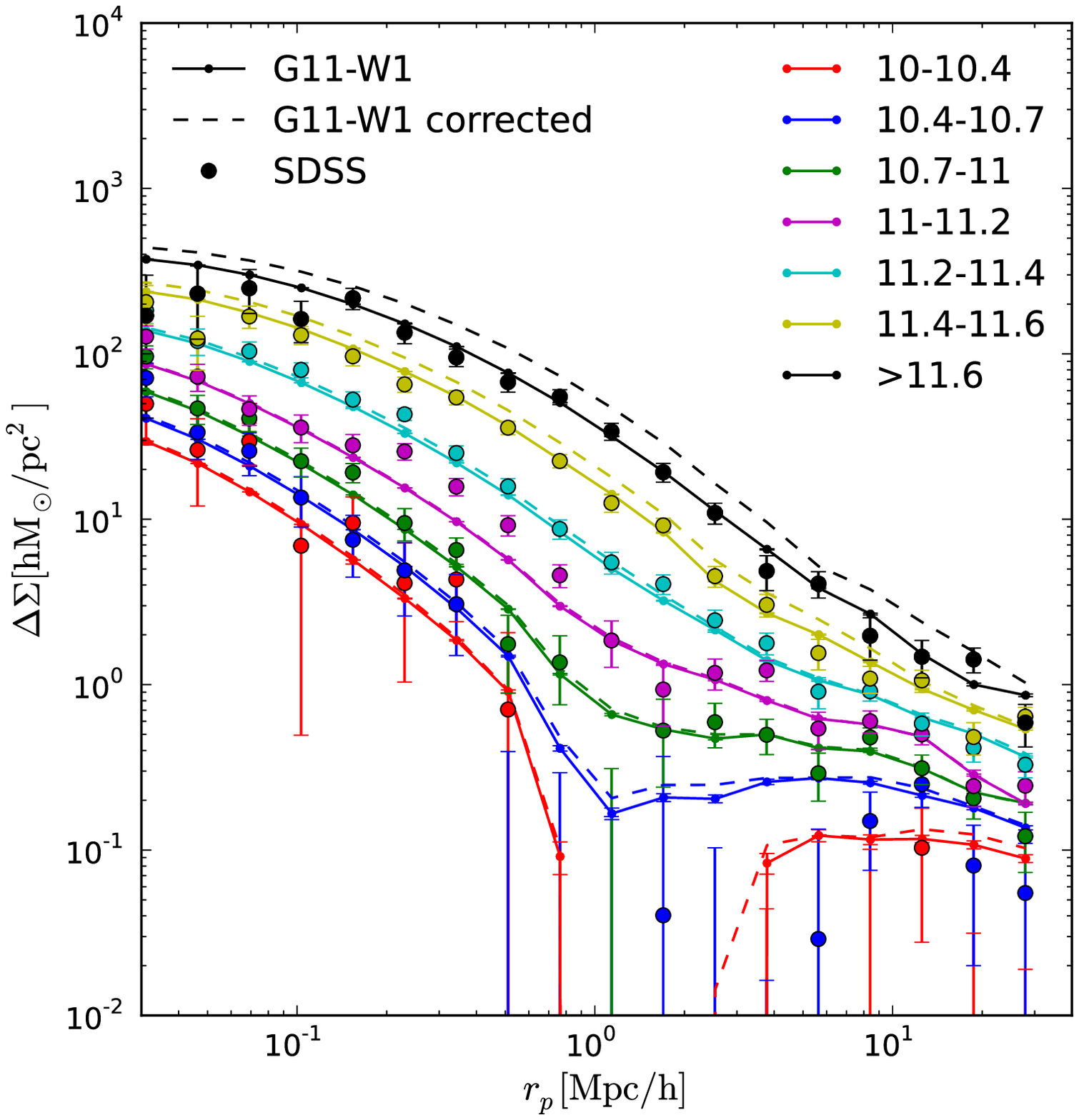,width=0.49\textwidth}%
\epsfig{figure=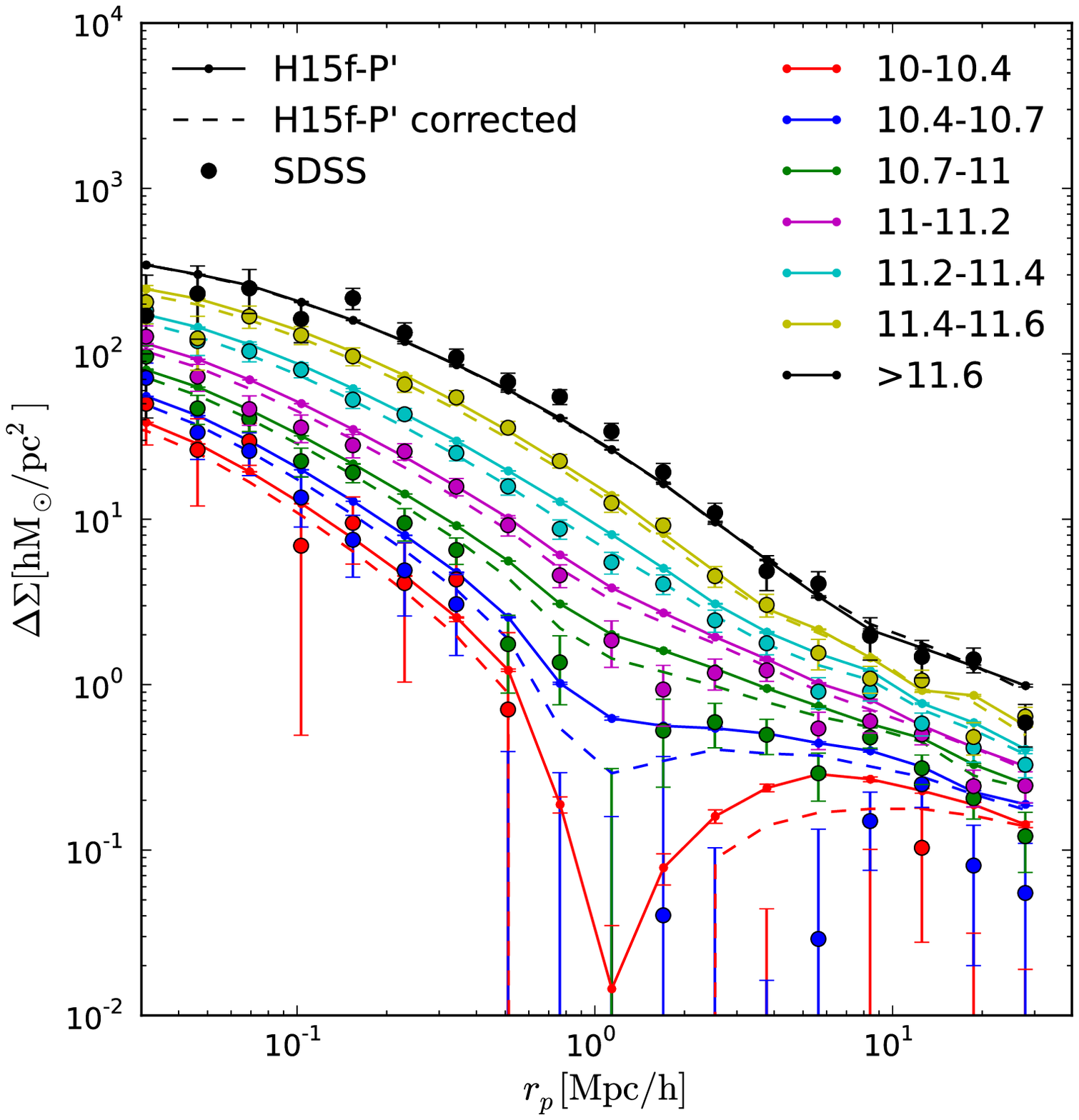,width=0.49\textwidth}\\
\epsfig{figure=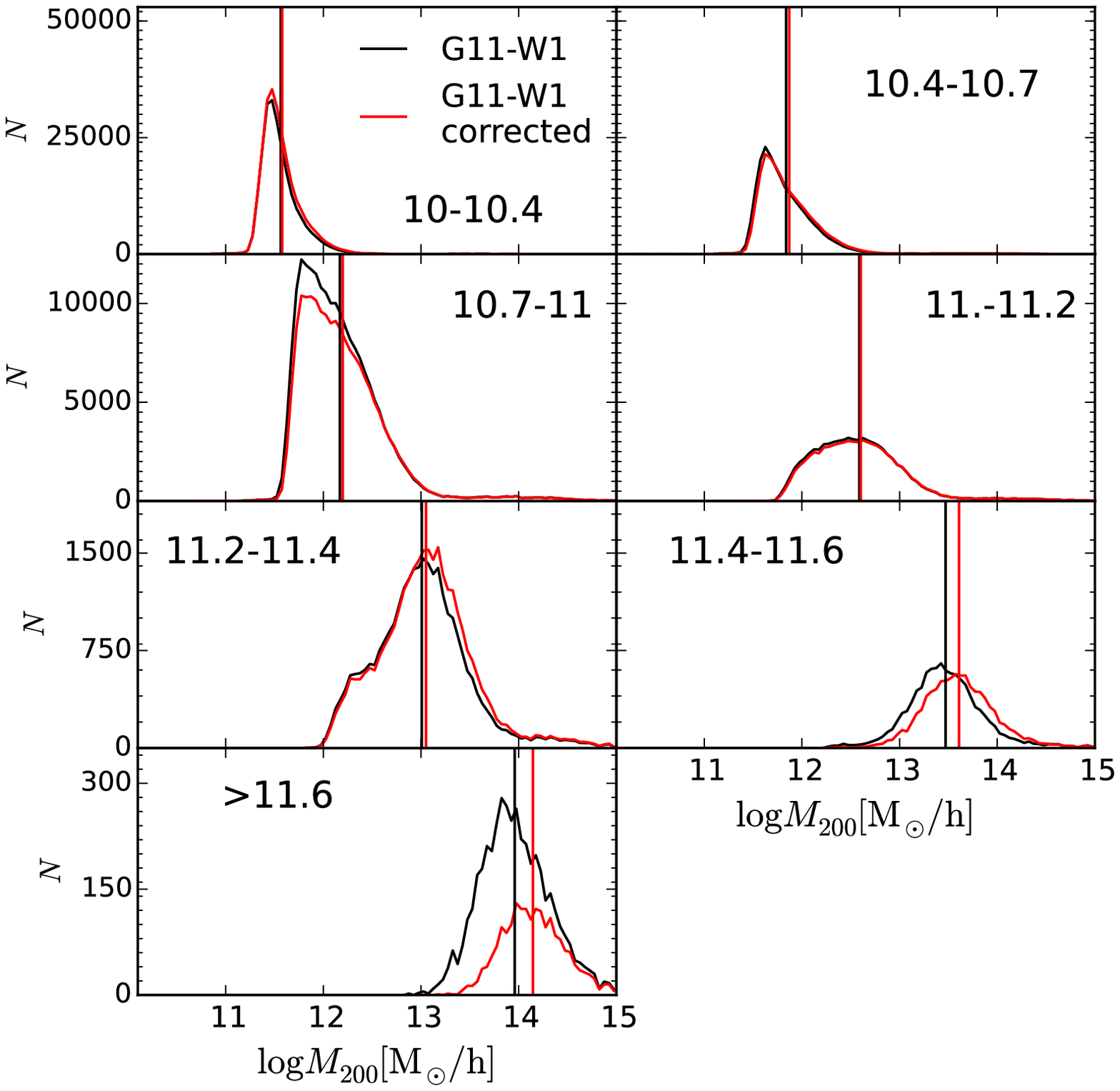,width=0.49\textwidth}%
\epsfig{figure=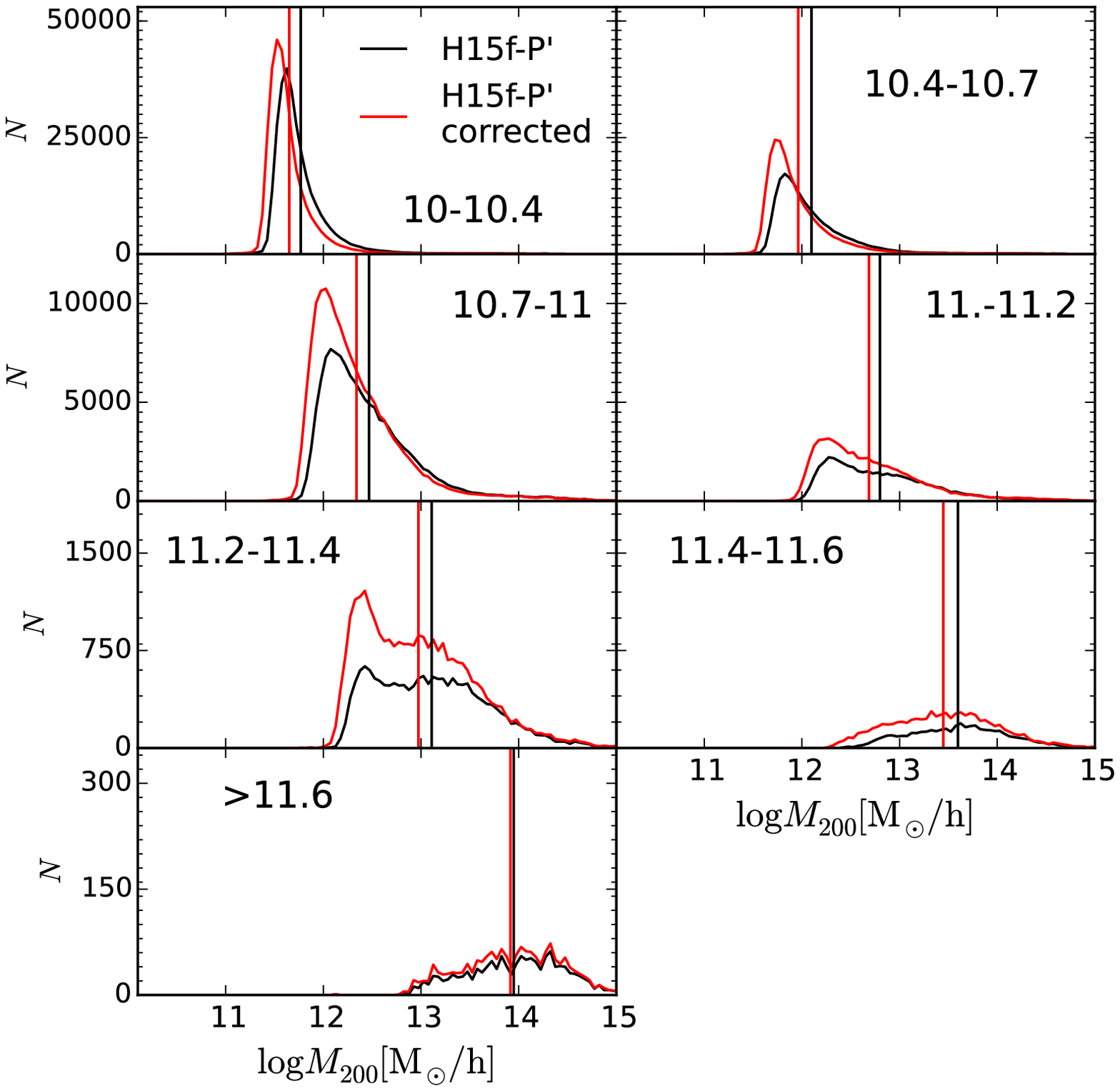,width=0.49\textwidth}
\caption{{\bf Top Left} and {\bf Bottom Left:} As
  Fig.~\ref{fig:g11planck} but for $\mathrm{G11-W1}$. {\bf Top Right}
  and {\bf Bottom Right:} As Fig.~\ref{fig:g11planck} but for $\mathrm{H15f-P'}$.} 
\label{fig:two}
\end{figure*}

\label{sec:cosm}
\begin{figure*}
\epsfig{figure=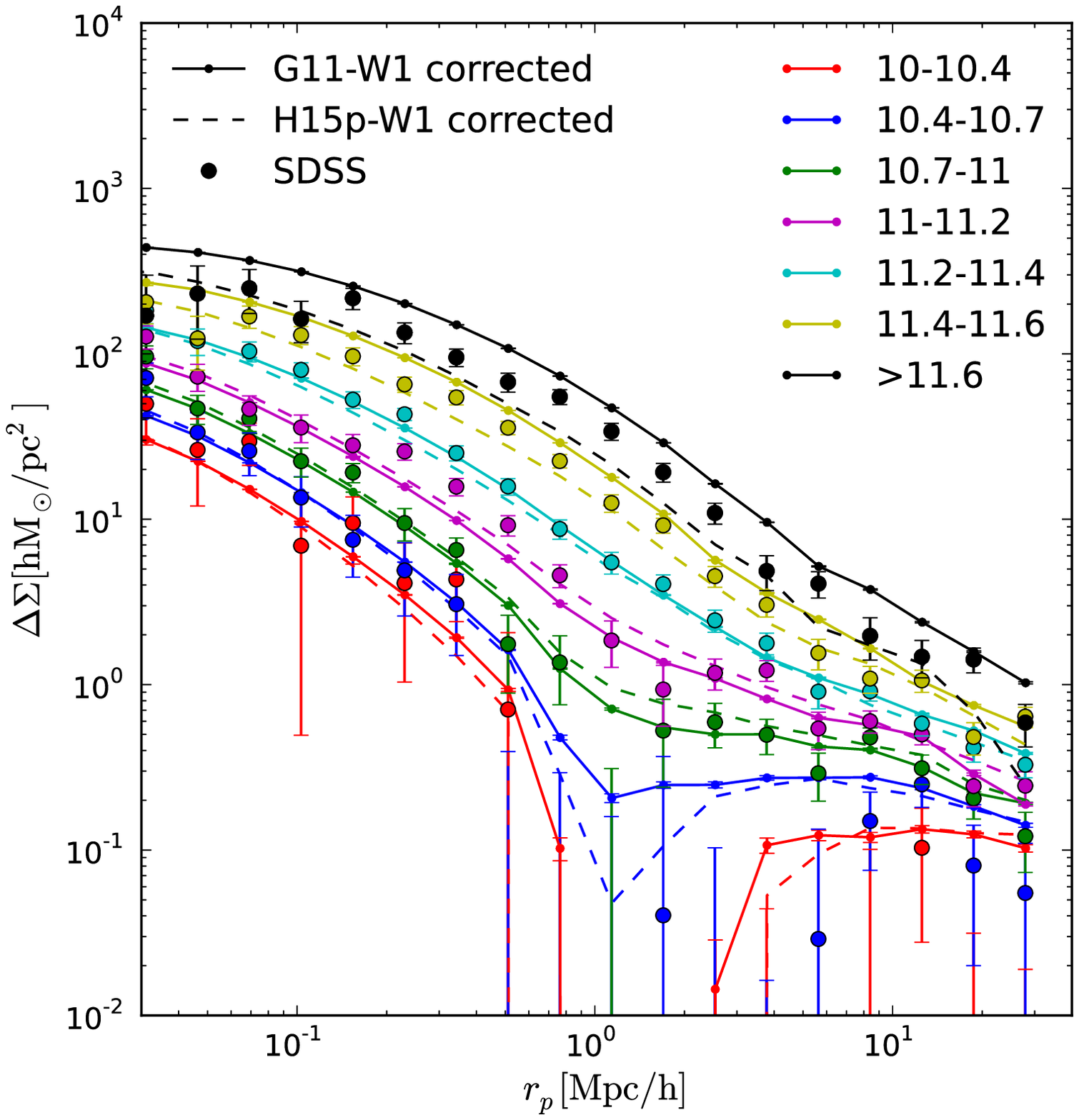,width=0.49\textwidth}%
\epsfig{figure=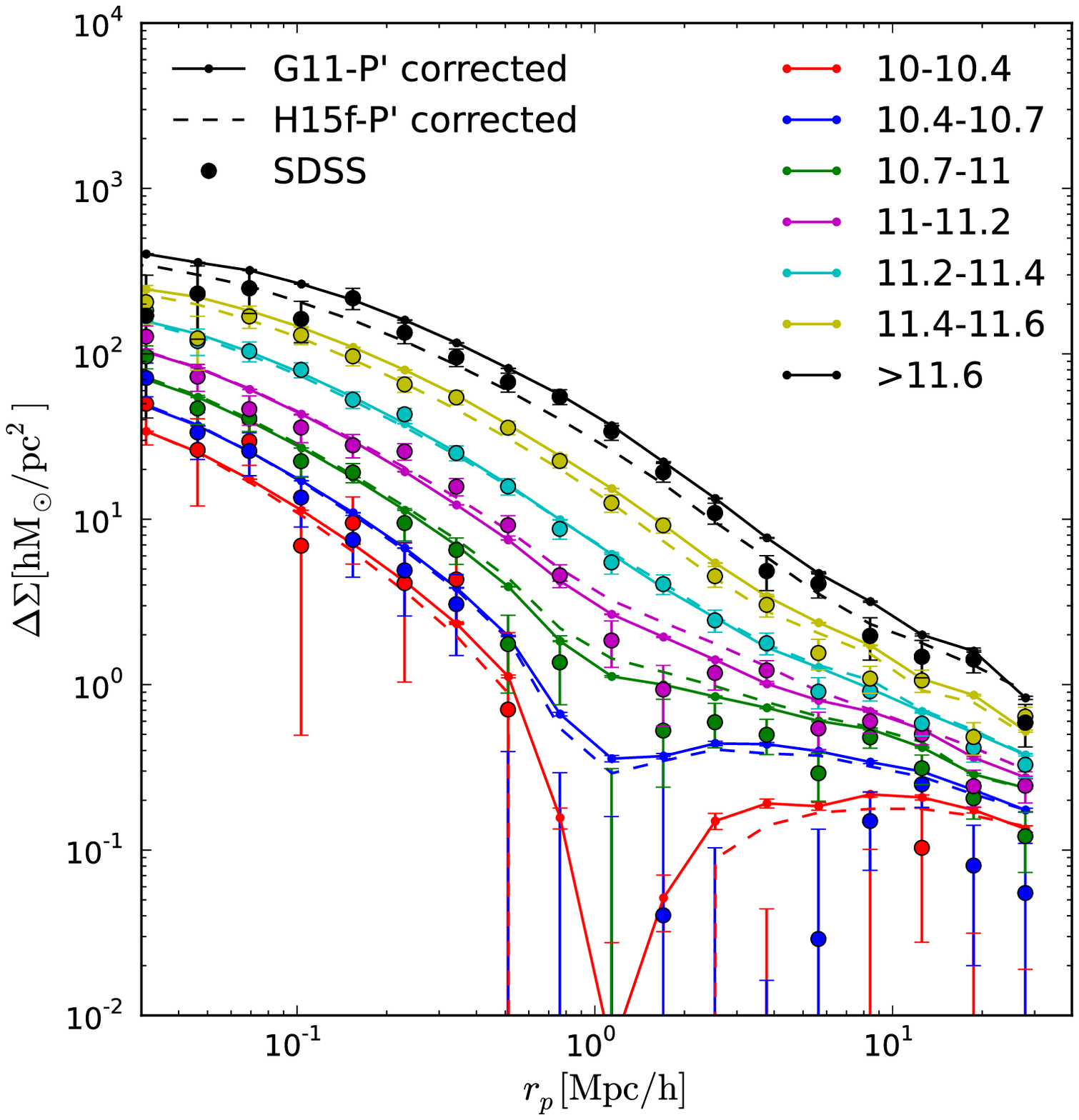,width=0.49\textwidth}\\
\epsfig{figure=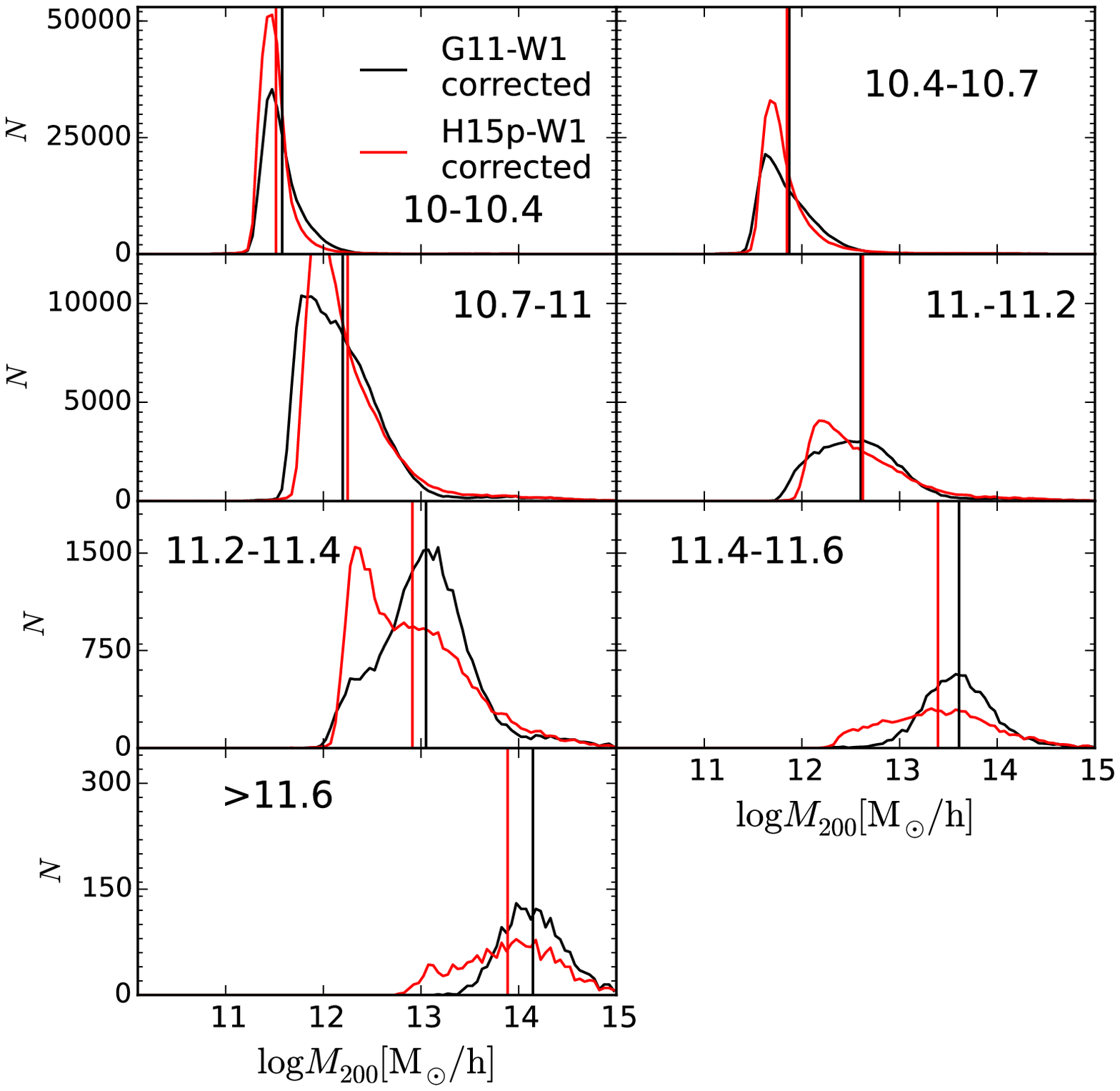,width=0.49\textwidth}%
\epsfig{figure=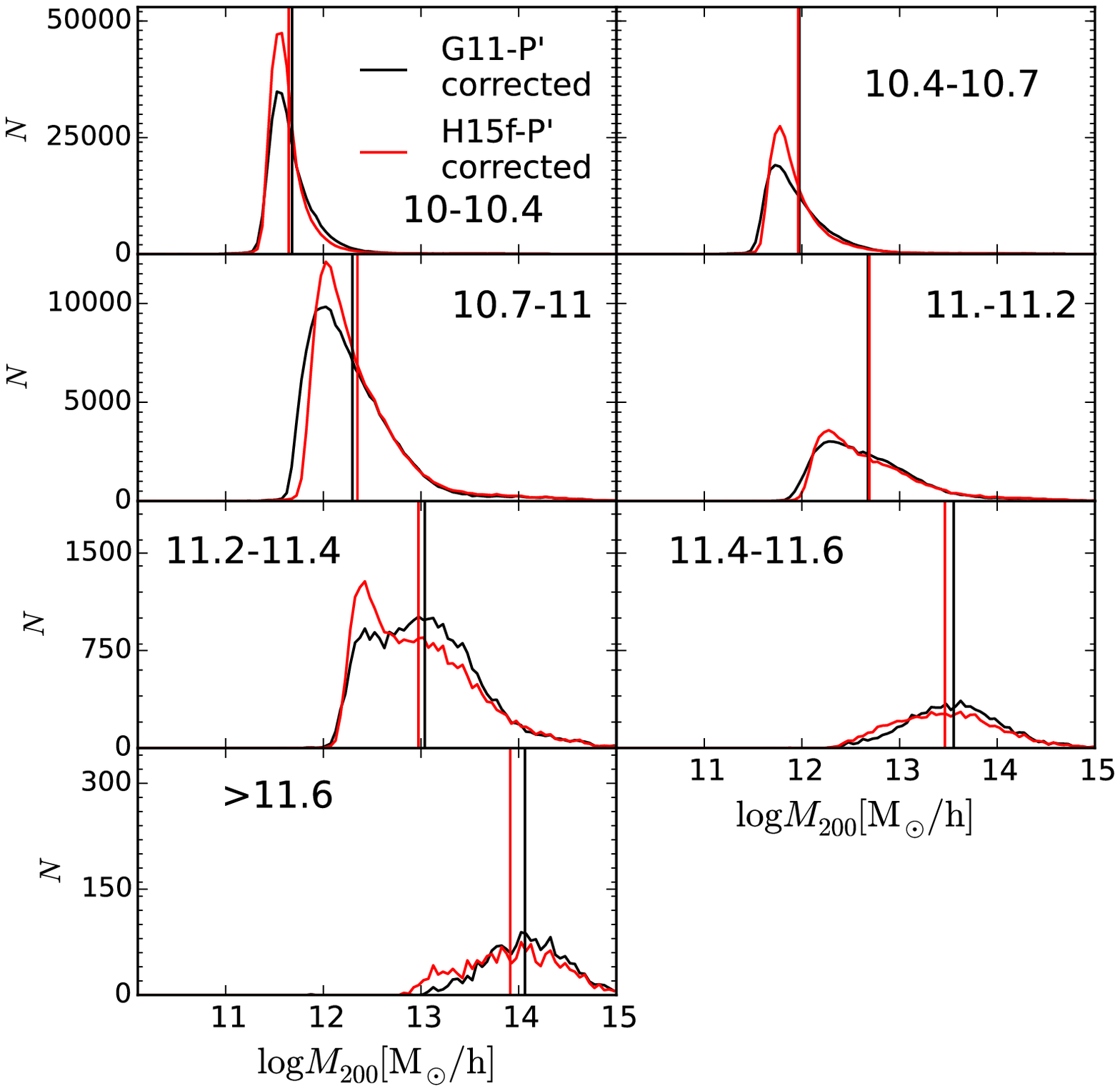,width=0.49\textwidth}
\caption{{\bf Top Left} and {\bf Bottom Left:} As
  Fig.~\ref{fig:g11planck} but now comparing $\mathrm{G11-W1}$ and
  $\mathrm{H15p-W1}$ after both have been corrected to reproduce the
  SDSS stellar mass function. These two models thus have exactly the
  same stellar mass function and cosmology (WMAP1) and are implemented
  on the same simulation (the original MS), yet they make
  substantially different predictions for the lensing profiles and
  have different halo mass distributions at given stellar mass. {\bf
    Top Right} and {\bf Bottom Right:} As the left-hand plots, but now
  for $\mathrm{G11-P'}$ and $\mathrm{H15f-P'}$, two simulations based
  on the Planck cosmology.}
\label{fig:cosmo}
\end{figure*}

\begin{figure*}
\epsfig{figure=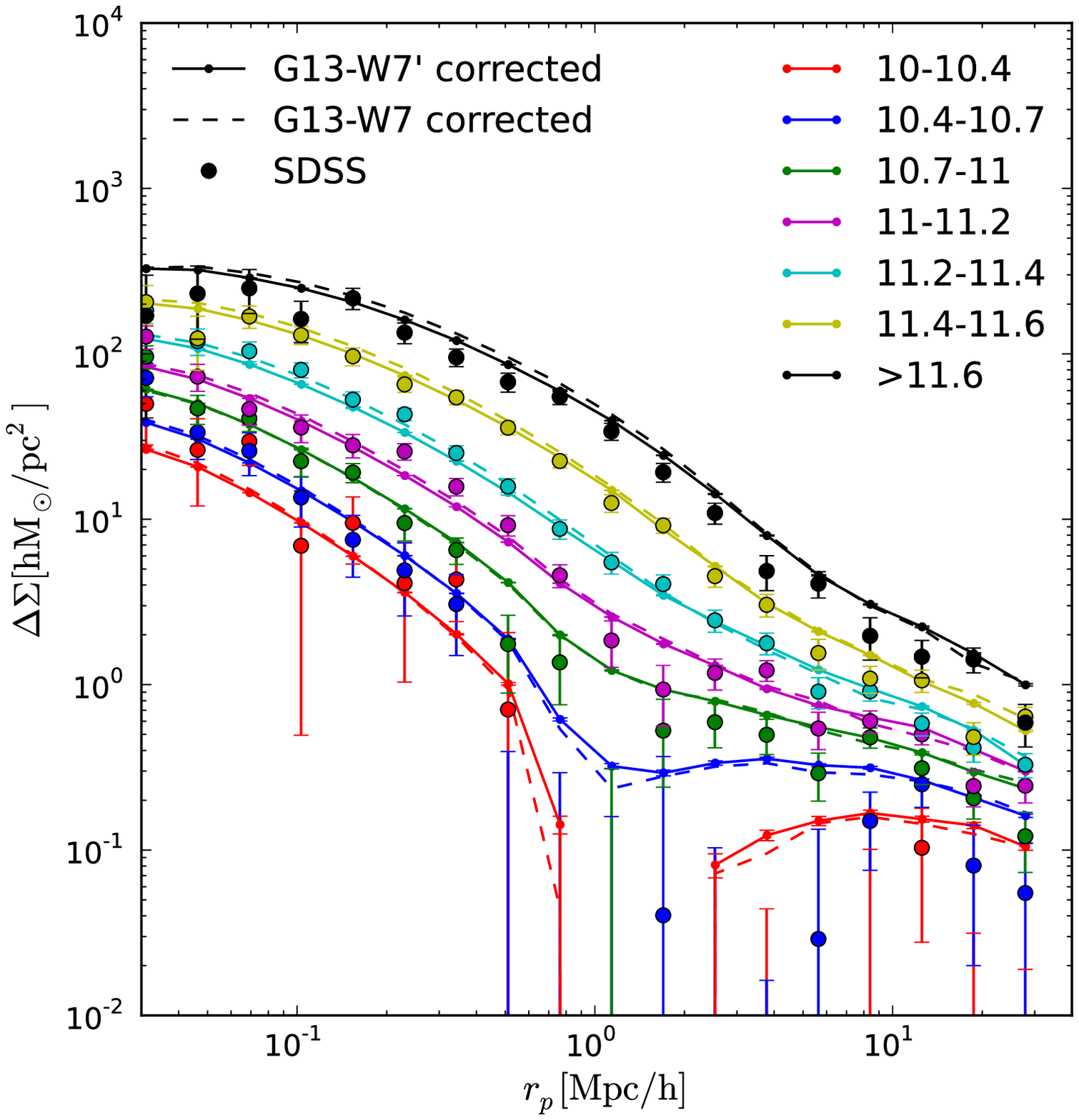,width=0.49\textwidth}%
\epsfig{figure=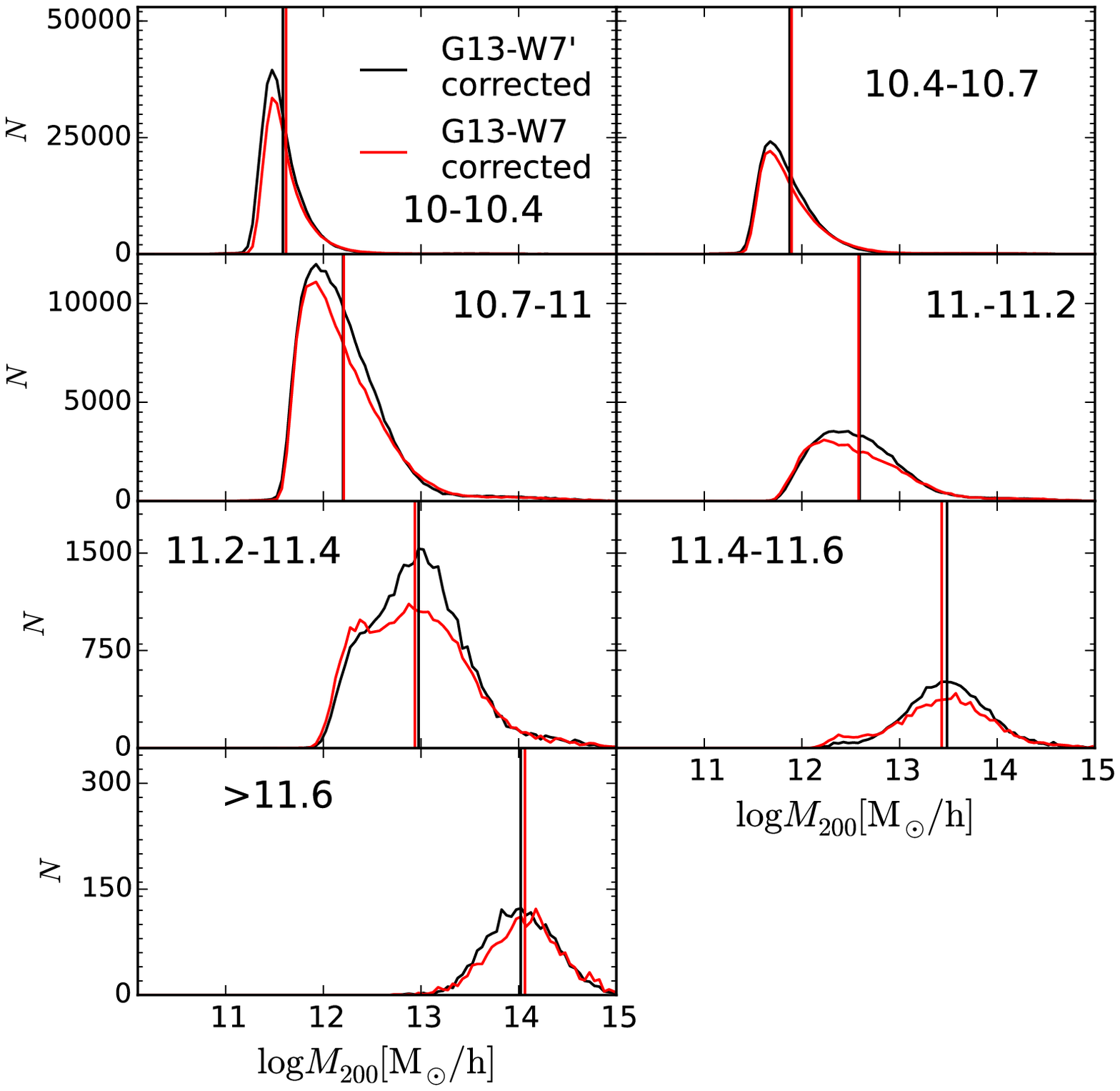,width=0.49\textwidth}
\caption{As Fig.~\ref{fig:g11planck}, but now comparing
  $\mathrm{G13-W7}$ and $\mathrm{G13-W7'}$, two models with the same
  galaxy formation physics and cosmology (WMAP7), one implemented on a
  simulation executed with that cosmology (MS-W7) and the other
  implemented on a scaled version of the MS.  Corrections to bring the
  stellar mass function into exact agreement with SDSS have been
  included in both cases. Differences here are due to cosmic variance
  and to inadequacies of the scaling procedure.}
\label{fig:scale}
\end{figure*}

The right-hand plot in Fig.~\ref{fig:g11planck} shows histograms of
host halo mass for LBGs in our seven stellar mass bins. For
simplicity, here and throughout the paper we use LBGs selected in the
$z=0$ snapshot to make these plots. For non-central LBGs, the
$M_{200}$ values adopted are those of the central galaxies of which
they are satellites. Black and red histograms in
Fig.~\ref{fig:g11planck} refer to the stellar mass uncorrected and
stellar mass corrected cases, respectively, as noted in the
legend. Vertical lines mark the characteristic halo mass, $\langle
\log M_\mathrm{200} \rangle$, for each bin. We adopt $\langle
\log M_{200} \rangle$ (instead of, for example, $\log
\langle M_{200} \rangle$) because the halo mass distributions are
broad and roughly lognormal so that $\langle \log M_{200}
\rangle$ values are in all cases close to the central values of the
distributions. They also turn out to be close to the ``effective''
halo masses defined by the SZ and X-ray stacking analyses of P13 and
A15.

The scatter in halo mass at fixed stellar mass is large and depends
both in width and in shape on stellar mass (and, as we will see below,
on galaxy formation model). For the four most massive bins, the
corrected and uncorrected models give almost the same distribution,
explaining why the solid and dashed lines are almost indistinguishable
in the left plot. For the three least massive bins, the distributions
shift slightly to the right for the mass corrected model, giving
higher characteristic halo masses and higher predictions in the
left-hand plot. The amplitude differences in the predicted lensing
signal agree qualitatively with the shifts in characteristic mass of the
halo mass distributions. Despite these small amplitude shifts, both
solid and dashed curves agree quite well with SDSS.

\subsection{Fit quality for two published Munich models}
\label{sec:two}

\begin{figure*}
\epsfig{figure=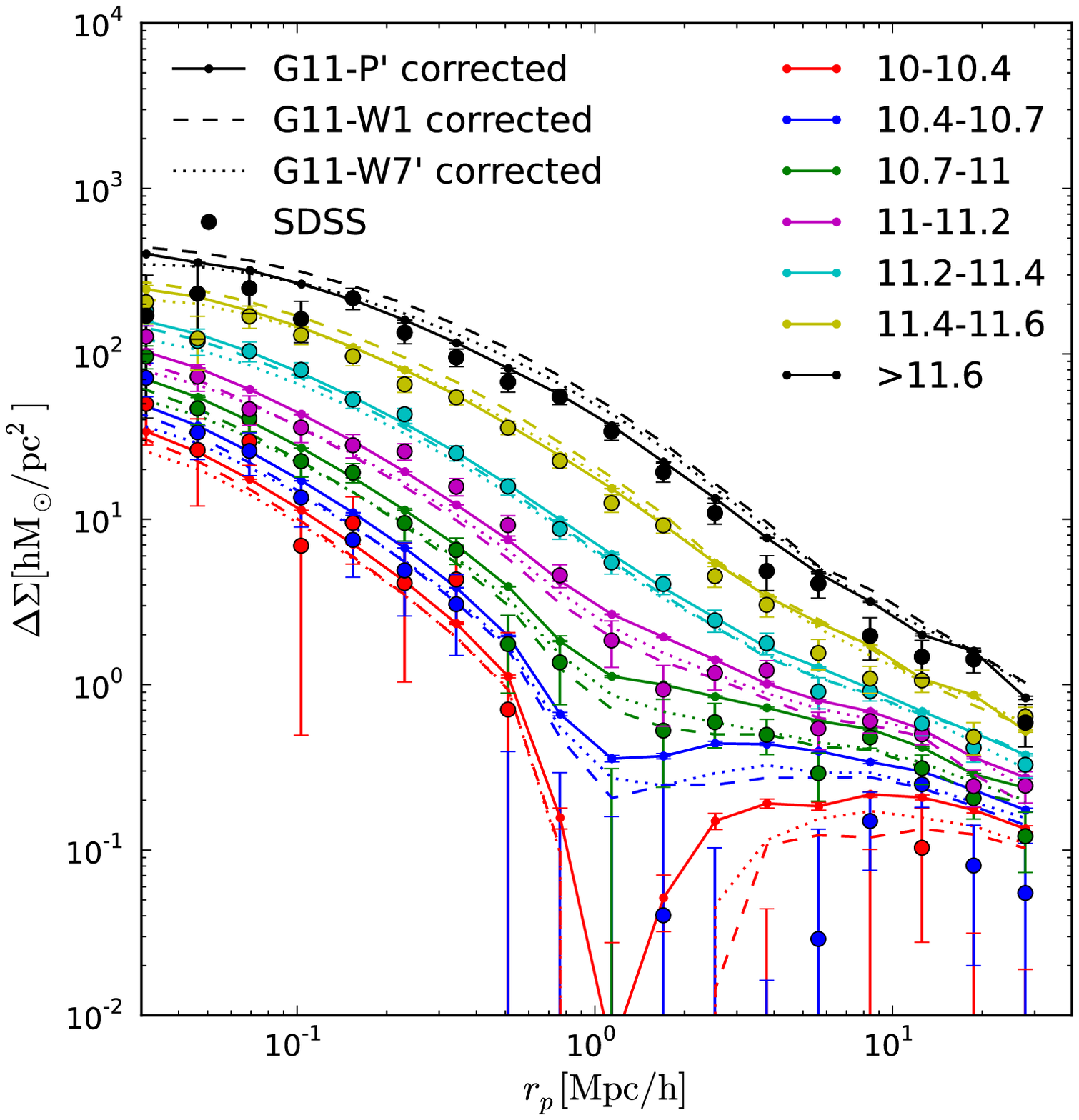,width=0.49\textwidth}%
\epsfig{figure=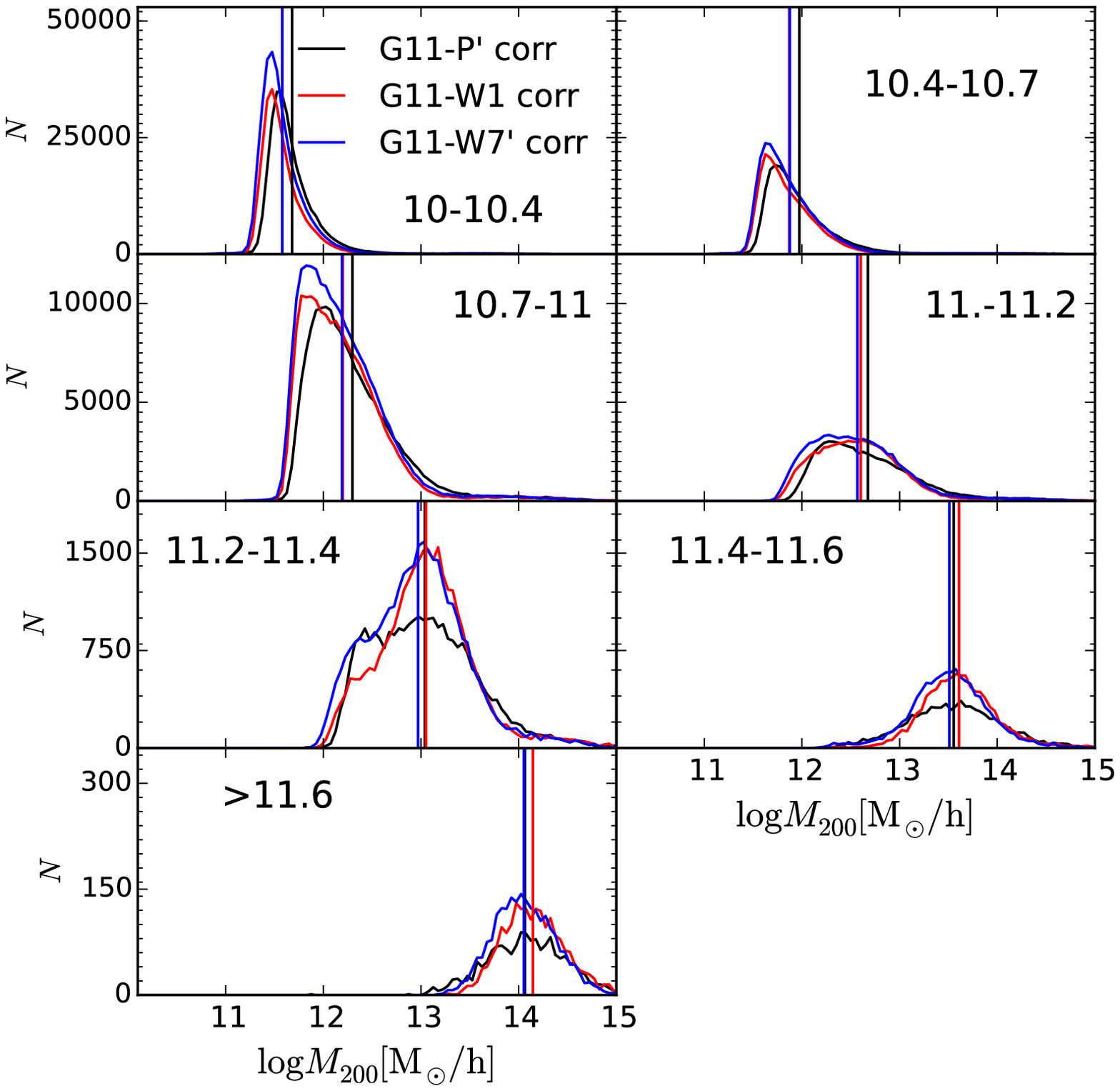,width=0.49\textwidth}
\caption{As Fig.~\ref{fig:g11planck}, but now comparing three
  simulations with the same galaxy formation physics and parameters,
  and implemented on the same N-body simulation (the MS) but with
  different cosmologies ($\mathrm{G11-P'}$, $\mathrm{G11-W1}$ and
  $\mathrm{G11-W7'}$). Corrections to bring the stellar mass function
  into exact agreement with SDSS have been included for all three
  models. Differences here are thus due almost entirely to the
  differing cosmologies.}
\label{fig:G11}
\end{figure*}

The model analysed in the last section, $\mathrm{G11-P'}$, implements
the galaxy formation physics and parameters of
\cite{2011MNRAS.413..101G} on a version of the MS scaled to the Planck
cosmology first used in \cite{2015MNRAS.451.2663H}. As a result, it
does not correspond to any previously published galaxy formation
simulation. In order to start our exploration of the model-dependence
of predicted lensing surface density profiles, we now present results
for two published and publicly available simulations,
$\mathrm{G11-W1}$, which is based on the WMAP1 cosmology and was
published by \cite{2011MNRAS.413..101G}, and $\mathrm{H15f-P'}$, which
is based on the Planck cosmology and was published by
\cite{2015MNRAS.451.2663H}. Fig.~\ref{fig:two} shows differential
lensing surface density profiles and halo mass distributions for these
two models in exactly the same format as Fig.~\ref{fig:g11planck}.
Both stellar mass corrected and stellar mass uncorrected results are
shown for both models. Note round dots with errors are lensing 
profiles based on SDSS, which are exactly the same in all relevant 
plots from Fig.~\ref{fig:g11planck} to Fig.~\ref{fig:G11} and also 
for Fig.~\ref{fig:correction}. For these figures we always only show 
errors for solid curves, while errors associated to dashed curves 
have similar size. In most cases the model predictions have 
extremely small errors comparable to the symbol size. 

The left side of Fig.~\ref{fig:two} shows results for
$\mathrm{G11-W1}$. As can be seen from Fig.~\ref{fig:cmpshiftMF}, this
model requires significant and negative corrections at high mass to
bring its stellar mass function into agreement with SDSS, i.e. it
overpredicts the abundance of high mass galaxies \citep[see Fig.~7
  of][]{2011MNRAS.413..101G}.  After correction, the predicted
surface density profile for the two highest mass bins lies
significantly above the observations, showing that the excellent
agreement found for the original model is, in fact, a coincidence. For
the lower stellar mass bins, the solid and dashed curves are very
similar since the required mass corrections are very small in this
stellar mass range. For all seven bins, the differences in
characteristic halo mass (the shifts between the red and black
vertical lines in the lower plot) agree quite well with the shifts in
amplitude between solid and dashed lines in the upper plot. This shows
that correcting the model stellar masses induces a (logarithmic) shift
in the ``effective halo mass'' (defined as the halo mass which would
produce the same signal when all LBGs are assumed to have identical
halos) which is very similar to that in the characteristic halo mass
$\langle\log M_{200}\rangle$. This implies that the {\it shape} of the
halo mass distributions at fixed stellar mass is not much affected 
by the correction.

There is quite good agreement with SDSS for the five lower mass bins,
except that magenta curves underpredict the SDSS signal at projected
radii of a few hundred kpc. $\chi^2$ values for the fit of the
uncorrected $\mathrm{G11-W1}$ model to the SDSS data are 18.05, 17.46,
22.85, 49.02, 24.86, 32.48 and 44.18 from most to least massive 
bin. For the mass-corrected model, the corresponding values are 157.40, 
98.22, 14.39, 46.55, 23.26, 35.56 and 47.75. It is clear that, after stellar
mass correction, the $\mathrm{G11-W1}$ model fits observation 
much less well than $\mathrm{G11-P'}$, at least for the two highest stellar 
mass bins.

The right side of Fig.~\ref{fig:two} shows similar results for
$\mathrm{H15f-P'}$. The SDSS signal is clearly underestimated 
between 200~kpc and 2~Mpc in the most massive bin, and 
this problem is not affected by the mass
correction because this particular model fits the high-mass tail of
the SDSS stellar mass function very well \citep[see Fig.~2 of][ note
also that the abundance at lower masses is underpredicted, giving
rise to the positive corrections seen for this model in
Fig.~\ref{fig:cmpshiftMF}]{2015MNRAS.451.2663H}. For the five lowest
stellar mass bins, significant stellar mass corrections cause both the
characteristic halo masses and the shape of the halo mass
distributions to shift. As a result, the lensing profile predictions
in the upper plot are lowered and come into better agreement with
SDSS. The $\chi^2$ values are 25.34, 25.64, 69.64, 94.26, 165.07,
96.06 and 78.50 for the mass-uncorrected model and 25.49, 28.66, 
18.17, 55.73, 78.25, 57.83 and 58.25 after mass correction (again from
most to least massive bin). Once more it is true that the changes in
amplitude of the surface density profiles (i.e. the shifts in the
effective halo mass for lensing) agree well with the shifts in
characteristic halo mass seen in the lower plot.

It is clear that overall $\mathrm{G11-W1}$ and $\mathrm{H15f-P'}$
agree less well with SDSS than $\mathrm{G11-P'}$, especially for the
few most massive bins and after correction to bring agreement with the
SDSS stellar mass function. For lower stellar mass, the
abundance-matching corrections improve the agreement with observation,
at least for these models.

\subsection{Models differing only in galaxy formation physics}

Fig.~\ref{fig:cosmo} compares differential surface density profiles
and halo mass distributions for two galaxy formation models
implemented on the original MS ($\mathrm{G11-W1}$ and
$\mathrm{H15p-W1}$ on the left side) and for two implemented on the MS
after scaling to the Planck cosmology ($\mathrm{G11-P'}$ and
$\mathrm{H15f-P'}$ on the right side). Thus the models in each pair
have the same evolving dark matter distribution but assume different
galaxy formation physics. All four models have been corrected to bring
their stellar mass distributions into exact agreement with SDSS and so
have identical stellar mass functions.  Differences within each pair
are thus due solely to differences in how galaxies of a given stellar
mass are assigned to (sub)halos.

A comparison of $\mathrm{G11-W1}$ and $\mathrm{H15p-W1}$ shows that
despite their identical mass distributions and (corrected) stellar
mass functions, their surface density profiles differ substantially in
the most massive bins, $\mathrm{G11-W1}$ overpredicting the observed
signal and $\mathrm{H15p-W1}$ underpredicting it. Corresponding
$\chi^2$ values are 73.34 and 43.42 for the two most massive bins in
$\mathrm{H15p-W1}$ and 157.40 and 98.22 for these same bins in
$\mathrm{G11-W1}$. The $\chi^2$ values for the five less massive bins
of $\mathrm{H15p-W1}$ are 35.44, 35.13, 29.02, 29.64 and 45.66, which
are closer to the corresponding $\chi^2$ values for $\mathrm{G11-W1}$
(see Sec.~\ref{sec:two}).

These substantial differences highlight the importance of scatter in
the stellar mass -- halo mass relation. This scatter is not, of
course, purely stochastic, but rather a consequence of the scatter in
assembly history among halos of given mass, which impacts the
properties of central galaxies differently in different galaxy
formation models.  Scatter effects are particularly important at high
mass where the dependence of central galaxy properties on halo mass is
weak \citep[e.g.][]{2008MNRAS.390.1157R,2010MNRAS.404.1111G,
2010ApJ...710..903M,2010MNRAS.402.1796W}. 
As a result, central galaxy stellar mass is a relatively poor
and model-dependent predictor of halo mass on the scale of rich
clusters. More precise mass proxies, perhaps, cluster richness, total
luminosity, X-ray or SZ properties \citep[e.g.][]{2010ApJ...708..645R,
  2014A&A...571A..20P, 2015MNRAS.449.1897O} will be needed
to derive robust and competitive cosmological constraints from the
cluster population.

The differences between models are very clear in the halo mass
distributions in the lower panels of Fig.~\ref{fig:cosmo}.  In a given
stellar mass bin, the distributions can differ substantially in shape,
width and characteristic mass, with effects being particularly large
in the higher stellar mass bins. $\mathrm{G11-W1}$ predicts less
scatter at the massive end of the halo mass -- stellar mass relation
than $\mathrm{H15p-W1}$.  In the lower stellar mass bins, the
characteristic halo masses agree better, but the shape and amplitude
of the halo mass distributions still differ noticeably.

Similar but less dramatic differences are seen for the two models
based on the MS scaled to the Planck cosmology. In the two most
massive bins, $\mathrm{G11-P'}$ predicts a significantly stronger
lensing signal than $\mathrm{H15f-P'}$, although the difference is not
as big as between $\mathrm{G11-W1}$ and $\mathrm{H15p-W1}$. The
$\chi^2$ values for the two most massive bins are 25.49 and 28.66 for
$\mathrm{H15f-P'}$ and 32.74 and 44.14 for $\mathrm{G11-P'}$. The
behaviour for the five less massive bins parallels that for the WMAP1
models, with $\chi^2$ values of 18.17, 55.73, 78.25, 57.82 and 58.25
for $\mathrm{H15f-P'}$.  Although the characteristic halo masses and
differential surface density profiles differ very little between the
two models at these stellar masses, the shape and amplitude of the
halo mass distributions again differ quite noticeably.

Despite the substantial shape and and scatter variations between these
models, the shifts in normalisation of the differential surface
density profiles correspond moderately well to the shifts in
characteristic halo mass, $\langle\log M_{200}\rangle$ for all stellar
mass bins and in both cosmologies. The relative amplitude of the
lensing profiles provides an estimate of the change in effective halo
mass between two models (see Sec.~\ref{sec:scaling} for details of how
we estimate this). The logarithmic shifts from $\mathrm{H15p-W1}$ to
$\mathrm{G11-W1}$ are estimated to be 0.469, 0.313, 0.074, -0.076, -0.058, 
-0.010 and 0.019. from the most to the least massive stellar
mass bin. For comparison, the logarithmic differences between the
characteristic halo masses indicated by the vertical lines in the
bottom left plot are 0.256, 0.216, 0.140, -0.023, -0.052, 0.024 and
0.065. Similarly, the shifts in effective halo mass from
$\mathrm{H15f-P'}$ to $\mathrm{G11-P'}$ estimated from their
differential lensing surface density profiles are  0.238, 0.111, 
0.027, -0.014, -0.025, 0.001 and 0.028, while the corresponding 
logarithmic shifts in characteristic mass between their halo mass 
distributions are 0.150, 0.104, 0.065, -0.012, -0.042, 0.012 and 0.028. 
The shifts in the effective halo mass for lensing and in $\langle\log
M_{200}\rangle$ are similar but certainly not identical. The detailed
{\it shape} of the halo mass distribution at fixed stellar mass thus
affects the ratio of these two masses, rendering it model-dependent.

\begin{table*}
\caption{Lensing calibrated characteristic halo masses and their
  systematic uncertainties as inferred using our full set of galaxy
  formation simulations. We use each simulation both in its original
  form and after correction to bring its stellar mass function into
  exact agreement with the SDSS, resulting in a total of sixteen
  simulations. For each simulation and each stellar mass bin, we
  calculate the mean and median values of $\log M_\mathrm{200}$.
  These are then calibrated by shifting logarithmically by the amount
  needed to bring the simulated differential surface density profile
  into agreement with the SDSS data (see
  Sec.~\ref{sec:scaling}). Column 1 lists the boundaries of the
  stellar mass bins; columns 2 and 4 provide lensing calibrated
  estimates of $\langle \log M_\mathrm{200} \rangle$ and
  $\mathrm{median}(\log M_\mathrm{200})$ for our SDSS sample, averaged
  over all 16 models, while columns 3 and 5 give the rms scatter among
  estimates for these quantities based on the individual models.
  $M_{200}$ is in units of $\mathrm{M_\odot h}^{-1}$ throughout.}
\begin{center}
\begin{tabular}{lcccccccccccc}
\hline
\hline
$\log M_\ast/\mathrm{M_\odot}$& $\langle\langle \log M_{200} \rangle\rangle $& 
$\sigma(\langle\log M_{200}\rangle)$&$\langle\mathrm{median}(\log M_{200})\rangle$& $\sigma(\mathrm{median}(\log M_\mathrm{200}))$\\
\hline
10.-10.4 & 11.837 & 0.027 & 11.778 & 0.026 \\
10.4-10.7 & 11.836 & 0.022 & 11.743 & 0.017 \\
10.7-11. & 12.234 & 0.054 & 12.131 & 0.051 \\
11.-11.2 & 12.709 & 0.043 & 12.620 & 0.056 \\
11.2-11.4 & 13.015 & 0.047 & 12.975 & 0.060 \\
11.4-11.6 & 13.475 & 0.064 & 13.471 & 0.060 \\
11.6-15. & 13.996 & 0.083 & 14.000 & 0.099 \\
\hline
\hline
\label{tbl:halomass}
\end{tabular}
\end{center}
\end{table*}

\subsection{True versus scaled N-body simulations}
\label{sec:scale}

In this subsection we compare $\mathrm{G13-W7}$ and
$\mathrm{G13-W7'}$. These two simulations have identical galaxy
formation physics, including all parameters, and assume the same
cosmology.  However, $\mathrm{G13-W7}$ is based on the MS-W7, which
was set up and carried out assuming a WMAP7 cosmology, whereas
$\mathrm{G13-W7'}$ is based on the MS (carried out assuming
WMAP1 parameters) after rescaling to a WMAP7 cosmology.  We can
already see from Fig.~\ref{fig:cmpshiftMF} that the two models do not
produce exactly the same stellar mass function; their mass correction
curves differ by about 0.05 dex on all scales (see also Fig.~5 of
\citealt{2013MNRAS.428.1351G}). Such differences could arise 
from cosmic variance (since the two simulations have independent 
initial fluctuation fields), from the (small) differences in time 
and mass resolution between MS-W7 and the scaled MS, and from 
inadequate accuracy in the scaling algorithm.

Fig.~\ref{fig:scale} presents a direct comparison between
$\mathrm{G13-W7}$ and $\mathrm{G13-W7'}$ in the same format as
Fig.~\ref{fig:g11planck}. For both simulations we have included the
corrections needed to bring their stellar mass functions into exact
agreement with SDSS. With this source of difference removed, they
predict very similar differential surface density profiles. For the
few most massive bins, $\mathrm{G13-W7}$ (dashed) is slightly higher
than $\mathrm{G13-W7'}$ (solid) on small scales. The difference shows
up in the $\chi^2$ values for the two models: 77.42, 38.36, 13.64,
35.69, 53.71, 47.40 and 50.98 for $\mathrm{G13-W7}$ and 47.02, 21.92,
23.69, 38.77, 51.74, 51.59 and 54.48 for $\mathrm{G13-W7'}$. Thus
$\mathrm{G13-W7'}$ agrees better with SDSS for the two most massive
bins, while for the remaining bins the two models have similar
$\chi^2$ values. We note that $\mathrm{G13-W7'}$ is the simulation
that was used by P13 and A15 to interpret their SZ and X-ray 
stacking analyses, and that its agreement with the SDSS lensing data 
is very good, only slightly worse than that of $\mathrm{G11-P'}$, 
the ``best'' model of Sect.~\ref{sec:g11planck}

The halo mass functions of the two simulations are compared in Fig.~1
of \cite{2013MNRAS.428.1351G}, where MS-W7 can be seen to have
slightly higher halo abundances at $z=0$ than the scaled MS. On the
other hand, Fig.~\ref{fig:cmpshiftMF} shows that at a given stellar
mass, $\mathrm{G13-W7}$ predicts slightly {\it lower} galaxy
abundances than $\mathrm{G13-W7'}$. Thus the halo mass -- stellar mass
relation shifts between the two models, with halos of a given mass
forming central galaxies of slightly lower stellar mass in the scaled
simulation, perhaps as a consequence of its slightly more widely
spaced sequence of output times. Correcting both simulations so that
they reproduce the SDSS stellar mass function removes much of this
difference.

Halo mass distributions are shown on the right-hand side of
Fig.~\ref{fig:scale}.  For every stellar mass bin, the two models have
similar characteristic halo masses, even though in some cases the
shapes of the distributions differ significantly.  Here again the
logarithmic shifts in characteristic halo mass are similar in size to
those in the effective mass corresponding to the lensing profiles. From
$\mathrm{G13-W7'}$ to $\mathrm{G13-W7}$ the latter shift by  -0.063, 
-0.080, -0.072, -0.053, 0.003, -0.028 and -0.036, while the former shift
by -0.040, 0.055, 0.039, 0.012, -0.008, -0.020 and -0.032.  The
two sets of numbers do not correspond exactly, showing once more a
dependence of the ratio of effective to characteristic halo mass on the
detailed shape of the halo mass distributions.

\subsection{Variations with cosmology}
\label{sec:G11}

Fig.~\ref{fig:G11} shows differential surface density profiles and
halo mass distributions for three stellar-mass-corrected models with
the same galaxy formation physics (including all parameters) but
different cosmologies, $\mathrm{G11-P'}$, $\mathrm{G11-W1}$ and
$\mathrm{G11-W7'}$.  The differential density profiles for these
models are inconsistent with each other, showing that cosmology also
has a significant effect on the lensing predictions. However,
understanding the source of these differences is not straightforward.
Given the extremely small errors of the model predictions 
(comparable to the symbol size), $\mathrm{G11-W1}$ predicts
significantly stronger lensing signals than the other two models 
for the two highest stellar mass bins. For most of the points 
over the whole radius range, the significance is a few tens of 
$\sigma$ with respect to the model errors. Although it is tempting 
to ascribe this to its higher value of $\sigma_8$, $\mathrm{G11-P'}$ 
has a higher value of $\Omega_m$ and simple considerations predict 
the abundance of massive halos at $z=0$ to depend on the combination 
$\sigma_8\Omega_m^{0.6}$ which is largest for $\mathrm{G11-P'}$ and 
smallest for $\mathrm{G11-W7'}$ (see Table~\ref{tbl:sim}). At 
lower stellar masses, $\mathrm{G11-W1}$ and $\mathrm{G11-W7'}$ both 
predict weaker lensing signals than $\mathrm{G11-P'}$. These 
mass-dependent differences hinder any simple interpretation of 
the cosmology dependence of the lensing signals.

Despite these complications, we again see a fair correspondence
between the differences in effective halo mass (inferred from the
relative amplitude of differential surface density profiles) and the
differences in characteristic halo mass (found directly from the halo mass
distributions). The logarithmic shifts in effective halo mass from
$\mathrm{G11-W1}$ to $\mathrm{G11-P'}$ are found from their lensing
profiles to be -0.156, -0.116, 0.069, 0.135, 0.122, 0.101 and 0.094.
The corresponding shifts in characteristic halo mass are -0.081,
-0.056, -0.014, 0.074, 0.099, 0.104 and 0.100. From $\mathrm{G11-W7'}$
to $\mathrm{G11-P'}$, the shifts in effective halo mass are 
 -0.038, 0.029, 0.123, 0.160, 0.165, 0.151 and 0.164, while the
shifts in characteristic halo mass are 0.013, 0.045, 0.070, 0.110,
0.106, 0.099 and 0.102. Again, the failure of these sets of numbers to
correspond perfectly indicates a model-dependence of the ratio between 
the effective halo mass for lensing and our characteristic halo mass,
$\langle\log M_{200}\rangle$. 

\subsection{The model dependence of effective halo masses}
\label{sec:medianscatter}

The differential surface density profiles that we have estimated for
the SDSS through gravitational lensing and calculated directly from 
our galaxy formation simulations are a weighted average of those of the
individual galaxies in each stellar mass bin. Similarly, the stacked SZ
and X-ray signals estimated from Planck and ROSAT data by P13 and A15
are weighted averages of the signals from individual galaxies, which
the modelling relates to halo mass through power-law scaling
relations. In all three cases, the stacked signal is very similar to
that which would be found if all galaxies had halos of one particular
mass, the effective halo mass for that stellar mass bin. The effective
halo masses will not be the same in the three cases, however, because
of differing weights and differing sensitivity of the observed signal
to redshift and halo mass.

If the distribution of halo masses at given stellar mass were always
the same shape, for example, lognormal with known width, then the
ratios of the effective halo masses to each other and to other
characteristic masses such as $\langle M_{200}\rangle$ or
$\mathrm{median}(M_{200})$ would always be the same. The direct
measurement of any one of these quantities could then be used to
calibrate the others. We have seen in previous sections, however, that
this is not the case; the shape of the halo mass distribution varies
between stellar mass bins and from model to model. This introduces
model-dependence into any gravitational lensing based recalibration
of the scaling relations linking the SZ and X-ray properties of halos
to their mass, a primary goal of this paper. In this subsection we use
our suite of galaxy formation simulations to estimate the level of
uncertainty resulting from this model-dependence.

In their Table~B.1, P13 give SZ effective halo masses together with
mean and median values of $M_{200}$ for each stellar mass bin used in
their SZ stacking analysis. A15 give X-ray effective halo masses for
these same bins in their Table A.1. In both cases, the effective halo
masses are estimated using $\mathrm{G13-W7'}$ and are much closer to
$\mathrm{median}(M_{200})$ than to $\langle M_{200}\rangle$ in each
bin. Since we do not wish to repeat the detailed modelling of the SZ
or X-ray analyses in this paper, we will assume that the model
dependence of the ratio of SZ (or X-ray) effective mass to
gravitational lensing effective mass is similar to that of the ratio
of $\mathrm{median}(M_{200})$ to gravitational lensing effective
mass. The rms scatter in (the logarithm of) this ratio over our set of
galaxy formation simulations then provides an estimate of the
systematic uncertainty induced by model-dependence when recalibrating
the scaling relations of P13 and A15 based on a single simulation,
$\mathrm{G13-W7'}$. Note that the accuracy of this estimate 
is not explicitly tested in our analysis. It can only be tested in 
detail by running the full observational modelling analysis of P13 
and A15 on all the other models, which is not possible within the 
author collaboration of this paper.
 
In this paper, we consider eight different galaxy formation
simulations (see Table~\ref{tbl:model}), for each of which we create
mock SDSS samples using both the stellar masses originally produced by
the simulations and stellar masses corrected so that the simulated
stellar mass function matches the SDSS function exactly. Thus, for
each of our seven stellar mass bins we end up with 16 different halo
mass distributions with corresponding lensing signal predictions.  For
each case and each bin, we find the shift in halo mass, $\Delta\log
M_{200}$, needed to bring the predicted lensing signal into agreement
with our SDSS measurement (see Section 6 for details) and we apply
this shift to $\mathrm{median}(\log M_{200})$ and $\langle\log
M_{200}\rangle$. If the halo mass distributions predicted for the bin
by the 16 different models were all the same shape, the 16 values
obtained for each characteristic mass would coincide. This is not the
case, showing that the ratio of lensing effective mass to
characteristic mass is varying from model to model.

Table~\ref{tbl:halomass} gives the results of this exercise. For each
stellar mass bin, column 2 gives the mean over our 16 simulations of
the recalibrated value of $\langle\log M_{200}\rangle$, while column 3
shows the rms scatter among these values. Columns 4 and 5 are similar
but for the recalibrated values of $\mathrm{median}(\log M_{200})$.
These means are thus our best lensing-based estimates of these
particular characteristic halo masses for real SDSS galaxies.
The scatter values give estimates of the systematic uncertainty in
these means due to their model-dependence. For both quantities the
model-dependence is quite small (less than about 0.1 dex) but is an
increasing function of stellar mass, reflecting the fact that central
galaxy stellar mass becomes a poorer proxy for halo mass in more
massive systems. As remarked above, we will later adopt the scatter
values given here as estimates of the systematic uncertainty due to
model-dependence when using our gravitational lensing observations to
recalibrate SZ and X-ray scaling relations for our LBGs.

\section{Galaxy Clustering}
\label{sec:clustering}

\begin{figure}
\epsfig{figure=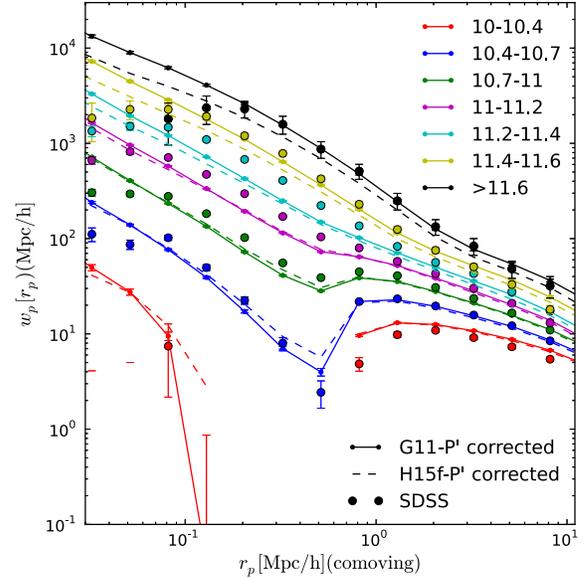,width=0.49\textwidth}
\caption{Projected cross correlation, $w_p(r_p)$, between LBGs and a
  reference galaxy sample with $\log M_\ast/\mathrm{M_\odot}>10$. Solid 
  and dashed lines are for $\mathrm{G11-P'}$ and $\mathrm{H15f-P'}$,
  respectively. We only show errors for solid lines. These errors 
  are comparable to the symbol size. In both cases stellar masses 
  have been corrected to bring the model stellar mass functions into 
  exact agreement with SDSS. Solid dots with error bars are measurements 
  of $w_p(r_p)$ for real galaxies based on SDSS (see 
  Sect.~\ref{sec:sdssclustering}). Lensing profiles and halo mass 
  distributions for these two models can be found on the right-hand 
  side of Fig.~\ref{fig:cosmo}. Note that in this plot $r_p$ denotes 
  comoving (rather than physical) separation. }
\label{fig:clustering}
\end{figure}

In previous subsections we have compared our SDSS measurements of
differential surface density profiles around LBGs to predictions from
a range of galaxy formation simulations, finding the latter to be
significantly affected by the assumed galaxy formation physics and by
the underlying cosmology, even after the model stellar mass functions
have been corrected to bring them into exact agreement with SDSS. Such
effects are particularly important for massive objects and reflect the
considerable (and model-dependent) scatter in the stellar mass -- halo
mass relation. In this subsection we consider clustering of other
galaxies around our LBGs and assess whether this can provide
additional information to help choose between models.


We measure the projected cross-correlation, $w_p(r_p)$, between LBGs
and a reference sample, chosen to be all galaxies more massive than
$10^{10}\mathrm {M_\odot}$ in the SDSS, and also in the simulations.
Cross-correlation with a dense reference sample enables
us to get better signal than for an autocorrelation, and the low mass
threshold reduces the strength of the feature introduced by our
isolation criterion. For SDSS, the correlation is measured as outlined
in Sec.~\ref{sec:sdssclustering}. The weighting results in a
relatively low effective redshift for the measured signal, with 50\%
coming from $z<0.08$ and 90\% from $z<0.15$ even for our highest LBG
stellar mass bin. This reduces evolutionary effects on the signal at
the price of increasing the cosmic variance in the measurement.
Corresponding cross-correlation functions for the simulations can be
measured directly by counting reference galaxies around LBGs in a
projection of the $z=0$ snapshot. Note that, in order to be consistent
with convention in this subfield, we use $r_p$ to denote comoving
(rather than physical) separation in this subsection.

In Fig.~\ref{fig:clustering} we compare our SDSS clustering
measurement with predictions from two representative models,
$\mathrm{G11-P'}$ (solid) and $\mathrm{H15f-P'}$ (dashed), both with
stellar masses corrected to bring their stellar mass function into
agreement with that in the SDSS. The lensing signal predicted by these
same models is compared with our SDSS results in the upper right panel
of Fig.~\ref{fig:cosmo}. The relative behaviour of the two models is
very similar for the two measurements. $\mathrm{G11-P'}$ predicts
significantly stronger signals than $\mathrm{H15f-P'}$ on all scales
for the two highest stellar mass bins, given the extremely small 
errors of the model predictions. Note we only show errors for the 
solid lines and these errors are mostly comparable to the symbol 
size. At lower stellar mass
their signals are very similar, both for clustering and for
lensing. The large-scale clustering amplitudes ($r_p\geq 2$~Mpc/h)
measure the biasing of LBGs relative to reference galaxies and hence
are primarily sensitive to their halo mass, which is measured directly
on smaller scales by the one-halo lensing signal. Fig.~\ref{fig:cosmo}
shows that $\mathrm{G11-P'}$ predicts effective halo masses for the
two highest stellar mass bins which are slightly above those observed,
while those for $\mathrm{H15f-P'}$ are clearly too low. The
large-scale clustering amplitudes of the two models in
Fig.~\ref{fig:clustering} mirror this behaviour relative to the
observations as expected. For lower LBG stellar masses, both models
fit the observed one-halo lensing and large-scale clustering signals
quite well, with the remaining differences probably within the cosmic
variance uncertainties of the clustering measurements.

On smaller scales and for intermediate LBG stellar masses ($10.7 <
\log M_\ast/\mathrm{M_\odot} < 11.4$) the observed clustering signal in 
SDSS is significantly larger than in either model, indicating that the 
number of satellites with $\log M_\ast/\mathrm{M_\odot} > 10.0$ is 
significantly underestimated (although, curiously, not around more or 
less massive LBGs). We have explicitly checked all the other models, and 
found such small scale excess persists across cosmologies. This is unexpected, 
both because Fig.~\ref{fig:cosmo} indicates that the models reproduce 
the {\it mass} distribution around these LBGs quite accurately on 
these same scales, and because \cite{2012MNRAS.424.2574W} found that 
the count of satellites more massive than $\log M_\ast/\mathrm{M_\odot} = 10.0$ 
projected within 300 kpc of a large sample of bright isolated galaxies 
with $10.8 < \log M_\ast/\mathrm{M_\odot} < 11.4$ is reproduced quite 
precisely by $\mathrm{G11-W1}$ (see their Fig.~6). We have not isolated 
the origin of this discrepancy, which may in part reflect the underestimation 
of cosmic variance uncertainties by our bootstrap error estimation
procedure, but it appears unlikely to affect the main topics of this
paper. Note the apparent turnover on very small scales and 
especially for the few most massive bins is due to the effect 
that two fibres cannot be assigned closer than 
55$\arcsec$.

The two models considered in this section are based on the same
simulation, assume the same (Planck) cosmology, and have identical
stellar mass functions. They, nevertheless, predict significantly
different lensing signals for the most massive LBGs because of
differing amounts of scatter in their stellar mass -- halo mass
relations. The main motivation for the present analysis of clustering
was the hope that it would provide an independent observational
indication of which model is more realistic. To the extent that it
provides a significantly better fit to SDSS measurements for both
kinds of data in both the one-halo and the two-halo regime 
of the two most massive bins, $\mathrm{G11-P'}$ appears to 
be a higher fidelity representation of the real stellar mass 
-- halo mass relation (including scatter) than $\mathrm{H15f-P'}$. 


\begin{figure*}
\epsfig{figure=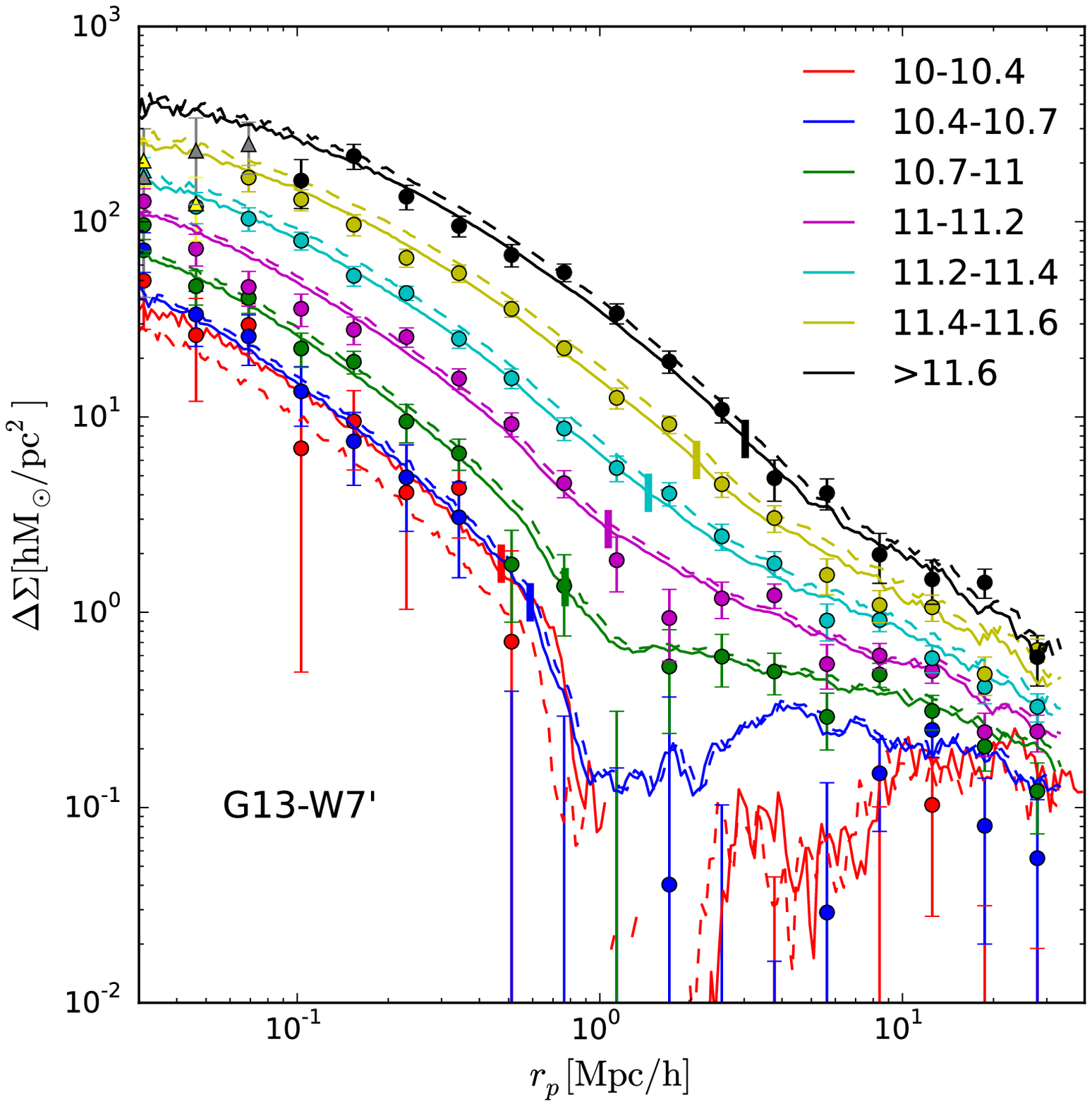,width=0.49\textwidth}
\epsfig{figure=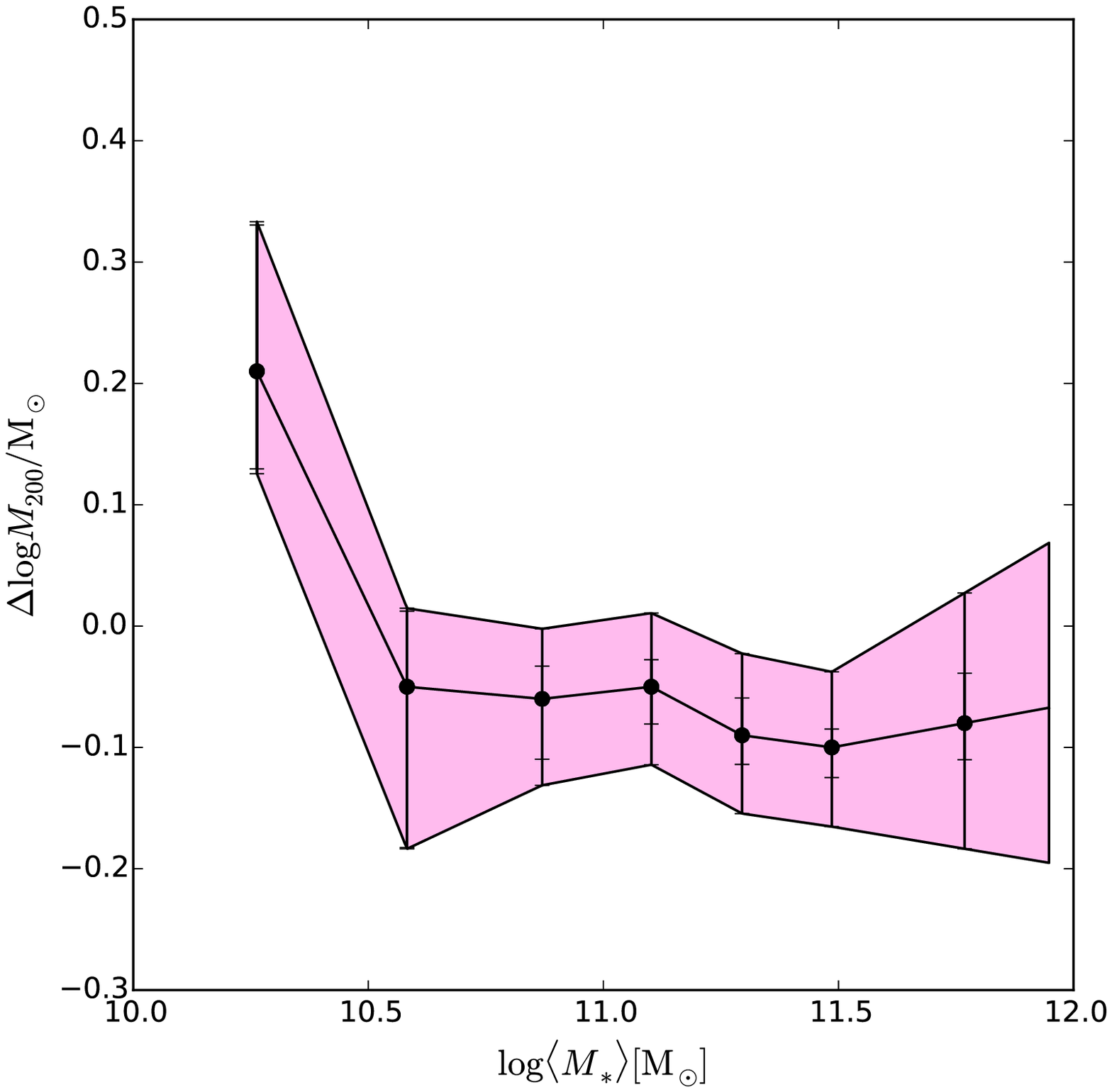,width=0.49\textwidth}
\caption{{\bf Left:} As Fig.~\ref{fig:g11planck}, but now comparing
  our SDSS lensing measurements (dots and triangles with errors) 
  to the predictions of $\mathrm{G13-W7'}$, the model used by P13 
  and A15. Dashed curves give predictions for the original model, 
  while solid curves are shifted upwards and to the right (or down 
  and to the left) by $(\Delta\log M_{200})/3$, the (potentially 
  negative) amount needed to bring them into optimal agreement with 
  the SDSS data in the one-halo regime, $r_p < 4 R_{200}$ (indicated 
  by a short vertical bar on each curve). Upper triangles are 
  shown in a lighter colour and are
  not used when fitting because of concerns that they may be affected
  by light from the LBG. To be consistent with P13 and A15, each LBG
  in the SDSS was matched in stellar mass alone to LBGs in the $z=0$
  output of the simulation when making this plot.  {\bf Right:}
  Recalibration shifts in $\log M_{200}$ are shown as filled symbols
  with error bars indicating: (a) statistical uncertainties in the
  shifts resulting from the error bars on the SDSS lensing
  measurements; and (b) systematic uncertainties (taken from
  Table~\ref{tbl:halomass}) in the conversion from effective lensing
  mass to $\langle\log M_{200}\rangle$.  The shorter error bar
  reflects (a) alone, whereas the longer includes (a) and (b) added in
  quadrature. The pink shaded region interpolates linearly both in the
  recalibration shift and in its uncertainty.  }
\label{fig:correction}
\end{figure*}

\section{Lensing calibration of SZ and X-ray scaling relations}
\label{sec:scaling}

\begin{figure*}
\epsfig{figure=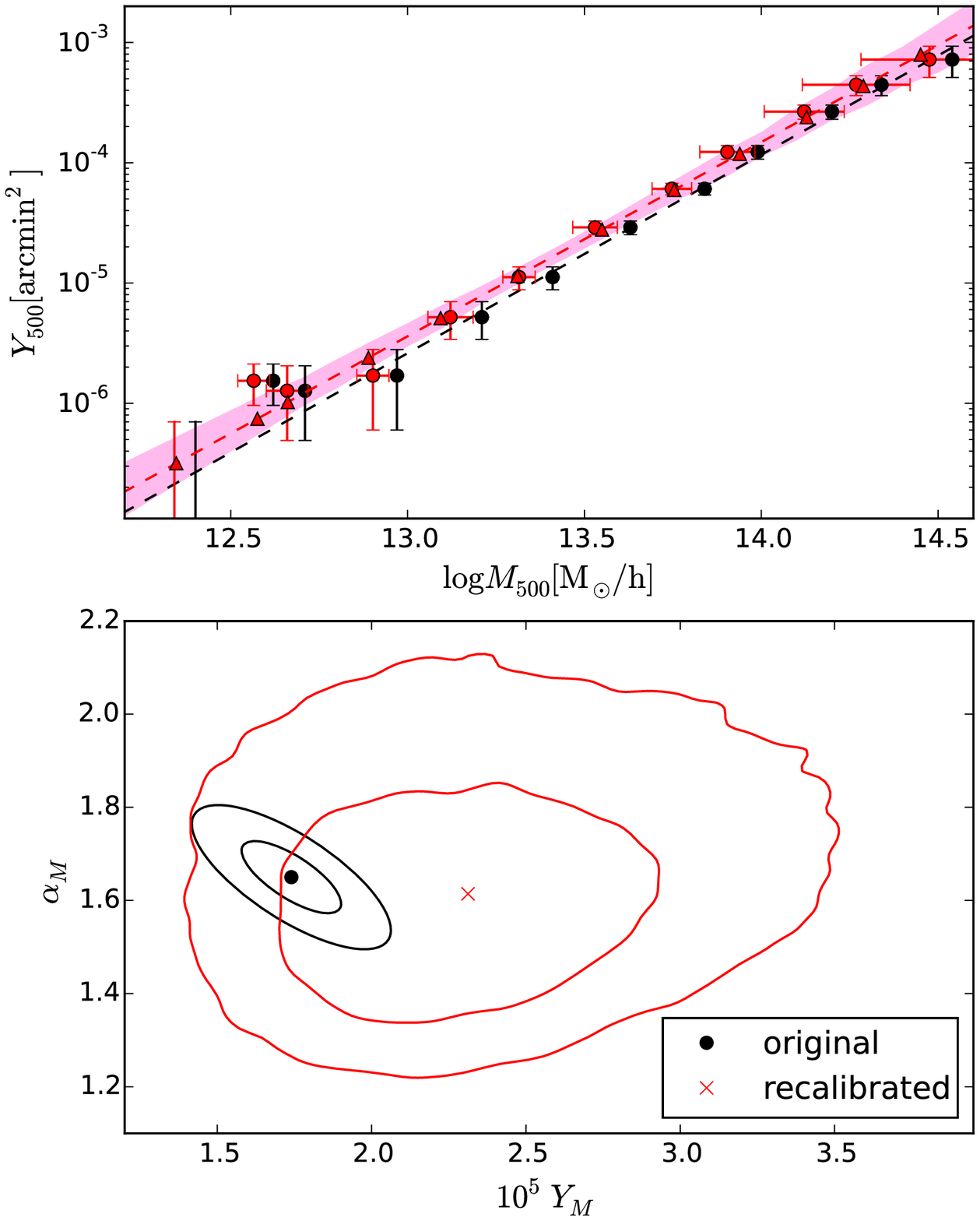,width=0.49\textwidth}
\epsfig{figure=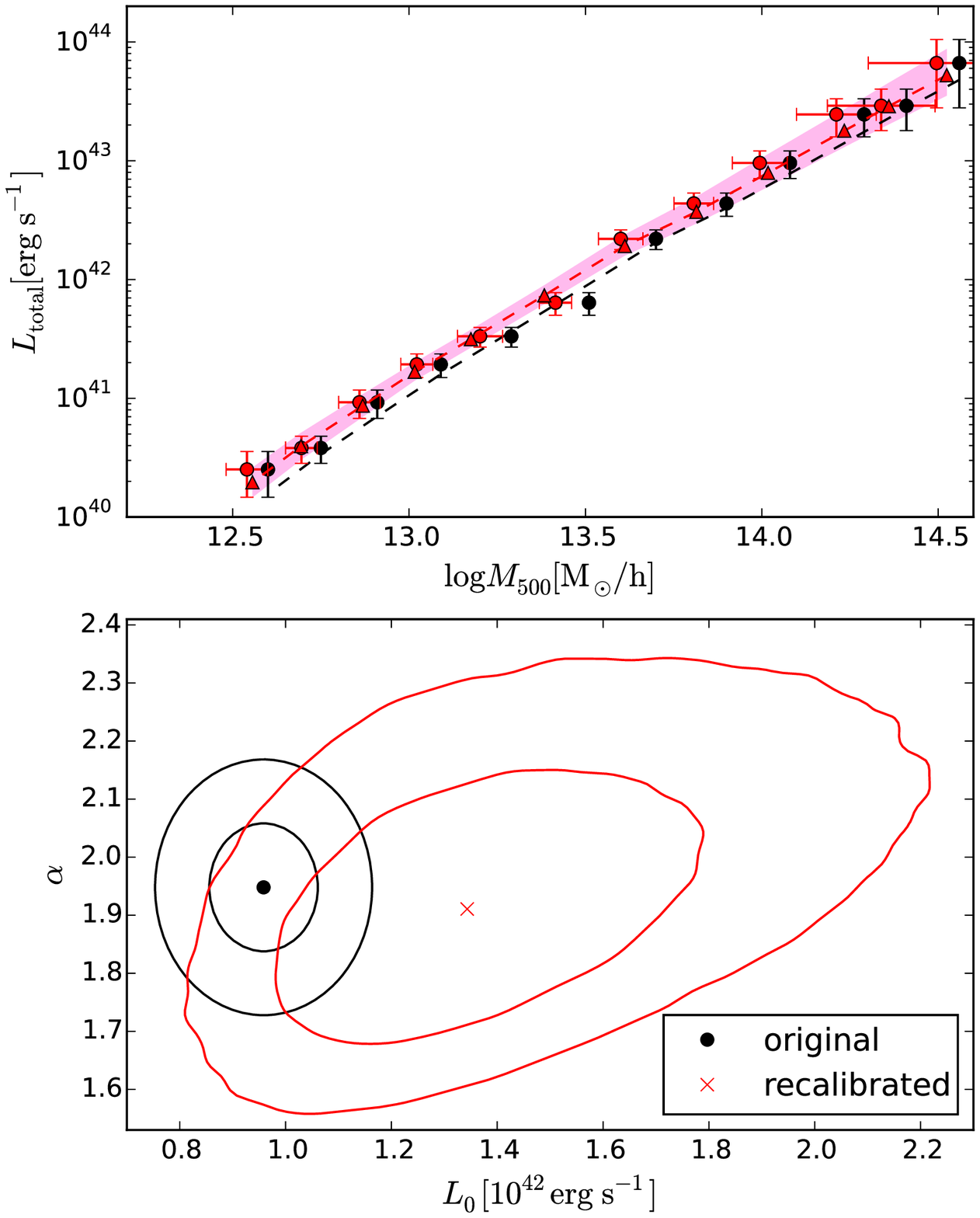,width=0.49\textwidth}
\caption{The SZ $Y_{500}$-$M_{500}$ relation (top left) and X-ray
  $L_\mathrm{total}$-$M_{500}$ relation (top right). Black dots with
  vertical error bars are the original relations given in P13 and A15. 
  Red dots with both vertical and horizontal error bars are the same
  measurements after recalibrating the effective halo masses using our
  lensing measurements.  The black and red dashed lines are best-fit
  power-law scaling relations for the original and the recalibrated
  data, respectively. Triangles along the red dashed lines indicate
  the maximum likelihood effective halo masses associated with each
  stellar mass bin. The bolometric correction, $C^{-1}_\mathrm{bolo}$,
  is included for the X-ray model. This varies between bins, so the
  model is not strictly a straight line. At each halo mass, the pink
  shaded regions enclose 68\% of the scaling relations proposed by our
  MCMC chains.  The bottom panels show the best-fit parameters from
  these chains as symbols surrounded by contours indicating their 1-
  and 2-$\sigma$ confidence regions.  Black and red again indicate the
  original and recalibrated cases, respectively.  The confidence
  regions are larger after recalibration because both systematic
  modelling uncertainties and random noise in the effective halo mass
  estimates are now consistently included.}
\label{fig:scaling}
\end{figure*}

P13 and A15 stacked the Planck and ROSAT sky maps around the LBG
sample analysed in this paper in order to measure mean
Sunyaev-Zeldovich signals and mean X-ray luminosities as functions of
stellar mass. The large sample size resulted in detections to
substantially lower masses and with higher precision than had been
possible in earlier work. In both papers, $\mathrm{G13-W7'}$ was
combined with simple assumptions about halo gas profiles and power-law
relations between halo mass and observable signal to carry out
detailed forward modelling of the stacking and detection
procedures. This allowed estimation both of effective halo mass as a
function of LBG stellar mass, and of the parameters of the scaling
relation between observable and halo mass.  Since $\mathrm{G13-W7'}$
was originally tuned to reproduce the stellar mass function of the
SDSS, this procedure was effectively an abundance-matching calibration
of the halo mass--stellar mass relation which, as we have seen in
earlier sections, is model-dependent because of: (i) residual mismatch
between model and observed stellar mass functions; (ii) uncertainties
in the cosmological parameters; and (iii) the model-dependence of the
detailed shape of the halo mass distribution at fixed LBG stellar mass
(including effects due to varying satellite fractions and offsets).

The goal of the current section is to remove most of this
model-dependence by recalibrating the scaling relations of P13 and A15
using our gravitational lensing measurements. The new calibrations
will be model-dependent only to the extent that the {\it ratio} of the
effective halo mass for gravitational lensing to that for SZ (or
X-ray) stacking varies between models. Since all three effective
masses correspond to rather similar moments of the halo mass
distribution, this dependence is quite weak, and we include an
estimate of the associated uncertainty in the analysis leading to our
final recalibrated scaling relations. When estimating halo masses, P13
and A15 matched the observed LBGs to simulated LBGs of the same
stellar mass in the $z=0$ output of $\mathrm{G13-W7'}$. For
consistency, we will match in the same way in this section, rather
than matching in stellar mass, luminosity and redshift, as in earlier
sections.

In the left panel of Fig.~\ref{fig:correction} we compare our SDSS
measurements of differential lensing surface density profiles with the
predictions of $\mathrm{G13-W7'}$. Dashed curves are for the original
model, as used by P13 and A15. These are close to the observations,
suggesting that the model already represents the halos of SDSS
galaxies well, and that the shifts needed to recalibrate to the
lensing data will be small. For each stellar mass bin, we estimate 
the shift (and its uncertainty) by assuming that the halos of all 
model LBGs change mass by $\Delta\log M_{200}$ without changing
concentration\footnote{$\Delta\log M_{200}$ is fairly small for 
$\mathrm{G13-W7'}$. Given the flat slope of concentration 
halo mass relations \citep[e.g.][where the slope is about -0.1]
{2001MNRAS.321..559B,2008MNRAS.390L..64D}, the assumption of 
unchanged concentration should be reasonable, i.e., 0.1 dex of 
change in halo masses corresponds to at most a few percent  
change in halo concentrations.}. In the one-halo regime, the 
differential density profiles then shift upwards by 
$\frac{1}{3}\Delta \log M_{200}$ and to the right by the same 
amount. We find the best shift and its uncertainty by calculating 
$\chi^2$ values for the model curves with respect to the SDSS 
data at $r_p<4 R_{200}$, the region over which 
\cite{2008MNRAS.388....2H} found the lensing profiles of central
galaxies to be well described by the one-halo term. For the three 
most massive stellar mass bins, we exclude the innermost three, 
two or one SDSS data points from the fit (indicated by a lighter 
colour using upper triangle symbols in Fig.~\ref{fig:correction}) 
because measurements on these scales may be affected by light from 
the central galaxy. We determine appropriate inner scales for 
exclusion using the criteria set out by \cite{2005MNRAS.361.1287M}. 

The solid curves in the left panel of Fig.~\ref{fig:correction} show
results for $\mathrm{G13-W7'}$ after applying these recalibration
shifts. A short vertical bar on each curve marks $r_p=4 R_{200}$. The
shifts themselves are shown as the black points in the right panel,
with the shorter error bars indicating uncertainties from the $\chi^2$
fitting. Note the quantity plotted for the $x$-axis, 
$\log \langle M_\ast \rangle$, is averaged by taking equation~\ref{eqn:weight} 
for each galaxy as weights. The SDSS measurements for the lowest 
stellar mass bin are very noisy, but this point is not needed for 
our recalibration below. At higher mass, both the shifts and their 
statistical uncertainties are quite small. Assuming that the {\it shape} 
of the halo mass distribution at given stellar mass is unchanged\footnote{
This should be a reasonable assumption given the fact that the curve 
in the right panel is close to flat at $\log \langle M_\ast \rangle>10.5$}, 
these shifts can be applied to the effective halo masses estimated by 
P13 and A15 in order to calibrate them to the lensing data. To obtain 
the overall uncertainties shown in Fig.~\ref{fig:correction}, we add 
in quadrature a systematic uncertainty due to the fact that the ratio 
of lensing to SZ (or X-ray) effective mass may differ between
$\mathrm{G13-W7'}$ and the real galaxies. This is estimated as the
larger of columns 3 and 5 in Table~\ref{tbl:halomass}, which give the
scatter in similar effective mass ratios over our 16 galaxy 
formation models. The black line and the surrounding red band in 
Fig.~\ref{fig:correction} then give the correction and associated 
uncertainty which we apply to the effective halo masses of P13 and 
A15 in order to recalibrate their scaling relations. Note that the 
uncertainties come predominantly from noise in the lensing results 
at small stellar mass, but from modelling systematics at large 
stellar mass.

In Fig.~\ref{fig:scaling} we show the effect of recalibration on the
scaling relations. The upper left panel replots the
$\tilde{Y}_{500}$--$M_{500}$ relation from Fig. 9 of P13 as black
symbols. As in the original paper, these have vertical error bars,
indicating flux uncertainties estimated from the stacking analysis,
but allow for no uncertainty in effective halo mass $M_{500}$. Red
symbols show the recalibrated relation.  The vertical error bar is
unchanged but the points now also have a horizontal error
bar\footnote{ Uncertainties for the twelve SZ/X-ray stellar mass bins
  are estimated by interpolating linearly between those for the
  neighbouring lensing bins. Since we have changed to the Planck
  cosmology, which increases the stellar mass of each SDSS galaxy by
  about 10\% over that used by P13 and A15, an additional correction
  of $\log (0.704/0.673)^2$ was added to the central stellar mass of
  each SZ/X-ray bins bin before interpolating.}, corresponding to the
pink band in Fig.~\ref{fig:correction}. The upper right panel is a
similar plot for the $L_{\rm total}$--$M_{500}$ relation from Tables 3
and A1 of A15. Exact definitions of the various observational
quantities plotted here can be found in the original papers. We fit
these relations to power-law expressions similar to those of Eqn.~1 in
P13 and Eqn.~3 in A15.  For the original data we minimise a $\chi^2$
estimated from the vertical offsets in $\tilde{Y}_{500}$ and
$L_\mathrm{total}$ and the bootstrap errors from the original papers.
For the recalibrated data, we take account of the uncertainties in
both directions. This is a complex statistical problem because of the
strongly correlated errors on neighbouring bins resulting from the fact
that our twelve correction factors are interpolated/extrapolated from
the seven lensing measurements of Fig.~\ref{fig:correction}. This
effect is particularly important for the highest stellar mass bins. We
have developed a fully Bayesian procedure similar to that of
\cite{2007ApJ...665.1489K} and presented in detail in the Appendix,
which accounts consistently for these effects.

We write our scaling relations as
\begin{equation}
\tilde{Y}_{500}=Y_M\big(\frac{M_{500}}
{10^{13.5}\mathrm{M_\odot}}\big)^{\alpha_M}
\label{eqn:SZscaling}
\end{equation}
and
\begin{equation}
L_{\rm total}=L_0 \times E(z)^{7/3} \times C^{-1}_\mathrm{bolo} 
\big(\frac{M_{500}}{10^{13.5} \mathrm{M_\odot}}\big)^\alpha.
\label{eqn:Xrayscaling}
\end{equation} 
These differ from the scaling relations of P13 and A15 only in that we
here adopt lower mass pivot-points than in the original papers in
order to decorrelate uncertainties on the two parameters. This is
necessary because our lensing data constrain halo masses most tightly
at a few times $10^{13} \mathrm{M_\odot}$.  In
Eqn.~\ref{eqn:Xrayscaling}, $C^{-1}_\mathrm{bolo}$ is a bolometric
correction factor to convert the 0.5-2.0~keV luminosities to total
luminosities, and is given in Table~2 of
\cite{2015MNRAS.449.3806A}. $E(z)=\sqrt{\Omega_m(1+z)^3+\Omega_\Lambda}$
is the dimensionless Hubble parameter in a flat cosmological model 
to allow for self-similar evolution, which is already included in the 
P13 definition of $\tilde{Y}_{500}$ (see the end of their Section 1).

The best-fit scaling relations are overplotted in the upper panels of
Fig.~\ref{fig:scaling} as black dashed lines for the original
measurements, and red dashed lines for the recalibrated
measurements. The lower panels show the amplitudes and slopes of these
fits, together with contours indicating 1 and 2-$\sigma$ confidence
regions. For the original data, these are ellipses obtained from the
derivatives of $\chi^2$ at its minimum. For the recalibrated data,
they are taken directly from the MCMC chains used to map out the
likelihood surfaces. Red triangles along the best fit relations show
the maximum likelihood effective halo masses associated with each
stellar mass bin. At each halo mass, the pink shaded regions enclose
68\% of the scaling relations proposed by our MCMC chains. The maximum
likelihood parameters for our recalibrated scaling relations and their
68\% confidence regions, both obtained from the one-dimensional
marginalised likelihood distributions generated by our MCMC chains,
are:


\begin{equation}
10^5\times Y_M = 2.31~~[1.93, 2.67],~~~\alpha_M = 1.61~~[1.43, 1.75] 
\end{equation}
and


\begin{equation}
L_0/10^{42}{\mathrm {erg~s^{-1}}} = 1.34~~[1.12, 1.67],~~~\alpha = 1.91~~[1.78, 2.07]
\end{equation}

The recalibration makes the slope of the best-fit SZ relation slightly
shallower, while its amplitude increases by about 30\%.  In addition,
the confidence region is significantly larger now that uncertainties
in the estimation of effective halo masses are consistently
included. For the best fit X-ray scaling relation, the slope
barely changes, but the amplitude increases by about 40\%. These
changes in amplitude are due primarily to the decrease in effective
halo mass for most stellar mass bins (corresponding to the
negative corrections in Fig.~\ref{fig:correction}). Note that, as
expected, the uncertainties in the amplitude and slope of the scaling
relations are only weakly correlated, reflecting the fact that mean SZ
and X-ray fluxes are best constrained (to better than $\pm 20\%$ at
one $\sigma$) for halo masses near our new pivot value. This is
perhaps the most interesting result of our paper, since the systematic
X-ray and SZ properties of halos in the range $10^{12.5}<
M_{500}/\mathrm{M_\odot}< 10^{14}$ have not previously been constrained 
at anything approaching this level of precision because of difficulties
in the selection of representative samples of objects, in the
estimation of their halo masses, and in the measurement of their X-ray
and SZ signals.


\section{Conclusions}

We have measured weak gravitational lensing profiles at high
signal-to-noise and as a function of stellar mass for stacks of
Locally Brightest Galaxies selected from SDSS/DR7, comparing the
observed signal over the projected radius range $30{\mathrm{h^{-1} kpc}}< r_p
< 30{\mathrm{h^{-1} Mpc}}$ to predictions from eight semi-analytic galaxy
formation simulations differing in underlying N-body simulation, in
assumed cosmological parameters and in the modelling of galaxy
formation processes. This LBG sample was previously used by
\cite{2013A&A...557A..52P} to measure stacked SZ signal and by
\cite{2015MNRAS.449.3806A} to measure stacked X-ray signal as
functions of LBG stellar mass. Both studies used a galaxy formation
simulation from the set in this paper to carry out forward modelling
of the sample selection and signal measurement processes, and hence to
derive scaling relations between the mass of dark matter halos and the
properties of their hot gas atmospheres. Here, we use our simulation
set to explore the model-dependence of such calibrations, before
recalibrating the SZ and X-ray scaling relations to a halo mass scale
based directly on our lensing results. This recalibration takes full
account of the observational uncertainties in the lensing, SZ and
X-ray observations, as well as residual modelling uncertainties
arising from the scatter in halo mass at given LBG stellar mass.


We compared the differential surface density profiles measured around
our stacks of SDSS LBGs to those around LBGs defined in an exactly
analogous way in our galaxy formation simulations, taking care to
reproduce both the redshift distribution of the observed galaxies at
each stellar mass and the redshift-dependent weighting of their
lensing signals. For each of our eight simulations we compared to
predictions from the simulation both in its original form and after
applying small stellar mass corrections to bring its stellar mass
function into exact agreement with that observed in the SDSS. Such
corrections eliminate inaccurate stellar mass functions as a possible
source of disagreement between the simulated and observed lensing
signals. From these comparisons we found:

\begin{itemize}
 \item All of our galaxy formation simulations predict lensing
   profiles that are in good qualitative agreement with the SDSS 
   data over two orders of magnitude in projected radius and a factor 
   of 30 in stellar mass. The quantitative level of agreement varies 
   between simulations, however, with amplitude offsets of up to a 
   few tenths of a dex in some cases. The specific model used by
   \cite{2013A&A...557A..52P} and \cite{2015MNRAS.449.3806A} to
   calibrate their scaling relations produces one of the best fits to
   the lensing data and requires very small corrections to bring its
   stellar mass function into agreement with SDSS.

 \item Including stellar mass corrections improves the agreement in
   mean halo mass as a function of stellar mass between simulations,
   but significant scatter remains, particularly for large stellar
   masses. In most cases, these corrections also improve agreement
   with the SDSS lensing measurements for $\log
   M_\ast<10^{11.2}\mathrm{M_\odot}$, but this is often not the case
   at high stellar mass.

 \item Models with identical stellar mass functions (as a result of
   the stellar mass corrections) and based on the same N-body
   simulation can produce significantly different lensing profiles and
   halo mass distributions at fixed stellar mass. This is especially
   noticeable for $\log M_\ast>10^{11.4}\mathrm{M_\odot}$. In general,
   shifts in the amplitude of the lensing profiles are matched, at
   least qualitatively, by shifts in mean host halo mass.  Because the
   scatter in halo mass is large and model-dependent, LBG stellar mass
   has large systematic uncertainties as a proxy for host halo mass,
   and so may not be useful for many cosmological applications.
 
 \item Simulations with the same stellar mass function, the same
   galaxy formation model, but different cosmologies can make
   different predictions both for lensing profiles and for halo mass
   distributions. The differences are not straightforward to
   interpret, with several factors playing a significant role. 
   Again, this is likely to complicate the drawing of conclusions 
   about cosmological parameters from lensing observations of the 
   type considered here.
\end{itemize}

Because the model-dependence of the scatter in halo mass at fixed LBG
stellar mass leads to systematic uncertainties in the interpretation
of the observed SDSS lensing profiles, we investigated whether
additional useful constraints could be obtained from the clustering of
other galaxies around LBGs, specifically a reference sample of all
objects more massive than $\log M_\ast>10^{10}\mathrm{M_\odot}$.  A
comparison of predictions from two of our simulations to clustering
measurements from SDSS showed that differences in effective LBG halo
mass inferred from lensing are also seen as differences in the
amplitude of the large-scale (two-halo) clustering of other galaxies
around LBGs, and that conclusions about the relative merits of the
models drawn from their lensing predictions are consistent with those
drawn from their large-scale clustering. On smaller scales, clustering
is sensitive to how reference galaxies populate LBG halos, and this
introduces additional model-dependencies. In our case, models which
predict well the observed mean distribution of dark matter around LBGs
on scales of a few hundred kpc do not, for LBGs of intermediate
stellar mass, predict well the distribution of reference galaxies on
these same scales.

By combining results from our eight simulations both with and without
stellar mass corrections we are able to estimate, for each LBG stellar
mass bin in our lensing analysis, the systematic uncertainty in the
ratio between the effective halo mass for gravitational lensing and
the mean and median halo masses. Such uncertainties reflect the
model-dependence of the detailed shape of the halo mass distribution
at given LBG stellar mass. In Section 6 we use them to estimate the
systematic uncertainty in the ratio of effective halo mass for
gravitational lensing to effective halo mass for SZ or X-ray stacking
when recalibrating the scaling relations of \cite{2013A&A...557A..52P}
and \cite{2015MNRAS.449.3806A}\footnote{P13 and A15 have made 
other assumptions which may introduce systematic uncertainties. 
For the SZ scaling relation, a pressure profile has to be assumed 
to estimate the effective halo mass. To obtain the X-ray scaling 
relation, A15 have made bolometric corrections to the observed 
X-ray flux, which might introduce additional uncertainties. Studying 
such effects is outside the scope of this paper.}. Because the lensing 
analysis is carried out in broader stellar mass bins than the earlier 
stacking analyses, interpolation of the lensing results is required 
when carrying through the recalibration. This complicates the formal
problem of accounting properly both for observational uncertainties in
the three stacking measurements and for systematic uncertainties from
residual model dependencies. We have developed a Bayesian analysis
technique for this problem which is presented in the Appendix. We
believe the resulting recalibrated scaling relations, given in
equations 8 to 10, to be robust and minimally model dependent and to
be quoted with realistic uncertainties. They should be fully
representative for halos in the mass range
$10^{12.5}<M_{500}/\mathrm{h^{-1}\mathrm{M_\odot}}<10^{14.5}$, a 
significantly broader and lower range than it has been possible to 
cover through observations of individual objects.

\section*{Acknowledgments}

This work was supported by the European Research Council [GA 267291] 
COSMIWAY and Science and Technology Facilities Council
Durham Consolidated Grant. WW acknowledges a Durham Junior Research 
Fellowship (RF040353). RM was supported by the US Department of 
Energy Early Career Award Program for the duration of this work.
BH was supported by the Advanced Grant 246797 ``GALFORMOD''
from the European Research Council and a Zwicky fellowship.
SW thanks Simona Vegetti for useful Bayesian conversations.
Much of the final editing of this paper was carried out at the Aspen
Center for Physics, which is supported by National Science Foundation
grant PHY-1066293. WW is grateful for useful discussions with 
Shaun Cole, Jun Hou, Idit Zehavi and Carlos Frenk. We thank 
the anomymous referee for his/her very careful and detailed 
comments.

\bibliography{master}

\begin{thebibliography}{}

\bibitem[\protect\citeauthoryear{{Abazajian}, {Adelman-McCarthy},
  {Ag{\"u}eros}, {Allam}, {Allende Prieto}, {An}, {Anderson}, {Anderson},
  {Annis}, {Bahcall} \& et al.}{{Abazajian} et~al.}{2009}]{2009ApJS..182..543A}
{Abazajian} K.~N.,  {Adelman-McCarthy} J.~K.,  {Ag{\"u}eros} M.~A.,  {Allam}
  S.~S.,  {Allende Prieto} C.,  {An} D.,  {Anderson} K.~S.~J.,  {Anderson}
  S.~F.,  {Annis} J.,  {Bahcall} N.~A.,    et al. 2009, \apjs, 182, 543

\bibitem[\protect\citeauthoryear{{Allen}, {Rapetti}, {Schmidt}, {Ebeling},
  {Morris} \& {Fabian}}{{Allen} et~al.}{2008}]{2008MNRAS.383..879A}
{Allen} S.~W.,  {Rapetti} D.~A.,  {Schmidt} R.~W.,  {Ebeling} H.,  {Morris}
  R.~G.,    {Fabian} A.~C.,  2008, \mnras, 383, 879

\bibitem[\protect\citeauthoryear{{Anderson}, {Gaspari}, {White}, {Wang} \&
  {Dai}}{{Anderson} et~al.}{2015}]{2015MNRAS.449.3806A}
{Anderson} M.~E.,  {Gaspari} M.,  {White} S.~D.~M.,  {Wang} W.,    {Dai} X.,
  2015, \mnras, 449, 3806

\bibitem[\protect\citeauthoryear{{Angulo} \& {Hilbert}}{{Angulo} \&
  {Hilbert}}{2015}]{2015MNRAS.448..364A}
{Angulo} R.~E.,  {Hilbert} S.,  2015, \mnras, 448, 364

\bibitem[\protect\citeauthoryear{{Angulo} \& {White}}{{Angulo} \&
  {White}}{2010}]{2010MNRAS.405..143A}
{Angulo} R.~E.,  {White} S.~D.~M.,  2010, \mnras, 405, 143

\bibitem[\protect\citeauthoryear{{Bernstein} \& {Jarvis}}{{Bernstein} \&
  {Jarvis}}{2002}]{2002AJ....123..583B}
{Bernstein} G.~M.,  {Jarvis} M.,  2002, \aj, 123, 583

\bibitem[\protect\citeauthoryear{{Blanton} \& {Roweis}}{{Blanton} \&
  {Roweis}}{2007}]{2007AJ....133..734B}
{Blanton} M.~R.,  {Roweis} S.,  2007, \aj, 133, 734

\bibitem[\protect\citeauthoryear{{Blanton}, {Schlegel}, {Strauss}, {Brinkmann},
  {Finkbeiner}, {Fukugita}, {Gunn}, {Hogg}, {Ivezi{\'c}}, {Knapp}, {Lupton},
  {Munn}, {Schneider}, {Tegmark} \& {Zehavi}}{{Blanton}
  et~al.}{2005}]{2005AJ....129.2562B}
{Blanton} M.~R.,  {Schlegel} D.~J.,  {Strauss} M.~A.,  {Brinkmann} J.,
  {Finkbeiner} D.,  {Fukugita} M.,  {Gunn} J.~E.,  {Hogg} D.~W.,  {Ivezi{\'c}}
  {\v Z}.,  {Knapp} G.~R.,  {Lupton} R.~H.,  {Munn} J.~A.,  {Schneider} D.~P.,
  {Tegmark} M.,    {Zehavi} I.,  2005, \aj, 129, 2562

\bibitem[\protect\citeauthoryear{{Bruzual} \& {Charlot}}{{Bruzual} \&
  {Charlot}}{2003}]{2003MNRAS.344.1000B}
{Bruzual} G.,  {Charlot} S.,  2003, \mnras, 344, 1000

\bibitem[\protect\citeauthoryear{{Bullock}, {Kolatt}, {Sigad}, {Somerville},
  {Kravtsov}, {Klypin}, {Primack} \& {Dekel}}{{Bullock}
  et~al.}{2001}]{2001MNRAS.321..559B}
{Bullock} J.~S.,  {Kolatt} T.~S.,  {Sigad} Y.,  {Somerville} R.~S.,  {Kravtsov}
  A.~V.,  {Klypin} A.~A.,  {Primack} J.~R.,    {Dekel} A.,  2001, \mnras, 321,
  559

\bibitem[\protect\citeauthoryear{{Cen} \& {Ostriker}}{{Cen} \&
  {Ostriker}}{1999}]{1999ApJ...514....1C}
{Cen} R.,  {Ostriker} J.~P.,  1999, \apj, 514, 1

\bibitem[\protect\citeauthoryear{{Chabrier}}{{Chabrier}}{2003}]{2003PASP..115..763C}
{Chabrier} G.,  2003, \pasp, 115, 763

\bibitem[\protect\citeauthoryear{{Cunha}, {Lima}, {Oyaizu}, {Frieman} \&
  {Lin}}{{Cunha} et~al.}{2009}]{2009MNRAS.396.2379C}
{Cunha} C.~E.,  {Lima} M.,  {Oyaizu} H.,  {Frieman} J.,    {Lin} H.,  2009,
  \mnras, 396, 2379

\bibitem[\protect\citeauthoryear{{Dai}, {Kochanek} \& {Morgan}}{{Dai}
  et~al.}{2007}]{2007ApJ...658..917D}
{Dai} X.,  {Kochanek} C.~S.,    {Morgan} N.~D.,  2007, \apj, 658, 917

\bibitem[\protect\citeauthoryear{{Duffy}, {Schaye}, {Kay} \& {Dalla
  Vecchia}}{{Duffy} et~al.}{2008}]{2008MNRAS.390L..64D}
{Duffy} A.~R.,  {Schaye} J.,  {Kay} S.~T.,    {Dalla Vecchia} C.,  2008,
  \mnras, 390, L64

\bibitem[\protect\citeauthoryear{{Fagotto}, {Bressan}, {Bertelli} \&
  {Chiosi}}{{Fagotto} et~al.}{1994a}]{1994A&AS..104..365F}
{Fagotto} F.,  {Bressan} A.,  {Bertelli} G.,    {Chiosi} C.,  1994a, AAPS, 104,
  365

\bibitem[\protect\citeauthoryear{{Fagotto}, {Bressan}, {Bertelli} \&
  {Chiosi}}{{Fagotto} et~al.}{1994b}]{1994A&AS..105...29F}
{Fagotto} F.,  {Bressan} A.,  {Bertelli} G.,    {Chiosi} C.,  1994b, AAPS, 105,
  29

\bibitem[\protect\citeauthoryear{{Fagotto}, {Bressan}, {Bertelli} \&
  {Chiosi}}{{Fagotto} et~al.}{1994c}]{1994A&AS..105...39F}
{Fagotto} F.,  {Bressan} A.,  {Bertelli} G.,    {Chiosi} C.,  1994c, AAPS, 105,
  39

\bibitem[\protect\citeauthoryear{{Feldmann}, {Carollo}, {Porciani}, {Lilly},
  {Capak}, {Taniguchi}, {Le F{\`e}vre} \& {Renzini}}{{Feldmann}
  et~al.}{2006}]{2006MNRAS.372..565F}
{Feldmann} R.,  {Carollo} C.~M.,  {Porciani} C.,  {Lilly} S.~J.,  {Capak} P.,
  {Taniguchi} Y.,  {Le F{\`e}vre} O.,    {Renzini} A. e.~a.,  2006, \mnras,
  372, 565

\bibitem[\protect\citeauthoryear{{Greco}, {Hill}, {Spergel} \&
  {Battaglia}}{{Greco} et~al.}{2015}]{2015ApJ...808..151G}
{Greco} J.~P.,  {Hill} J.~C.,  {Spergel} D.~N.,    {Battaglia} N.,  2015, \apj,
  808, 151

\bibitem[\protect\citeauthoryear{{Guo}, {White}, {Angulo}, {Henriques},
  {Lemson}, {Boylan-Kolchin}, {Thomas} \& {Short}}{{Guo}
  et~al.}{2013}]{2013MNRAS.428.1351G}
{Guo} Q.,  {White} S.,  {Angulo} R.~E.,  {Henriques} B.,  {Lemson} G.,
  {Boylan-Kolchin} M.,  {Thomas} P.,    {Short} C.,  2013, \mnras, 428, 1351

\bibitem[\protect\citeauthoryear{{Guo}, {White}, {Boylan-Kolchin}, {De Lucia},
  {Kauffmann}, {Lemson}, {Li}, {Springel} \& {Weinmann}}{{Guo}
  et~al.}{2011}]{2011MNRAS.413..101G}
{Guo} Q.,  {White} S.,  {Boylan-Kolchin} M.,  {De Lucia} G.,  {Kauffmann} G.,
  {Lemson} G.,  {Li} C.,  {Springel} V.,    {Weinmann} S.,  2011, \mnras, 413,
  101

\bibitem[\protect\citeauthoryear{{Guo}, {White}, {Li} \&
  {Boylan-Kolchin}}{{Guo} et~al.}{2010}]{2010MNRAS.404.1111G}
{Guo} Q.,  {White} S.,  {Li} C.,    {Boylan-Kolchin} M.,  2010, \mnras, 404,
  1111

\bibitem[\protect\citeauthoryear{{Han}, {Eke}, {Frenk}, {Mandelbaum},
  {Norberg}, {Schneider}, {Peacock}, {Jing}, {Baldry}, {Bland-Hawthorn},
  {Brough}, {Brown}, {Liske}, {Loveday} \& {Robotham}}{{Han}
  et~al.}{2015}]{2015MNRAS.446.1356H}
{Han} J.,  {Eke} V.~R.,  {Frenk} C.~S.,  {Mandelbaum} R.,  {Norberg} P.,
  {Schneider} M.~D.,  {Peacock} J.~A.,  {Jing} Y.,  {Baldry} I.,
  {Bland-Hawthorn} J.,  {Brough} S.,  {Brown} M.~J.~I.,  {Liske} J.,  {Loveday}
  J.,    {Robotham} A.~S.~G.,  2015, \mnras, 446, 1356

\bibitem[\protect\citeauthoryear{{Hayashi} \& {White}}{{Hayashi} \&
  {White}}{2008}]{2008MNRAS.388....2H}
{Hayashi} E.,  {White} S.~D.~M.,  2008, \mnras, 388, 2

\bibitem[\protect\citeauthoryear{{Henriques}, {White}, {Thomas}, {Angulo},
  {Guo}, {Lemson}, {Springel} \& {Overzier}}{{Henriques}
  et~al.}{2015}]{2015MNRAS.451.2663H}
{Henriques} B.~M.~B.,  {White} S.~D.~M.,  {Thomas} P.~A.,  {Angulo} R.,  {Guo}
  Q.,  {Lemson} G.,  {Springel} V.,    {Overzier} R.,  2015, \mnras, 451, 2663

\bibitem[\protect\citeauthoryear{{Hirata} \& {Seljak}}{{Hirata} \&
  {Seljak}}{2003}]{2003MNRAS.343..459H}
{Hirata} C.,  {Seljak} U.,  2003, \mnras, 343, 459

\bibitem[\protect\citeauthoryear{{Jenkins}}{{Jenkins}}{2013}]{2013MNRAS.434.2094J}
{Jenkins} A.,  2013, \mnras, 434, 2094

\bibitem[\protect\citeauthoryear{{Kauffmann}, {Heckman}, {White}, {Charlot},
  {Tremonti}, {Brinchmann}, {Bruzual} \& et al.}{{Kauffmann}
  et~al.}{2003}]{2003MNRAS.341...33K}
{Kauffmann} G.,  {Heckman} T.~M.,  {White} S.~D.~M.,  {Charlot} S.,  {Tremonti}
  C.,  {Brinchmann} J.,  {Bruzual} G.,    et al. P.,  2003, \mnras, 341, 33

\bibitem[\protect\citeauthoryear{{Kelly}}{{Kelly}}{2007}]{2007ApJ...665.1489K}
{Kelly} B.~C.,  2007, \apj, 665, 1489

\bibitem[\protect\citeauthoryear{{Komatsu}, {Smith}, {Dunkley}, {Bennett},
  {Gold}, {Hinshaw}, {Jarosik} \& {Larson}}{{Komatsu}
  et~al.}{2011a}]{2011ApJS..192...18K}
{Komatsu} E.,  {Smith} K.~M.,  {Dunkley} J.,  {Bennett} C.~L.,  {Gold} B.,
  {Hinshaw} G.,  {Jarosik} N.,    {Larson} 2011a, \apjs, 192, 18

\bibitem[\protect\citeauthoryear{{Komatsu}, {Smith}, {Dunkley}, {Bennett},
  {Gold}, {Hinshaw}, {Jarosik} \& {Larson}}{{Komatsu}
  et~al.}{2011b}]{Komatsu2011}
{Komatsu} E.,  {Smith} K.~M.,  {Dunkley} J.,  {Bennett} C.~L.,  {Gold} B.,
  {Hinshaw} G.,  {Jarosik} N.,    {Larson} D. e.~a.,  2011b, \apjs, 192, 18

\bibitem[\protect\citeauthoryear{{Landy} \& {Szalay}}{{Landy} \&
  {Szalay}}{1993}]{1993ApJ...412...64L}
{Landy} S.~D.,  {Szalay} A.~S.,  1993, \apj, 412, 64

\bibitem[\protect\citeauthoryear{{Le Brun}, {McCarthy} \& {Melin}}{{Le Brun}
  et~al.}{2015}]{2015MNRAS.451.3868L}
{Le Brun} A.~M.~C.,  {McCarthy} I.~G.,    {Melin} J.-B.,  2015, \mnras, 451,
  3868

\bibitem[\protect\citeauthoryear{{Li} \& {White}}{{Li} \&
  {White}}{2009}]{2009MNRAS.398.2177L}
{Li} C.,  {White} S.~D.~M.,  2009, \mnras, 398, 2177

\bibitem[\protect\citeauthoryear{{Mandelbaum}, {Hirata}, {Broderick}, {Seljak}
  \& {Brinkmann}}{{Mandelbaum} et~al.}{2006}]{2006MNRAS.370.1008M}
{Mandelbaum} R.,  {Hirata} C.~M.,  {Broderick} T.,  {Seljak} U.,    {Brinkmann}
  J.,  2006, \mnras, 370, 1008

\bibitem[\protect\citeauthoryear{{Mandelbaum}, {Hirata}, {Leauthaud}, {Massey}
  \& {Rhodes}}{{Mandelbaum} et~al.}{2012}]{2012MNRAS.420.1518M}
{Mandelbaum} R.,  {Hirata} C.~M.,  {Leauthaud} A.,  {Massey} R.~J.,    {Rhodes}
  J.,  2012, \mnras, 420, 1518

\bibitem[\protect\citeauthoryear{{Mandelbaum}, {Hirata}, {Seljak}, {Guzik},
  {Padmanabhan}, {Blake}, {Blanton}, {Lupton} \& {Brinkmann}}{{Mandelbaum}
  et~al.}{2005}]{2005MNRAS.361.1287M}
{Mandelbaum} R.,  {Hirata} C.~M.,  {Seljak} U.,  {Guzik} J.,  {Padmanabhan} N.,
   {Blake} C.,  {Blanton} M.~R.,  {Lupton} R.,    {Brinkmann} J.,  2005,
  \mnras, 361, 1287

\bibitem[\protect\citeauthoryear{{Mandelbaum}, {Seljak}, {Kauffmann}, {Hirata}
  \& {Brinkmann}}{{Mandelbaum} et~al.}{2006}]{Mandelbaum2006a}
{Mandelbaum} R.,  {Seljak} U.,  {Kauffmann} G.,  {Hirata} C.~M.,    {Brinkmann}
  J.,  2006, \mnras, 368, 715

\bibitem[\protect\citeauthoryear{{Mandelbaum}, {Slosar}, {Baldauf}, {Seljak},
  {Hirata}, {Nakajima}, {Reyes} \& {Smith}}{{Mandelbaum}
  et~al.}{2013}]{2013MNRAS.432.1544M}
{Mandelbaum} R.,  {Slosar} A.,  {Baldauf} T.,  {Seljak} U.,  {Hirata} C.~M.,
  {Nakajima} R.,  {Reyes} R.,    {Smith} R.~E.,  2013, \mnras, 432, 1544

\bibitem[\protect\citeauthoryear{{Mandelbaum}, {Wang}, {Zu}, {White},
  {Henriques} \& {More}}{{Mandelbaum} et~al.}{2015}]{2015M}
{Mandelbaum} R.,  {Wang} W.,  {Zu} Y.,  {White} S.,  {Henriques} B.,    {More}
  S.,  2015, ArXiv 1509.06762

\bibitem[\protect\citeauthoryear{{Marrone}, {Smith}, {Okabe}, {Bonamente},
  {Carlstrom}, {Culverhouse}, {Gralla} \& {Greer}}{{Marrone}
  et~al.}{2012}]{2012ApJ...754..119M}
{Marrone} D.~P.,  {Smith} G.~P.,  {Okabe} N.,  {Bonamente} M.,  {Carlstrom}
  J.~E.,  {Culverhouse} T.~L.,  {Gralla} M.,    {Greer} C.~H.,  2012, \apj,
  754, 119

\bibitem[\protect\citeauthoryear{{M{\'e}nard} \& {Fukugita}}{{M{\'e}nard} \&
  {Fukugita}}{2012}]{2012ApJ...754..116M}
{M{\'e}nard} B.,  {Fukugita} M.,  2012, \apj, 754, 116

\bibitem[\protect\citeauthoryear{{M{\'e}nard}, {Scranton}, {Fukugita} \&
  {Richards}}{{M{\'e}nard} et~al.}{2010}]{2010MNRAS.405.1025M}
{M{\'e}nard} B.,  {Scranton} R.,  {Fukugita} M.,    {Richards} G.,  2010,
  \mnras, 405, 1025

\bibitem[\protect\citeauthoryear{{Moster}, {Somerville}, {Maulbetsch}, {van den
  Bosch}, {Macci{\`o}}, {Naab} \& {Oser}}{{Moster}
  et~al.}{2010}]{2010ApJ...710..903M}
{Moster} B.~P.,  {Somerville} R.~S.,  {Maulbetsch} C.,  {van den Bosch} F.~C.,
  {Macci{\`o}} A.~V.,  {Naab} T.,    {Oser} L.,  2010, \apj, 710, 903

\bibitem[\protect\citeauthoryear{{Moustakas}, {Coil}, {Aird}, {Blanton},
  {Cool}, {Eisenstein}, {Mendez} \& {Wong}}{{Moustakas}
  et~al.}{2013}]{2013ApJ...767...50M}
{Moustakas} J.,  {Coil} A.~L.,  {Aird} J.,  {Blanton} M.~R.,  {Cool} R.~J.,
  {Eisenstein} D.~J.,  {Mendez} A.~J.,    {Wong} K.~C. e.~a.,  2013, \apj, 767,
  50

\bibitem[\protect\citeauthoryear{{Nakajima}, {Mandelbaum}, {Seljak}, {Cohn},
  {Reyes} \& {Cool}}{{Nakajima} et~al.}{2012}]{2012MNRAS.420.3240N}
{Nakajima} R.,  {Mandelbaum} R.,  {Seljak} U.,  {Cohn} J.~D.,  {Reyes} R.,
  {Cool} R.,  2012, \mnras, 420, 3240

\bibitem[\protect\citeauthoryear{{Old}, {Wojtak}, {Mamon}, {Skibba}, {Pearce},
  {Croton}, {Bamford} \& {Behroozi}}{{Old} et~al.}{2015}]{2015MNRAS.449.1897O}
{Old} L.,  {Wojtak} R.,  {Mamon} G.~A.,  {Skibba} R.~A.,  {Pearce} F.~R.,
  {Croton} D.,  {Bamford} S.,    {Behroozi} P.,  2015, \mnras, 449, 1897

\bibitem[\protect\citeauthoryear{{Padmanabhan}, {White}, {Norberg} \&
  {Porciani}}{{Padmanabhan} et~al.}{2009}]{2009MNRAS.397.1862P}
{Padmanabhan} N.,  {White} M.,  {Norberg} P.,    {Porciani} C.,  2009, \mnras,
  397, 1862

\bibitem[\protect\citeauthoryear{{Planck Collaboration}}{{Planck
  Collaboration}}{2011}]{2011A&A...536A..11P}
{Planck Collaboration} 2011, \aap, 536, A11

\bibitem[\protect\citeauthoryear{{Planck Collaboration}}{{Planck
  Collaboration}}{2013}]{2013A&A...557A..52P}
{Planck Collaboration} 2013, \aap, 557, A52

\bibitem[\protect\citeauthoryear{{Planck Collaboration}}{{Planck
  Collaboration}}{2014a}]{2014A&A...571A..16P}
{Planck Collaboration} 2014a, \aap, 571, A16

\bibitem[\protect\citeauthoryear{{Planck Collaboration}}{{Planck
  Collaboration}}{2014b}]{2014A&A...571A..20P}
{Planck Collaboration} 2014b, \aap, 571, A20

\bibitem[\protect\citeauthoryear{{Planck Collaboration}}{{Planck
  Collaboration}}{2015}]{2015arXiv150201589P}
{Planck Collaboration} 2015, ArXiv 1502.01589

\bibitem[\protect\citeauthoryear{{Reyes}, {Mandelbaum}, {Gunn}, {Nakajima},
  {Seljak} \& {Hirata}}{{Reyes} et~al.}{2012}]{2012MNRAS.425.2610R}
{Reyes} R.,  {Mandelbaum} R.,  {Gunn} J.~E.,  {Nakajima} R.,  {Seljak} U.,
  {Hirata} C.~M.,  2012, \mnras, 425, 2610

\bibitem[\protect\citeauthoryear{{Reyes}, {Mandelbaum}, {Hirata}, {Bahcall} \&
  {Seljak}}{{Reyes} et~al.}{2008}]{2008MNRAS.390.1157R}
{Reyes} R.,  {Mandelbaum} R.,  {Hirata} C.,  {Bahcall} N.,    {Seljak} U.,
  2008, \mnras, 390, 1157

\bibitem[\protect\citeauthoryear{{Rozo}, {Wechsler}, {Rykoff}, {Annis},
  {Becker}, {Evrard}, {Frieman}, {Hansen}, {Hao}, {Johnston}, {Koester},
  {McKay}, {Sheldon} \& {Weinberg}}{{Rozo} et~al.}{2010}]{2010ApJ...708..645R}
{Rozo} E.,  {Wechsler} R.~H.,  {Rykoff} E.~S.,  {Annis} J.~T.,  {Becker} M.~R.,
   {Evrard} A.~E.,  {Frieman} J.~A.,  {Hansen} S.~M.,  {Hao} J.,  {Johnston}
  D.~E.,  {Koester} B.~P.,  {McKay} T.~A.,  {Sheldon} E.~S.,    {Weinberg}
  D.~H.,  2010, \apj, 708, 645

\bibitem[\protect\citeauthoryear{{Rykoff}, {Evrard}, {McKay}, {Becker},
  {Johnston}, {Koester}, {Nord}, {Rozo}, {Sheldon}, {Stanek} \&
  {Wechsler}}{{Rykoff} et~al.}{2008}]{2008MNRAS.387L..28R}
{Rykoff} E.~S.,  {Evrard} A.~E.,  {McKay} T.~A.,  {Becker} M.~R.,  {Johnston}
  D.~E.,  {Koester} B.~P.,  {Nord} B.,  {Rozo} E.,  {Sheldon} E.~S.,  {Stanek}
  R.,    {Wechsler} R.~H.,  2008, \mnras, 387, L28

\bibitem[\protect\citeauthoryear{{Salim}, {Rich}, {Charlot}, {Brinchmann},
  {Johnson}, {Schiminovich}, {Seibert} \& {Mallery}}{{Salim}
  et~al.}{2007}]{2007ApJS..173..267S}
{Salim} S.,  {Rich} R.~M.,  {Charlot} S.,  {Brinchmann} J.,  {Johnson} B.~D.,
  {Schiminovich} D.,  {Seibert} M.,    {Mallery} R. e.~a.,  2007, \apjs, 173,
  267

\bibitem[\protect\citeauthoryear{{Schaye}, {Crain}, {Bower}, {Furlong},
  {Schaller}, {Theuns}, {Dalla Vecchia} \& {Frenk}}{{Schaye}
  et~al.}{2015}]{2015MNRAS.446..521S}
{Schaye} J.,  {Crain} R.~A.,  {Bower} R.~G.,  {Furlong} M.,  {Schaller} M.,
  {Theuns} T.,  {Dalla Vecchia} C.,    {Frenk} 2015, \mnras, 446, 521

\bibitem[\protect\citeauthoryear{{Spergel}, {Verde}, {Peiris}, {Komatsu},
  {Nolta}, {Bennett}, {Halpern}, {Hinshaw}, {Jarosik}, {Kogut}, {Limon},
  {Meyer}, {Page}, {Tucker}, {Weiland}, {Wollack} \& {Wright}}{{Spergel}
  et~al.}{2003}]{2003ApJS..148..175S}
{Spergel} D.~N.,  {Verde} L.,  {Peiris} H.~V.,  {Komatsu} E.,  {Nolta} M.~R.,
  {Bennett} C.~L.,  {Halpern} M.,  {Hinshaw} G.,  {Jarosik} N.,  {Kogut} A.,
  {Limon} M.,  {Meyer} S.~S.,  {Page} L.,  {Tucker} G.~S.,  {Weiland} J.~L.,
  {Wollack} E.,    {Wright} E.~L.,  2003, \apjs, 148, 175

\bibitem[\protect\citeauthoryear{{Springel}, {White}, {Jenkins}, {Frenk},
  {Yoshida}, {Gao}, {Navarro}, {Thacker}, {Croton}, {Helly}, {Peacock}, {Cole},
  {Thomas}, {Couchman}, {Evrard}, {Colberg} \& {Pearce}}{{Springel}
  et~al.}{2005}]{2005Natur.435..629S}
{Springel} V.,  {White} S.~D.~M.,  {Jenkins} A.,  {Frenk} C.~S.,  {Yoshida} N.,
   {Gao} L.,  {Navarro} J.,  {Thacker} R.,  {Croton} D.,  {Helly} J.,
  {Peacock} J.~A.,  {Cole} S.,  {Thomas} P.,  {Couchman} H.,  {Evrard} A.,
  {Colberg} J.,    {Pearce} F.,  2005, \nat, 435, 629

\bibitem[\protect\citeauthoryear{{van den Bosch}, {More}, {Cacciato}, {Mo} \&
  {Yang}}{{van den Bosch} et~al.}{2013}]{2013MNRAS.430..725V}
{van den Bosch} F.~C.,  {More} S.,  {Cacciato} M.,  {Mo} H.,    {Yang} X.,
  2013, \mnras, 430, 725

\bibitem[\protect\citeauthoryear{{Vikhlinin}, {Burenin}, {Ebeling}, {Forman},
  {Hornstrup}, {Jones}, {Kravtsov}, {Murray}, {Nagai}, {Quintana} \&
  {Voevodkin}}{{Vikhlinin} et~al.}{2009}]{2009ApJ...692.1033V}
{Vikhlinin} A.,  {Burenin} R.~A.,  {Ebeling} H.,  {Forman} W.~R.,  {Hornstrup}
  A.,  {Jones} C.,  {Kravtsov} A.~V.,  {Murray} S.~S.,  {Nagai} D.,  {Quintana}
  H.,    {Voevodkin} A.,  2009, \apj, 692, 1033

\bibitem[\protect\citeauthoryear{{Vogelsberger}, {Genel}, {Springel}, {Torrey},
  {Sijacki}, {Xu}, {Snyder} \& {Nelson}}{{Vogelsberger}
  et~al.}{2014}]{2014MNRAS.444.1518V}
{Vogelsberger} M.,  {Genel} S.,  {Springel} V.,  {Torrey} P.,  {Sijacki} D.,
  {Xu} D.,  {Snyder} G.,    {Nelson} D.,  2014, \mnras, 444, 1518

\bibitem[\protect\citeauthoryear{{Wang} \& {Jing}}{{Wang} \&
  {Jing}}{2010}]{2010MNRAS.402.1796W}
{Wang} L.,  {Jing} Y.~P.,  2010, \mnras, 402, 1796

\bibitem[\protect\citeauthoryear{{Wang}, {Yang}, {Shen}, {Mo}, {van den Bosch},
  {Luo}, {Wang}, {Lau}, {Wang}, {Kang} \& {Li}}{{Wang}
  et~al.}{2014}]{2014MNRAS.439..611W}
{Wang} L.,  {Yang} X.,  {Shen} S.,  {Mo} H.~J.,  {van den Bosch} F.~C.,  {Luo}
  W.,  {Wang} Y.,  {Lau} E.~T.,  {Wang} Q.~D.,  {Kang} X.,    {Li} R.,  2014,
  \mnras, 439, 611

\bibitem[\protect\citeauthoryear{{Wang}, {Sales}, {Henriques} \&
  {White}}{{Wang} et~al.}{2014}]{2014MNRAS.442.1363W}
{Wang} W.,  {Sales} L.~V.,  {Henriques} B.~M.~B.,    {White} S.~D.~M.,  2014,
  \mnras, 442, 1363

\bibitem[\protect\citeauthoryear{{Wang} \& {White}}{{Wang} \&
  {White}}{2012}]{2012MNRAS.424.2574W}
{Wang} W.,  {White} S.~D.~M.,  2012, \mnras, 424, 2574

\bibitem[\protect\citeauthoryear{{Zehavi}, {Zheng}, {Weinberg}, {Blanton},
  {Bahcall}, {Berlind}, {Brinkmann}, {Frieman}, {Gunn}, {Lupton}, {Nichol},
  {Percival}, {Schneider}, {Skibba}, {Strauss}, {Tegmark} \& {York}}{{Zehavi}
  et~al.}{2011}]{2011ApJ...736...59Z}
{Zehavi} I.,  {Zheng} Z.,  {Weinberg} D.~H.,  {Blanton} M.~R.,  {Bahcall}
  N.~A.,  {Berlind} A.~A.,  {Brinkmann} J.,  {Frieman} J.~A.,  {Gunn} J.~E.,
  {Lupton} R.~H.,  {Nichol} R.~C.,  {Percival} W.~J.,  {Schneider} D.~P.,
  {Skibba} R.~A.,  {Strauss} M.~A.,  {Tegmark} M.,    {York} D.~G.,  2011,
  \apj, 736, 59

\bibitem[\protect\citeauthoryear{{Zu} \& {Mandelbaum}}{{Zu} \&
  {Mandelbaum}}{2015a}]{2015MNRAS.454.1161Z}
{Zu} Y.,  {Mandelbaum} R.,  2015a, \mnras, 454, 1161

\bibitem[\protect\citeauthoryear{{Zu} \& {Mandelbaum}}{{Zu} \&
  {Mandelbaum}}{2015b}]{ZandM2015b}
{Zu} Y.,  {Mandelbaum} R.,  2015b, ArXiv 1509.06758

\end{thebibliography}


\renewcommand{\theequation}{A-\arabic{equation}}
\setcounter{equation}{0}  
\section*{APPENDIX}  

This appendix describes how we fit the power-law scaling relations of
Equations~\ref{eqn:SZscaling} and~\ref{eqn:Xrayscaling} to the
recalibrated data shown as red points with both horizontal and
vertical error bars in the upper panels of Fig.\ref{fig:scaling}. This
fitting problem is difficult because of the need to deal with
uncertainties on both variables which vary from point to point, and
are strongly correlated between neighbouring points. Since the
recalibrated scaling relations are the principal quantitative result
of our paper, we considered it important to devise a well defined and
rigorous statistical procedure to derive them. For simplicity, the
following exposition considers the SZ case and refers to the two
variables as $Y$ and $M$ rather than as $Y_{500}$ and $M_{500}.$ The
X-ray case can be treated in an exactly analogous way. The method
we propose is close to those developed by \cite{2007ApJ...665.1489K}.

The data to be fitted consist of twelve estimates, $(Y_i, M_i)$, one
for each of the twelve stellar mass bins we consider. The $Y_i$ values
come directly from P13, whereas the P13 $M_i$ values are recalibrated
using the correction indicated by a solid black line in
Fig.~\ref{fig:correction} (right hand panel). Let us denote 
as $y_i$ and $m_i$ the true mean SZ flux and the true effective halo 
mass for the population of galaxies with stellar masses in the range 
of bin $i$.  Our stochastic model may be written as
\begin{equation}
Y_i = y_i + \delta_i
\end{equation}
and
\begin{equation}
\log M_i = \log m_i + \epsilon_i + \eta_i, 
\end{equation}
where $\delta_i$, $\epsilon_i$ and $\eta_i$ are zero-mean,
Gaussian deviates. The first, $\delta_i$, reflects the observational
uncertainty in the stacked SZ flux that P13 measured for bin $i$, and
should be uncorrelated both between bins and with uncertainties in
effective halo mass. Hence
\begin{equation}
\langle\delta_i\delta_j\rangle \equiv C_{y,ij},~~\langle\delta_i\epsilon_j\rangle=
\langle\delta_i\eta_j\rangle = 0,
\end{equation}
where the covariance matrix $C_{y,ij}$ is diagonal with elements given
by the (square of the) bootstrap errors estimated by P13. We take the
distribution of SZ flux to be Gaussian in $Y$ rather than in $\log Y$ in
order to deal with lower signal-to-noise bins where the estimated flux
turns out to be negative in one case.

In contrast, for the effective mass it is more natural to assume the
uncertainties to be Gaussian in $\log M$. We distinguish two noise
sources.  The larger, $\epsilon_i$, reflects uncertainty in the
correction from the effective mass estimated by P13 to that implied by
our lensing measurements. The twelve $\epsilon_i$ are thus linearly
related to the seven $\theta_\alpha$ which represent the differences
between the seven correction estimates of Fig.~\ref{fig:correction}
and their true population values. These $\theta_\alpha$ represent both
systematic modelling uncertainties and observational noise in the
lensing data. They can be taken to be independent, zero-mean, Gaussian
variates with variances equal to the (squares of the) error bars in
Fig.~\ref{fig:correction}.  However, the $\epsilon_i =
D_{i\alpha}\theta_\alpha$ will have a nondiagonal (and indeed,
singular) covariance matrix.

Finally, the $\eta_i$ are introduced as a mathematical device to
make $C_{m,ij}$ well behaved and invertible. Additionally, the $\eta_i$ 
can be thought of as accounting for errors resulting from the 
interpolation of correction factors derived for our relatively broad 
lensing stellar mass bins to the narrower stellar mass bins used by P13 
and A15. This is expected to induce relatively small additional
errors, so for simplicity, we just take all the $\eta_i$ to be independent, 
zero-mean Gaussian variates with {\it rms} $\sigma=$~0.03dex. 
With these assumptions we have
\begin{equation}
\langle\epsilon_i\epsilon_j\rangle \equiv C_{m,ij},~~
\langle\eta_i\eta_j\rangle = \sigma^2 I_{ij},~~
\langle\epsilon_i\eta_j\rangle=0,
\end{equation}
where $I_{ij}$ is the identity matrix. We can then define a nonsingular
covariance matrix
\begin{equation}
C^\prime_{m,ij} \equiv \langle(\log M_i - \log m_i)(\log M_j - \log m_j)\rangle
 = C_{m,ij} + \sigma^2 I_{ij}.
\end{equation}

For a power-law scaling relation of the form
\begin{equation}
\log y = a\log m + b,
\end{equation}
we can now write the likelihood of a specific model (specified by $a$,
$b$ and the twelve $m_i$) given the data [the twelve $(M_i, Y_i)$]
as that corresponding to a multivariate Gaussian
\begin{eqnarray}
 2 \log {\cal L} =  - (\log M_i - \log m_i)C^{\prime -1}_{m,ij}(\log M_j- \log m_j)\\
  - (Y_i -y_i)C^{-1}_{y,ij}(Y_j-y_j),\nonumber
\end{eqnarray}
where $\log y_i = a\log m_i - b$, the $m_i$ can be thought of as
nuisance parameters, and we have dropped a constant which is
independent of the model parameters. Note that it was important to
include the additional stochastic element represented by the $\eta_i$
in order to ensure that the inverse covariance matrix in the first
term is well defined.  This likelihood function can be fed to a Monte
Carlo Markov Chain program in order to explore the likelihood surface,
and, in particular, to project it onto the $(a,b)$ plane where we can
identify a best-fit (i.e. maximum likelihood) scaling relation and
define confidence intervals for its parameters.
\end{document}